\documentclass[useAMS,usenatbib,usegraphicx]{mn2e}

\usepackage{times}
\usepackage{totpages}
\usepackage{amsmath}
\usepackage{amssymb}
\usepackage{xcolor}

\newcommand{\myemail}{chales@aoc.nrao.edu}

\newcommand{\farcsecd}{\mbox{\ensuremath{.\!\!^{\prime\prime}}}}

\newcommand{\degree}{\ensuremath{^\circ}}
\newcommand{\ms}{\scriptscriptstyle}
\newcommand{\tnm}{\textnormal}
\newcommand{\trm}{\textrm}
\renewcommand{\S}{Section}

\title[PROPERTIES OF THE POLARIZED 1.4~GHZ SKY]{ATLAS 1.4~GHz Data Release 2 -- II.
Properties of the faint polarized sky}
\author[HALES ET Al.]{
C.~A. Hales,$^{1,2}$\thanks{E-mail: \myemail}\thanks{Current address:
National Radio Astronomy Observatory, P.O. Box 0, Socorro, NM 87801,
USA; Jansky Fellow of the National Radio Astronomy Observatory.}
R. P. Norris,$^{2,3}$
B. M. Gaensler,$^{1,3}$
and
E. Middelberg$^{4}$\\
$^{1}$Sydney Institute for Astronomy, School of Physics,
The University of Sydney, NSW 2006, Australia\\
$^{2}$Australia Telescope National Facility, CSIRO Astronomy and Space Science,
P.O. Box 76, Epping, NSW 1710, Australia\\
$^{3}$ARC Centre of Excellence for All-sky Astrophysics (CAASTRO)\\
$^{4}$Astronomisches Institut, Ruhr-Universit\"{a}t, Universit\"{a}tsstr. 150,
44801 Bochum, Germany
}
\begin{document}

%\date{Accepted 1988 December 15. Received 1988 December 14; in original form 1988 October 11}
\date{Draft version \today}

\pagerange{\pageref{firstpage}--\ref{TotPages}} \pubyear{\the\year}
% note that \pageref{lastpage} won't count float pages beyond the last page of text

\maketitle

\label{firstpage}

\begin{abstract}
This is the second of two papers describing the second data release (DR2) of the
Australia Telescope Large Area Survey (ATLAS) at 1.4~GHz. In Paper I we detailed our
data reduction and analysis procedures, and presented catalogues of
components (discrete regions of radio emission) and sources (groups of physically
associated radio components). In this paper we present our key observational results.
We find that the 1.4~GHz Euclidean normalised differential number counts for ATLAS components
exhibit monotonic declines in both total intensity and linear polarization from millijansky levels
down to the survey limit of $\sim100$~$\mu$Jy. We discuss the parameter space in which component
counts may suitably proxy source counts. We do not detect any components or sources
with fractional polarization levels greater than 24\%. The ATLAS data are consistent with a
lognormal distribution of fractional polarization with median level 4\% that is independent of
flux density down to total intensity $\sim10$~mJy and perhaps even 1~mJy. Each of these
findings are in contrast to previous studies; we attribute these new results to improved data
analysis procedures. We find that polarized emission from 1.4~GHz millijansky sources originates
from the jets or lobes of extended sources that are powered by an active galactic nucleus,
consistent with previous findings in the literature. We provide estimates for the sky density
of linearly polarized components and sources in 1.4~GHz surveys with $\sim10\arcsec$ resolution.
\end{abstract}


\section{Introduction}\label{sec:1}

A number of studies have reported an anti-correlation between fractional linear
polarization and total intensity flux density for extragalactic 1.4~GHz sources;
faint sources were found to be more highly polarized \citep{2002A&A...396..463M,
2004MNRAS.349.1267T,2007ApJ...666..201T,2010ApJ...714.1689G,2010MNRAS.402.2792S}.
As a result, the Euclidean-normalised differential number-counts of polarized sources
have been observed to flatten at linearly polarized flux densities
$L$~{\footnotesize $\lesssim$}~1~mJy to levels greater than those expected from
convolving the known total intensity source counts with plausible distributions
for fractional polarization \citep{2008evn..confE.107O}. The flattening suggests
that faint polarized sources may exhibit more highly ordered magnetic fields than
bright sources, or may instead suggest the emergence of an unexpected faint population.
The anti-correlation trend for fractional linear polarization is not observed
at higher frequencies \citep[$\ge4.8$~GHz;][]{2011ApJ...732...45S,2011MNRAS.413..132B,2013MNRAS.436.2915M}.

To investigate possible explanations for the fractional polarization trend seen in previous studies,
we have produced the second data release of the Australia Telescope Large Area Survey (ATLAS DR2)
as described in Paper I \citep{halesPI} of this two paper series. ATLAS DR2 comprises
reprocessed and new 1.4~GHz observations with the Australia Telescope Compact Array (ATCA)
about the {\it Chandra} Deep Field-South \citep[CDF-S; Galactic coordinates $l\approx224\degree$,
$b\approx-55\degree$;][]{2006AJ....132.2409N} and European Large Area {\it Infrared Space Observatory}
Survey-South 1 \citep[ELAIS-S1; $l\approx314\degree$,
$b\approx-73\degree$;][]{2008AJ....135.1276M} regions in total intensity, linear polarization, and
circular polarization. The mosaicked multi-pointing survey areas for ATLAS DR2 are 3.626~deg$^2$
and 2.766~deg$^2$ for the CDF-S and ELAIS-S1 regions, respectively, imaged at approximately
$12\arcsec\times6\arcsec$ resolution. Typical source detection thresholds are 200~$\mu$Jy
in total intensity and polarization.

In Paper I we presented our data reduction and analysis prescriptions for ATLAS DR2.
We presented a catalogue of components (discrete regions of radio emission) comprising
2416 detections in total intensity and 172 independent detections in linear polarization.
No components were detected in circular polarization. We presented a catalogue of 2221
sources (groups of physically associated radio components; grouping scheme based on
total intensity properties alone, as described below), of which 130 were found to exhibit
linearly polarized emission. We described procedures to account for instrumental and
observational effects, including spatial variations in each of image sensitivity,
bandwidth smearing with a non-circular beam, and instrumental polarization leakage, clean bias,
the division between peak and integrated flux densities for unresolved and resolved components,
and noise biases in both total intensity and linear polarization. Analytic correction schemes were
developed to account for incompleteness in differential component number counts due to resolution
and Eddington biases. We cross-identified and classified sources according to two schemes, summarized
as follows.

In the first scheme, described in \S~6.1 of Paper I, we grouped total intensity radio components
into sources, associated these with infrared sources from the {\it Spitzer} Wide-Area Infrared
Extragalactic Survey \citep[SWIRE;][]{2003PASP..115..897L} and optical sources from \citet{2012MNRAS.426.3334M},
then classified them according to whether their energetics were likely to be driven by an active galactic
nucleus (AGN), star formation (SF) within a star-forming galaxy (SFG), or a radio star. Due to the limited
angular resolution of the ATLAS data, in Paper I we adopted the term {\it lobe} to describe both jets and
lobes in sources with radio double or triple morphologies. The term {\it core} was similarly defined in
a generic manner to indicate the central component in a radio triple source. Under this
terminology, a core does not indicate a compact, flat-spectrum region of emission; restarted
AGN jets or lobes may contribute or even dominate the emission observed in the regions we have
designated as cores. AGNs were identified using four selection criteria: radio morphologies,
24~$\mu$m to 1.4~GHz flux density ratios, mid-infrared colours, and optical spectral
characteristics. SFGs and stars were identified solely by their optical spectra. Of the 2221
ATLAS DR2 sources, 1169 were classified as AGNs, 126 as SFGs, and 4 as radio stars. We note
that our classification system was biased in favour of AGNs. As a result, the ATLAS DR2 data
are in general unsuited for statistical comparisons between star formation and AGN activity. 

In the second scheme, described in \S~6.2 of Paper I, we associated linearly polarized components,
or polarization upper limits, with total intensity counterparts. In most cases it was possible to
match a single linearly polarized component with a single total intensity component, forming a
one-to-one match. In other cases this was not possible, due to ambiguities posed by the blending
of adjacent components; for example, a polarized component situated mid-way between two
closely-separated total intensity components. In these cases, we formed group associations to
avoid biasing measurements of fractional polarization. We classified the polarization--total
intensity associations according to the following scheme, which we designed to account for
differing (de-)polarized morphologies (see Paper~I for graphical examples):
\begin{itemize}

\item[] \noindent {\it Type 0} -- A one-to-one or group association identified as a lobe
of a double or triple radio source. Both lobes of the source are clearly
polarized, having linearly polarized flux densities within a factor of 3.
(The ratio between lobe total intensity flux densities was found to be
within a factor of 3 for all double or triple ATLAS DR2 sources.)

\item[] \noindent {\it Types 1/2} -- A one-to-one or group association identified as a
lobe of a double or triple radio source that does not meet the criteria for Type 0.
A lobe classified as Type 1 indicates that the ratio of polarized flux densities
between lobes is greater than 3. A lobe classified as Type 2 indicates that the
opposing lobe is undetected in polarization and that the polarization ratio may
be less than 3, in which case it is possible that more sensitive observations
may lead to re-classification as Type 0. Sources with lobes classified as Type 1
exhibit asymmetric depolarization in a manner qualitatively consistent with the
Laing-Garrington effect \citep{1988Natur.331..149L,1988Natur.331..147G}, where
one lobe appears more fractionally polarized than the opposite lobe.

\item[] \noindent {\it Type 3} -- A group association representing a source, involving
a linearly polarized component situated midway between two total intensity components.
It is not clear whether such associations represent two polarized lobes, a polarized
lobe adjacent to a depolarized lobe, or a polarized core.

\item[] \noindent {\it Type 4} -- An unclassified one-to-one or group association
representing a source.

\item[] \noindent {\it Type 5} -- A one-to-one association clearly identified as the
core of a triple radio source (where outer lobes are clearly distinct from
the core).

\item[] \noindent {\it Type 6} -- A source comprising two Type 0 associations, or
a group association representing a non-depolarized double or triple radio source
where blended total intensity and linear polarization components have prevented
clear subdivision into two Type 0 associations.

\item[] \noindent {\it Type 7} -- A source comprising one or two Type 1 associations.

\item[] \noindent {\it Type 8} -- A source comprising one Type 2 association.

\item[] \noindent {\it Type 9} -- An unpolarized component or source.

\end{itemize}

In this work (Paper II) we present the key observational results from ATLAS DR2, with particular
focus on the nature of faint polarized sources. This paper is organised as follows. In \S~\ref{ch5:SecRes}
we present the ATLAS DR2 source diagnostics resulting from infrared and optical cross-identifications
and classifications, diagnostics resulting from polarization--total intensity cross-identifications
and classifications, differential component number-counts, and our model for the distribution of
fractional polarization. In \S~\ref{ch5:SecDisc} we compare the ATLAS DR2 differential counts in
both total intensity and linear polarization to those from other 1.4~GHz surveys, and discuss
asymmetric depolarization of classical double radio sources. We present our conclusions in
\S~\ref{ch5:SecConc}. This paper follows the notation introduced in Paper I. We typically denote
flux density by $S$, but split into $I$ for total intensity and $L$ for linearly polarized flux
density when needed for clarity.

\section{Results}\label{ch5:SecRes}

\subsection{Multiwavelength Cross-Identifications and Classifications}

In the following sections we present diagnostics of ATLAS DR2 sources resulting from the infrared
and optical cross-identification and classification schemes described in \S~6.1 of Paper I
(summarised in \S~\ref{sec:1} of this work). We focus on three parameter spaces formed by comparing
flux densities between different wavelength bands: radio to mid-infrared, mid-infrared colours,
and radio to far-infrared.

\subsubsection{Radio vs Mid-Infrared}

\begin{figure*}
 \centering
 \includegraphics[trim = 0mm 0mm 0mm 0mm, clip, angle=-90, width=0.75\textwidth]{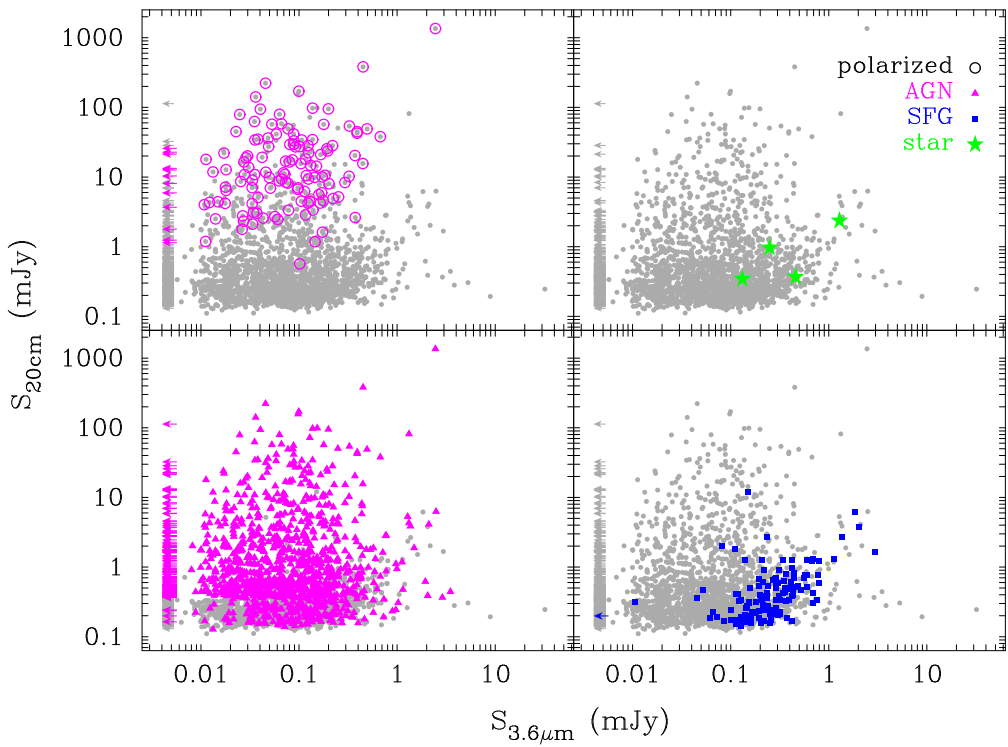}
 \caption{	Comparison of radio to mid-infrared flux densities for all ATLAS DR2
 		sources, as indicated by the grey points and grey upper bounds in each
		panel. For clarity, multiple panels are used to focus on different source
		populations. In each panel, the highlighted population includes those
		sources represented by upper bounds. The
		bottom-left panel highlights all sources classified as AGNs in ATLAS
		DR2. The bottom-right panel highlights all sources classified as SFGs.
		The top-left panel highlights all sources exhibiting linearly
		polarized emission, where the colour of the highlighted points and
		upper bounds indicates whether the source is an AGN, SFG, or star
		(i.e. they are all AGNs). The top-right panel highlights all sources
		identified as stars.
	}
 \label{ch5:fig:rNIR}
\end{figure*}

In Fig.~\ref{ch5:fig:rNIR} we compare the total intensity 1.4~GHz radio to 3.6~$\mu$m
mid-infrared flux densities for all 2221 ATLAS DR2 sources, taking into account
infrared upper bounds for the 298 radio sources without detected infrared counterparts.
The bottom-right panel of Fig.~\ref{ch5:fig:rNIR} indicates that the ATLAS sources
classified as stars or SFGs typically exhibit radio flux
densities {\footnotesize $\lesssim$}~1~mJy. The paucity of ATLAS sources with
$S_{\tnm{\scriptsize 3.6}\mu\tnm{\scriptsize m}}$~{\footnotesize $\lesssim$}~0.1~mJy
and star or SFG classifications likely represents a selection bias, in which only
those sources with relatively bright optical counterparts could be classified
spectroscopically. The top-left panel highlights all 130 polarized ATLAS sources,
12 of which are represented by upper bounds. The paucity of polarized sources
with $S_\tnm{\scriptsize 20cm}$~{\footnotesize $\lesssim$}~1~mJy is due to the
limited sensitivity of our linear polarization data; fractional polarization trends
will be presented in \S~\ref{ch5:SecResIdentM}.

\subsubsection{Mid-Infrared Colours}\label{ch5:SecRes:2}

\begin{figure*}
 \centering
 \includegraphics[trim = 0mm 0mm 0mm 0mm, clip, angle=-90, width=0.75\textwidth]{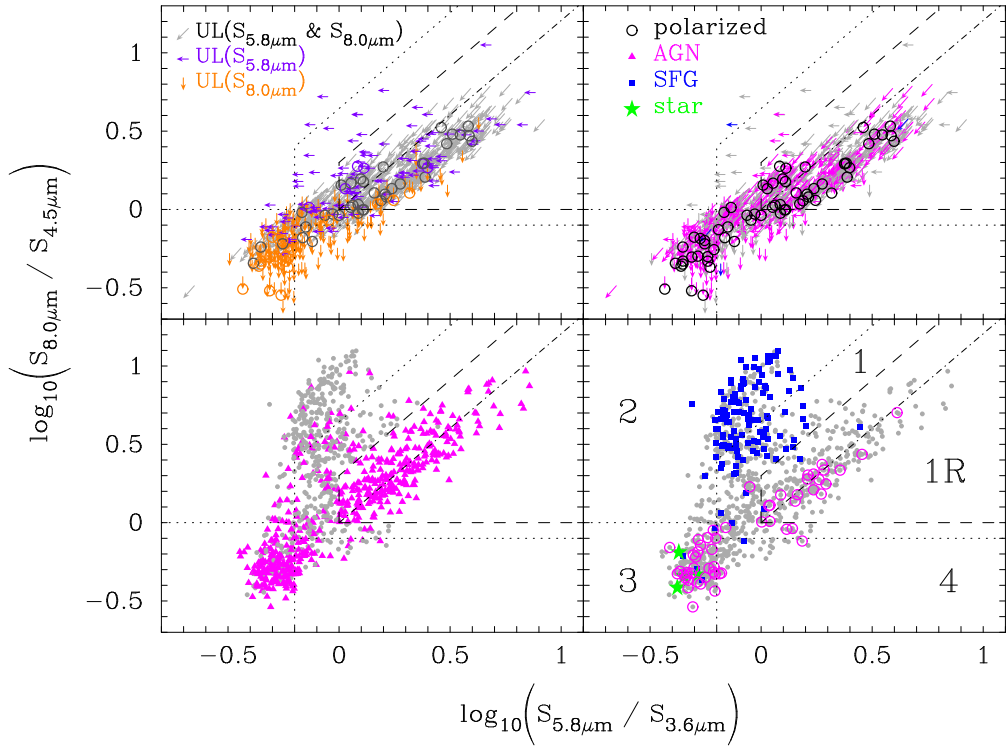}
 \caption{	Mid-infrared colour-colour diagrams in which ATLAS DR2 sources with
 		an infrared counterpart detected in at least two IRAC bands are displayed
		(84\% of all ATLAS DR2 sources). Radio sources with
		counterparts detected in all four IRAC bands are displayed in the lower
		panels. Those with infrared counterparts detected in only two or three
		IRAC bands are displayed in the upper panels. Data are replicated within
		the panels in each row to highlight different source populations. In each
		panel except the top-left, grey points or grey upper bounds indicate
		unclassified sources. The bottom-left panel highlights all sources
		classified as AGNs. The bottom-right panel highlights all sources
		classified as SFGs, stars, and those exhibiting polarized emission;
		each polarized source has been coloured according to its AGN/SFG/star
		classification. The top-right panel displays upper bounds, colour-coded
		according to their AGN/SFG/star classification. The 63 polarized
		sources indicated within this panel are all classified as
		AGNs, but shown black in this panel for visual contrast. The top-left
		panel shows upper bounds, colour-coded according to the non-detected
		bands. The dotted lines in each panel outline the four regions
		identified by \citet{2005ApJ...621..256S}, as described in
		\S~\ref{ch5:SecRes:2}. The dashed lines outlining region 1R indicate
		the restricted locus adopted in this parameter space for classifying
		AGNs, as described in \S~\ref{ch5:SecRes:2}. The dot-dashed
		line in each panel indicates the locus of sources exhibiting
		red power-law spectra over the four IRAC bands.
}
 \label{ch5:fig:NIRCC}
\end{figure*}

In Fig.~\ref{ch5:fig:NIRCC} we present mid-infrared colour-colour diagrams
in which the IRAC flux density ratios
$S_{8.0\,\mu\trm{\scriptsize m}}/S_{4.5\,\mu\trm{\scriptsize m}}$ and
$S_{5.8\,\mu\trm{\scriptsize m}}/S_{3.6\,\mu\trm{\scriptsize m}}$ have been
compared for ATLAS DR2 sources.
Of the 2221 ATLAS sources, 988 were detected in all four IRAC bands, while
935 were detected in only 2 or 3 bands; the remaining 298 sources were not detected
in any band, and have not been shown in Fig.~\ref{ch5:fig:NIRCC}. Regarding the 130
polarized ATLAS sources, 55 were detected in all four IRAC bands, 63 were
detected in only 2 or 3 bands, and 12 were not detected in any IRAC band; thus
118 polarized sources are indicated in Fig.~\ref{ch5:fig:NIRCC}.

The dotted lines in each panel of
Fig.~\ref{ch5:fig:NIRCC} represent the divisions identified through simulations by
\citet{2005ApJ...621..256S}. By considering the effects of redshift evolution on
the observed mid-infrared colours of three general source classes with spectral
characteristics dominated by old-population (10~Gyr) starlight, polycyclic aromatic hydrocarbon (PAH)
emission, or a power-law continuum, \citet{2005ApJ...621..256S} identified four regions
that could be used to preferentially select different source classes at
different redshifts within the $z=0-2$ range simulated.
Region 1 was found to preferentially select sources with spectra dominated by continuum
emission, likely produced by dust tori associated with AGNs
\citep{1993ApJ...418..673P,2008ApJ...685..147N}, over the full redshift range.
Region 2 was found to preferentially select PAH-dominated sources, indicative of
intense star formation, over the full redshift range. Region 3 was found to
preferentially select sources with spectra dominated by direct stellar light, but only for
sources with $z$~{\footnotesize $\lesssim$}~0.4. For increasing redshifts, region 3
was found to comprise a mixture of stellar- and PAH-dominated sources as the
latter migrated from region 2. However, beyond $z$~{\footnotesize $\gtrsim$}~1.6,
region 3 was found to be largely void of sources following the migrations
of both stellar- and PAH-dominated sources towards region 4. Region 4 was found
to be largely void of sources for $z$~{\footnotesize $\lesssim$}~0.4. For
increasing redshifts, PAH-dominated sources were found to migrate from region 2
into region 4. For $z$~{\footnotesize $\gtrsim$}~0.9, stellar-dominated sources
from region 3 were also found to migrate into region 4. \citet{2005ApJ...621..256S}
found that at all redshifts, sources dominated by PAH emission were located
slightly within the boundaries of region 1. In order to classify as AGNs only
those sources most likely to be such in Fig.~\ref{ch5:fig:NIRCC}, we constructed
the restricted locus indicated by the dashed lines; we label this region 1R. In
Paper~I we defined this locus \citep[following][]{2011ApJ...740...20P} as the union of
$\log_{\ms 10}[S_{8.0\,\mu\trm{\scriptsize m}}/S_{4.5\,\mu\trm{\scriptsize m}}]>0$,
$\log_{\ms 10}[S_{5.8\,\mu\trm{\scriptsize m}}/S_{3.6\,\mu\trm{\scriptsize m}}]>0$,
and $\log_{\ms 10}[S_{8.0\,\mu\trm{\scriptsize m}}/S_{4.5\,\mu\trm{\scriptsize m}}]<
11\log_{\ms 10}[S_{5.8\,\mu\trm{\scriptsize m}}/S_{3.6\,\mu\trm{\scriptsize m}}]/9+0.3$.
Continuum-dominated sources are expected to exhibit power-law spectra,
given by the dot-dashed locus in each panel. As noted by
\citet{2005ApJ...621..256S}, the spectra of continuum-dominated sources
are only expected to exhibit blue IRAC colours for largely unobscured AGNs,
in cases for which their rest-frame optical wavelengths are redshifted into the
mid-infrared band for sources at $z>2$. Thus the ATLAS sources with blue IRAC
colours in Fig.~\ref{ch5:fig:NIRCC} are unlikely to be
represented by continuum-dominated sources as defined by \citet{2005ApJ...621..256S}.
However, this does not imply that a source observed with blue IRAC colours at
$z<2$ cannot be an AGN, because sources with mid-infrared spectra dominated
by old stellar light may yet exhibit stronger signs of AGN activity at other
wavelengths.

The bottom-left panel of Fig.~\ref{ch5:fig:NIRCC} indicates that ATLAS sources
classified as AGNs are predominantly located in regions 1R and 3. The sources
classified as AGNs in region 2 perhaps suggest combinations of star formation
and AGN activity, or perhaps misclassifications due to the
largely statistical nature of our classification system. The upper bounds
classified as AGNs in the top-right panel are consistent with the observed
distribution of AGNs presented in the bottom-left panel. These upper bounds
suggest that additional AGNs are situated in region 4, though likely in proportion
with the additional AGNs remaining undetected in regions 1 and 3. The
bottom-right panel indicates that ATLAS sources classified as SFGs are
predominantly located in region 2, as well as between the boundaries of regions
1 and 1R, as expected. A small number of ATLAS sources classified as SFGs are
located in regions 1R and 3, consistent with the migratory paths of
PAH-dominated sources. The upper bounds classified as SFGs in the top-right
panel are consistent with the majority of SFGs being located in region 2. All
4 ATLAS sources classified as stars are located in region 3. The polarized
ATLAS sources detected in all four IRAC bands follow the distribution of
AGNs, situated predominantly in regions 1 and 3 in almost equal proportions.
The upper bounds for polarized sources presented in the top-right panel
are consistent with this finding. The lack of polarized sources in region 2
suggests that the polarized ATLAS sources observed in region 3 are unlikely
to be SFGs with rest-frame colours located in region 2 (i.e. if SFGs are
migrating from region 2 to region 3 with redshift, then a trail of sources
would be expected). Instead, we find two concentrations of polarized ATLAS
sources, highly coincident with the regions of parameter space identified by
\citet{2005ApJ...621..256S} in which starlight- and continuum-dominated
sources were preferentially located.

Thus we find that the radio emission from polarized ATLAS sources is most
likely powered by AGNs, where the active nuclei are embedded within host galaxies
with mid-infrared spectra dominated by old-population stellar light (blue IRAC
colours) or continuum likely produced by dusty tori (red IRAC colours).
This finding is in general agreement with the results from the ELAIS-North 1 (ELAIS-N1)
region presented by both \citet{2007ApJ...666..201T} and \citet{2011ApJ...733...69B}, but with
the following two notable exceptions. First, both these works identified radio sources (both
polarized and unpolarized) that were concentrated in region 3 about 
$\log_{\ms 10}\left(S_{5.8\,\mu\trm{\scriptsize m}}/S_{3.6\,\mu\trm{\scriptsize m}}\right)=-0.6$,
$\log_{\ms 10}\left(S_{8.0\,\mu\trm{\scriptsize m}}/S_{4.5\,\mu\trm{\scriptsize m}}\right)=-0.7$,
well beyond the parameter space typically occupied by the three generic source
classes investigated by \citet{2005ApJ...621..256S}. \citet{2007ApJ...666..201T}
reported that these sources were associated with elliptical galaxies dominated by
old-population starlight. However, Fig.~11 from \citet{2005ApJ...621..256S}
indicates that these sources are located within a region of parameter space
occupied by individual stars. It is not clear why
the IRAC colours of so many of the radio sources presented by
\citet{2007ApJ...666..201T} and \citet{2011ApJ...733...69B} were found to
occupy this region of parameter space, though it is possible that their selection
of isophotal flux densities for unresolved infrared sources may have biased
their colour ratios (aperture values are more appropriate for point sources).
And second, unlike these previous works, we do not find any polarized ATLAS sources in which the
radio emission is likely to be powered by star formation \citep[i.e. we do not see any polarized
sources in region 2; cf.][]{2011ApJ...733...69B}; we cannot conclude that any polarized
sources have infrared colours suggestive of significant PAH emission \citep[cf.][]{2007ApJ...666..201T}.
The fractional polarization properties of ATLAS AGNs and SFGs are described in
\S~\ref{ch5:SecResIdentM} and modelled in \S~\ref{ch5:SecResSubPI}.

\subsubsection{Radio vs Far-Infrared}

\begin{figure*}
 \centering
 \includegraphics[trim = 0mm 0mm 0mm 0mm, clip, angle=-90, width=0.75\textwidth]{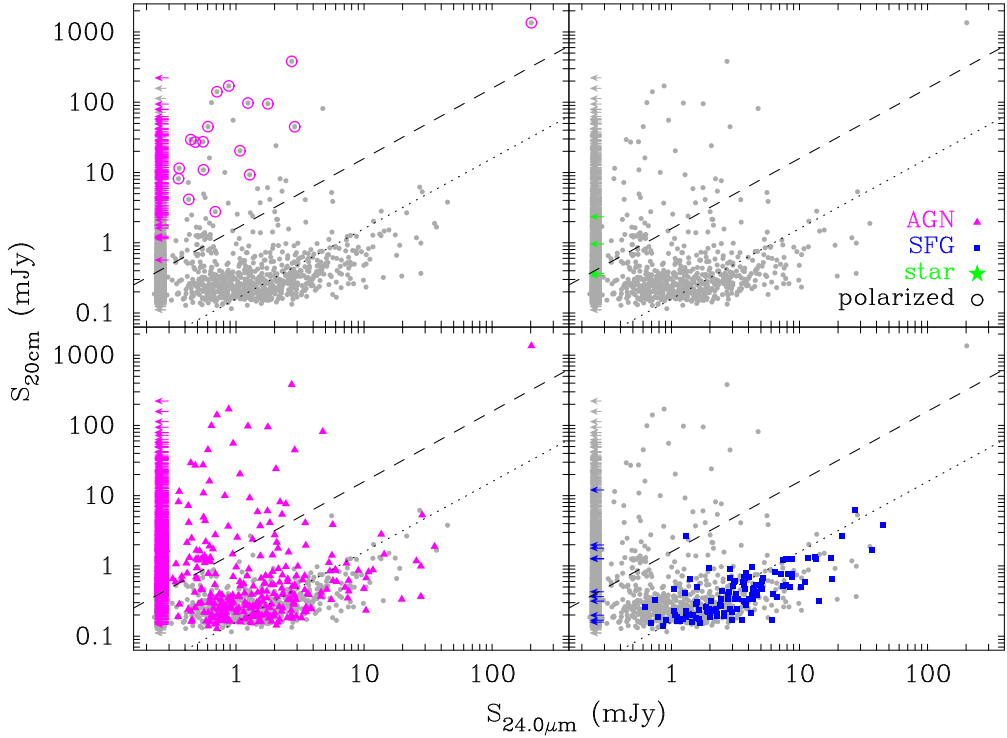}
 \caption{	Comparison of radio to far-infrared flux densities for all ATLAS DR2
 		sources observed at $24\mu$m, as indicated by the grey points and grey
		upper bounds in each panel. For clarity, multiple panels are used to focus
		on different source populations. The dotted line in each panel indicates the
		FRC, defined by \citet{2004ApJS..154..147A} as $q_{\ms 24}=0.8$. The dashed
		lines indicate our adopted criteria for classifying AGNs, based on radio
		flux densities 10 times greater than the FRC. In each panel, the
		highlighted population includes those sources represented by upper
		bounds. The bottom-left panel highlights all sources classified as
		AGNs in ATLAS DR2. The bottom-right and top-right panels highlight all
		sources classified as SFGs and stars, respectively.
		The top-left panel highlights all sources
		exhibiting linearly polarized emission, where the colours of the
		highlighted points and upper bounds indicate whether a source has
		been classified as an AGN, SFG, or star (i.e. they are all AGNs).
}
 \label{ch5:fig:rFIR}
\end{figure*}

In Fig.~\ref{ch5:fig:rFIR} we compare the total intensity 1.4~GHz radio to 24~$\mu$m
infrared flux densities for all ATLAS DR2 sources, taking into account infrared upper bounds
for all radio sources without detected infrared counterparts.
As noted in Paper~I, we use 24~$\mu$m flux density as a proxy for far-infrared (FIR) flux density.
The bottom-left panel indicates that ATLAS sources classified as AGNs are prevalent
both away from and on the FIR-radio correlation (FRC). The presence of a substantial number of AGNs below the
dashed line demonstrates the value of using multiple diagnostic criteria to classify
sources; only sources above the dashed line have been classified as AGNs using the
radio to far-infrared diagnostic. The bottom-right panel indicates that, as expected,
ATLAS sources classified as SFGs typically cluster along the FRC and have radio flux
densities {\footnotesize $\lesssim$}~1~mJy. However, a small number of sources classified
as SFGs (and stars) are observed with upper bounds clearly located within the AGN parameter
space. The top-left panel highlights all 130 polarized ATLAS sources, indicating that
each of these was classified as an AGN. No polarized stars or SFGs were detected in our data.

\subsection{Polarization Cross-Identifications and Classifications}\label{ch5:SecResIdentM}

We now present diagnostics of components, groups, and sources in ATLAS DR2, resulting
from the linear polarization$-$total intensity cross-identification and classification
procedures described in \S~6.2 of Paper I (summarised in \S~\ref{sec:1} of this work).
In this section we focus on a number of parameter spaces in which we detail relationships
between the polarized flux densities, fractional polarizations, classifications,
and angular sizes of sources and their constituents.

\subsubsection{Components, Groups, and Sources}

\begin{figure*}
 \centering
 \includegraphics[trim = 0mm 0mm 0mm 0mm, clip, angle=-90, width=0.75\textwidth]{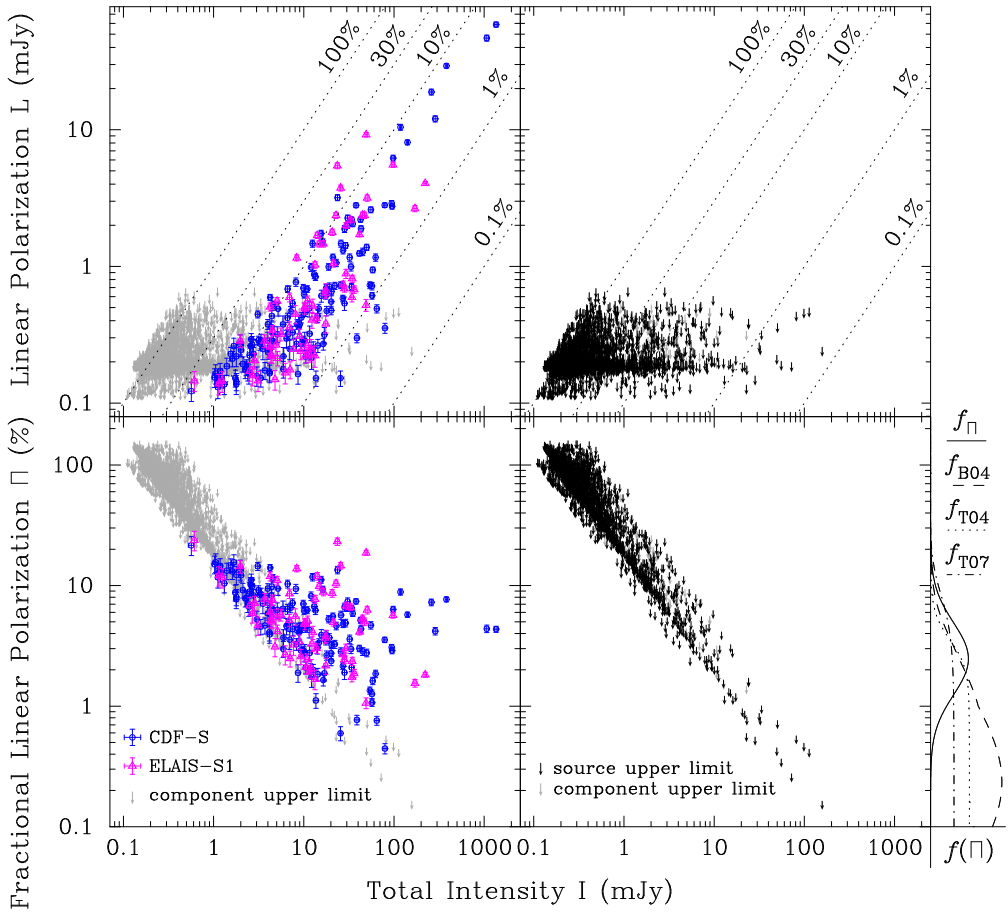}
 \caption{	Linearly polarized flux density (upper panels) and corresponding
		fractional linear polarization (lower panels) versus total intensity
		flux density for all ATLAS DR2 components, groups and sources. Dotted
		lines in the upper panels indicate fractional polarizations of 0.1\%, 1\%,
		10\%, 30\%, and 100\%. Upper limits for unpolarized components are
		shown in the left panels and replicated in the right panels.
		Upper limits for unpolarized sources are shown only in the right panels.
		Polarization detections are shown only in the left panels.
		Polarized components (i.e. one-to-one associations), groups,
		and sources, which are not differentiated in this Figure, are
		indicated in the left panels and colour-coded according to their
		respective ATLAS field. The curves adjacent to the lower-right panel
		replicate the fractional polarization distributions presented in
		Fig.~\ref{ch5:fig:PImodels} (see \S~\ref{ch5:SecResSubPI}).
}
 \label{ch5:fig:fracpolRaw}
\end{figure*}

In Fig.~\ref{ch5:fig:fracpolRaw} we plot the polarized flux densities and fractional
polarizations for all ATLAS DR2 components, groups, and sources versus their total
intensity flux densities, taking into account polarization upper limits.
The fractional polarization uncertainties displayed in the lower-left panel
were estimated following standard error propagation as
\begin{equation}\label{ch5:eqn:Merr}
	\sigma_{\ms \Pi} \approx
	\Pi\,\sqrt{
	\left(\frac{\sigma_\tnm{\tiny I}}{I}\right)^{\!2} +
	\left(\frac{\sigma_\tnm{\tiny L}}{L}\right)^{\!2}
	} \,.
\end{equation}
Fig.~\ref{ch5:fig:fracpolRaw} shows that the polarization upper limits for components
and sources are distributed almost identically, the reason being that the majority
of unpolarized sources comprise a single component (relevant statistics are detailed
toward the end of this section). Regarding polarization detections, we find that all
components, groups, and sources exhibit $\Pi<24\%$. This finding is in
contrast to the data from other 1.4~GHz polarization surveys. \citet{2010MNRAS.409..821S}
found that 1\% (381/38454) of polarized sources in the NRAO VLA Sky Survey (NVSS) exhibited
$\Pi\ge30\%$. \citet{2007ApJ...666..201T} and \citet{2010ApJ...714.1689G} found that 10\% (8/83)
and 7\% (10/136) of polarized sources in the ELAIS-N1 field exhibited $\Pi\ge30\%$,
respectively. \citet{2010MNRAS.402.2792S} found that 10\% (84/869) of polarized
sources throughout the two Australia Telescope Low-Brightness Survey (ATLBS) fields
exhibited $\Pi\ge30\%$. If a population of extragalactic sources with high 1.4~GHz
fractional polarizations were to exist, then it would be unexpected for such sources
to be detected in the surveys above [with $\sim50\arcsec$ full-width at half-maximum
(FWHM) beam sizes] yet undetected in this work (with $\sim10\arcsec$ FWHM beam size),
because the former are more susceptible to both beam and bandwidth depolarization.
Instead, we attribute the lack of ATLAS sources with $\Pi>30\%$ to our careful treatment
of local (rather than global) root mean square (rms) noise estimates, in particular to our
employment of {\tt BLOBCAT}'s flood-fill technique for extracting polarized flux densities
(see \citealt{2012MNRAS.425..979H} for details regarding biases introduced through
Gaussian fitting), and to our statistical classifications of unresolved and resolved
components. We found through testing that components with abnormally high levels of
fractional polarization (up to and even beyond 100\%) could be obtained if the
features above were not taken into account.

\subsubsection{Multiwavelength Classifications}

\begin{figure*}
 \centering
 \includegraphics[trim = 0mm 0mm 0mm 0mm, clip, angle=-90, width=0.75\textwidth]{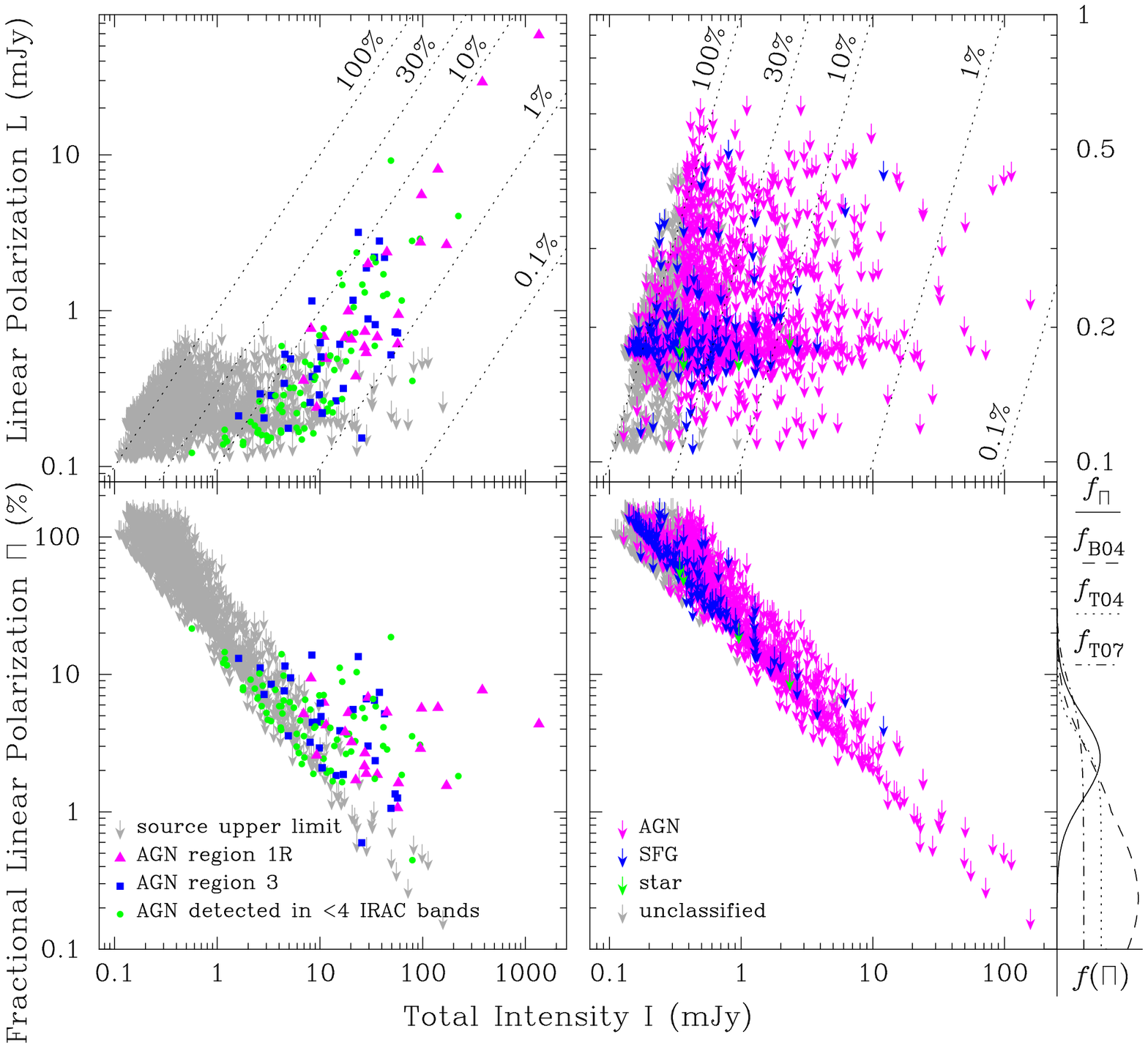}
 \caption{	Linearly polarized flux density (upper panels) and corresponding
		fractional linear polarization (lower panels) versus total intensity
		flux density for all ATLAS DR2 sources. The panel layout is similar
		to Fig.~\ref{ch5:fig:fracpolRaw}; the legends are independent for each column.
		Panels in the left column highlight polarized sources with infrared
		counterparts that exhibit blue [sources located within or just beyond region 3 with
		$\log_{\ms 10}\left(S_{5.8\,\mu\trm{\scriptsize m}}/S_{3.6\,\mu\trm{\scriptsize m}}\right)<-0.1$]
		or red [sources located within or just beyond region 1R with
		$\log_{\ms 10}\left(S_{5.8\,\mu\trm{\scriptsize m}}/S_{3.6\,\mu\trm{\scriptsize m}}\right)>-0.1$]
		mid-infrared colours according to the colour-colour diagram
		in the lower-right panel of Fig.~\ref{ch5:fig:NIRCC}.
		Note that all polarized ATLAS DR2 sources have been classified as AGNs.
		For comparison, upper limits for all unpolarized sources are presented in
		the background. Panels in the right column highlight AGN/SFG/star
		classifications for these unpolarized sources. Note that x-axis scaling differs between
		columns, and that y-axis scaling differs between the two upper panels.
}
 \label{ch5:fig:fracpolClass}
\end{figure*}

In Fig.~\ref{ch5:fig:fracpolClass} we plot the polarized flux densities and fractional
polarizations for all ATLAS DR2 sources only, indicating their infrared/optical
classifications.
Panels in the left column highlight polarized sources with infrared counterparts
detected in all four IRAC bands, or otherwise. We split those detected in all four
bands into sources located within or just beyond region 3 in the lower-right panel of
Fig.~\ref{ch5:fig:NIRCC} [i.e. polarized sources with
$\log_{\ms 10}\left(S_{5.8\,\mu\trm{\scriptsize m}}/S_{3.6\,\mu\trm{\scriptsize m}}\right)<-0.1$]
and those located within or just beyond region 1R
[i.e. $\log_{\ms 10}\left(S_{5.8\,\mu\trm{\scriptsize m}}/S_{3.6\,\mu\trm{\scriptsize m}}\right)>-0.1$].
We find no clear distinction between the fractional polarization properties of
sources with blue (region 3) or red (region 1R) mid-infrared colours. It is possible
that the region 3 polarized sources exhibit a larger dispersion in fractional
polarization than the region 1R polarized sources (compare range of observed fractional
polarizations in lower-left panel of Fig.~\ref{ch5:fig:fracpolClass}), though given our
sample size this marginal effect may be attributed to sampling variance. Using data from
\citet{2010ApJ...714.1689G}, \citet{2011ApJ...733...69B} found that polarized
sources in region 3 were more highly polarized than those in region 1R; the ATLAS
data do not support this result. The distributions of upper limits presented in the
right-column panels of Fig.~\ref{ch5:fig:fracpolClass} indicate that all sources with
$\Pi<1\%$ have been classified as AGNs. The fractional polarization upper limits
for sources classified as SFGs are not particularly restrictive, as their total
intensity flux densities are typically {\footnotesize $\lesssim$}~1~mJy. 
Characteristic $\Pi$ levels for the sub-millijansky SFG population are
{\footnotesize $\lesssim$}~$60\%$.

\subsubsection{Distribution of Fractional Linear Polarization}

Focusing on the lower-left panel of Fig.~\ref{ch5:fig:fracpolClass}, we note that a
general observational consequence of the rising distribution of fractional polarization
upper limits with decreasing total intensity flux density is that the mean
or median fractional polarization of {\it detected} polarized sources will always
{\it appear} to increase with decreasing flux density. This increase represents a selection bias;
it is not possible to detect low levels of fractional polarization for the faintest total
intensity sources. Any changes to the underlying distribution of fractional
polarization with decreasing total intensity flux density will be masked, and thus
dominated, by this selection bias. Therefore, it is not possible to investigate the
distribution of fractional polarization at faint flux densities without accounting
for polarization non-detections. Recently, 1.4~GHz polarimetric studies of the
ELAIS-N1 field \citep{2007ApJ...666..201T,2010ApJ...714.1689G} and ATLBS fields
\citep{2010MNRAS.402.2792S} concluded that their observational data demonstrated
an anti-correlation between fractional polarization and total intensity flux
density. These studies found that sources with $I$~{\footnotesize $\lesssim$}~$10$~mJy
were more highly polarized than stronger sources. However, \citet{2010MNRAS.402.2792S}
did not account for polarization upper limits, leading to their misinterpretation of
selection bias as an indication of true anti-correlation. \citet{2007ApJ...666..201T}
accounted for selection bias using Monte Carlo analysis, effectively incorporating
polarization upper limits. \citet{2010ApJ...714.1689G} accounted for selection bias by
comparing samples of sources in bins of polarized flux density rather than total flux
density, at sufficient polarized flux densities to neglect upper limits. However, the
findings of increased fractional polarization at faint total flux densities by
\citet{2007ApJ...666..201T} and \citet{2010ApJ...714.1689G} appear to be
reliant on the increasing number of sources observed with $\Pi>30\%$ at these faint levels.
For example, both studies reported extreme sources with $\Pi>60\%$, but only at faint total
intensities. Both \citet{2007ApJ...666..201T} and \citet{2010ApJ...714.1689G} found that
$\sim13\%$ of polarized sources with linearly polarized flux densities $L<2$~mJy (i.e. a
significant proportion of these sources) exhibited $\Pi>30\%$, while no sources with such high
levels of fractional
polarization were found for $L>2$~mJy. As described earlier, the $\Pi>30\%$ sources (and perhaps
many with lower $\Pi$) are likely to reflect rms noise estimation and source extraction errors.
The analytic form assumed by \citet{2007ApJ...666..201T} for the distribution of fractional
polarization (which will be described in \S~\ref{ch5:SecResSubPI}) may have also contributed to
their conclusion regarding increased fractional polarization; spurious conclusions may
be obtained if the observed fractional polarization data do not follow the assumed analytic
form of the fit. The arguments above suggest that existing evidence for an anti-correlation
between fractional polarization and total flux density may not be robust.

Similar to the studies above, earlier works by \citet{2002A&A...396..463M} and
\citet{2004MNRAS.349.1267T} concluded that NVSS \citep{1998AJ....115.1693C} sources
exhibited an anti-correlation between fractional linear polarization and total
intensity flux density. These analyses in effect
incorporated polarization upper bounds \citep[though not upper {\it limits};
see][]{2010ApJ...719..900K} because \citet{1998AJ....115.1693C} recorded a linearly
polarized flux density for each NVSS source, regardless of the statistical significance
of the polarization measurement. To determine the significance of their findings
and thus form a conclusion regarding evidence for anti-correlation, which we use to
justify our own fractional polarization model presented in \S~\ref{ch5:SecResSubPI},
we need to examine their works in more detail.

\citet{2002A&A...396..463M} and \citet{2004MNRAS.349.1267T} presented fractional
polarization distributions for steep- and flat/inverted-spectrum NVSS sources in four
flux density intervals: $100-200$, $200-400$, $400-800$, and $>800$~mJy. Their
distributions are remarkably consistent for $\Pi>1\%$, exhibiting a log-normal form
with approximately equal dispersion and a peak at $\sim2.5\%$. A separate component
with a peak at $\sim0.2\%$ is also present in each distribution, representing sources
with polarization dominated by instrumental leakage. We observe that the dispersions
of their distributions broaden with increasing flux density, solely due to broadening at
$\Pi<1\%$. \citet{2002A&A...396..463M} found that the median fractional polarization
was larger for the $100-200$~mJy data than for the $>800$~mJy data, for both steep-
and flat-spectrum sources. \citet{2004MNRAS.349.1267T} found the same result but for
steep-spectrum sources only. These results were essentially based on the lack of sources
with $\Pi<0.1\%$ in the $100-200$~mJy data when compared with the increased presence
of such sources at higher flux densities; a proportional increase in the number of
sources with $\Pi>1\%$ for decreasing flux density was not observed. However, the
presence of sources with $\Pi<0.1\%$ (i.e. less than the typical leakage level of
$\sim0.2\%$), and more generally the slight changes in distribution shape observed
for $\Pi<1\%$ between different flux density intervals, may be more appropriately
explained by the influence of noise on polarized flux densities rather than by
variation in the underlying distribution of fractional polarization. To
demonstrate, we first note that the expectation value of $L$ for an unpolarized
NVSS source is given by the mean of a Rayleigh distribution (i.e. a Ricean
distribution with no underlying polarized signal), which is
$\sqrt{\pi\sigma_{\ms Q,U}^{\ms 2}/2}\approx0.36$~mJy for
$\sigma_{\ms Q,U}\approx0.29$~mJy. This value is also characteristic of the
expected observed polarized flux density for a source with true underlying
polarized signal $L_{\ms 0}$~{\footnotesize $\lesssim$}~$\sigma_{\ms Q,U}$
(e.g. see the upper panel of Fig.~1 from \citealt{2012MNRAS.424.2160H}).
Thus, a tail of sources with true polarization $L_{\ms 0}$~{\footnotesize $\lesssim$}~0.29~mJy
will appear in the fractional polarization distribution at
$\Pi$~{\footnotesize $\gtrsim$}~0.18\% for total intensity sources with
$I<200$~mJy, and at $\Pi$~{\footnotesize $\lesssim$}~0.05\% for $I>800$~mJy.
These estimates are consistent with the distributions presented by
\citet{2002A&A...396..463M} and \citet{2004MNRAS.349.1267T}; a tail of sources
with $\Pi<0.1\%$ was observed for the $>800$~mJy data but not for the
$100-200$~mJy data. We therefore conclude that the results presented by
\citet{2002A&A...396..463M} and \citet{2004MNRAS.349.1267T} do not
demonstrate a statistically significant anti-correlation between fractional
linear polarization and total intensity flux density. Furthermore, we note
that the fractional polarization distributions presented in these works
are likely to overestimate the population of sources with
$\Pi<1\%$, even for the $100-200$~mJy data, for the following two reasons.
First, all catalogued NVSS measurements of polarized flux density were debiased
using a modified version of the expectation value for a Ricean distribution
\citep{1998AJ....115.1693C}. This debiasing scheme is known to impart a
significant overcorrection (i.e. negative bias) at low SNR (e.g. see
\citealt{1985A&A...142..100S}; the relevant scheme is labelled in reference
to its application by \citealt{1958AcA.....8..135S}). Thus
measurements of fractional polarization obtained using the NVSS catalogue
are likely to be negatively biased. And second, raw polarization measurements
for NVSS sources were obtained by interpolation at the total intensity centre
position. Therefore, polarized flux densities were underestimated for each
source in which the spatial peak of polarized emission was located in an
adjacent pixel to the total intensity peak. Both of these effects
could have been largely mitigated by obtaining polarization upper limits for
sources, rather than upper bounds.

Returning to the lower-left panel of Fig.~\ref{ch5:fig:fracpolClass}, we find that
the maximum level of fractional polarization exhibited by ATLAS sources does
not appear to be correlated with total intensity flux density. The maximum level
appears to be limited to $\Pi$~{\footnotesize $\lesssim$}~20\% for
$I$~{\footnotesize $\gtrsim$}~1~mJy, which becomes a strict limit for $I>4$~mJy
when accounting for the presence of all upper limits. Furthermore, we find
$\Pi$~{\footnotesize $\gtrsim$}~0.4\% for sources with
$I$~{\footnotesize $\gtrsim$}~10~mJy, where sources exhibiting higher levels of
fractional polarization significantly outnumber those potentially exhibiting
$\Pi<0.4\%$ as indicated by the upper limits.

\subsubsection{Polarization Classifications}

In Paper~I we found that 138 of the total 172 catalogued linearly polarized components
exhibited a clear one-to-one match with individual total intensity components. The
remaining 34 polarized components required grouping in order to be associated with
total intensity counterparts. Of the one-to-one associations, we classified 58 as
Type 0, 4 as Type 1, 25 as Type 2, 48 as Type 4, and 3 as Type 5. All 3 sources
containing Type 5 core associations were found to exhibit unpolarized lobes. Of
the group associations comprising a total of 34 polarized components, 2 groups
were classified as Type 0, 14 as Type 3, 1 as Type 4, and 8 as Type 6. There were
29 sources classified as Type 6, 2 as Type 7, and 25 as Type 8. These classifications
are catalogued in Appendix~B of Paper~I.

In Fig.~\ref{ch5:fig:fracpolTypes} we indicate the polarization$-$total intensity
classifications for all polarized ATLAS DR2 components, groups, and sources.
\begin{figure*}
 \centering
 \includegraphics[trim = 0mm 0mm 0mm 0mm, clip, angle=-90, width=0.75\textwidth]{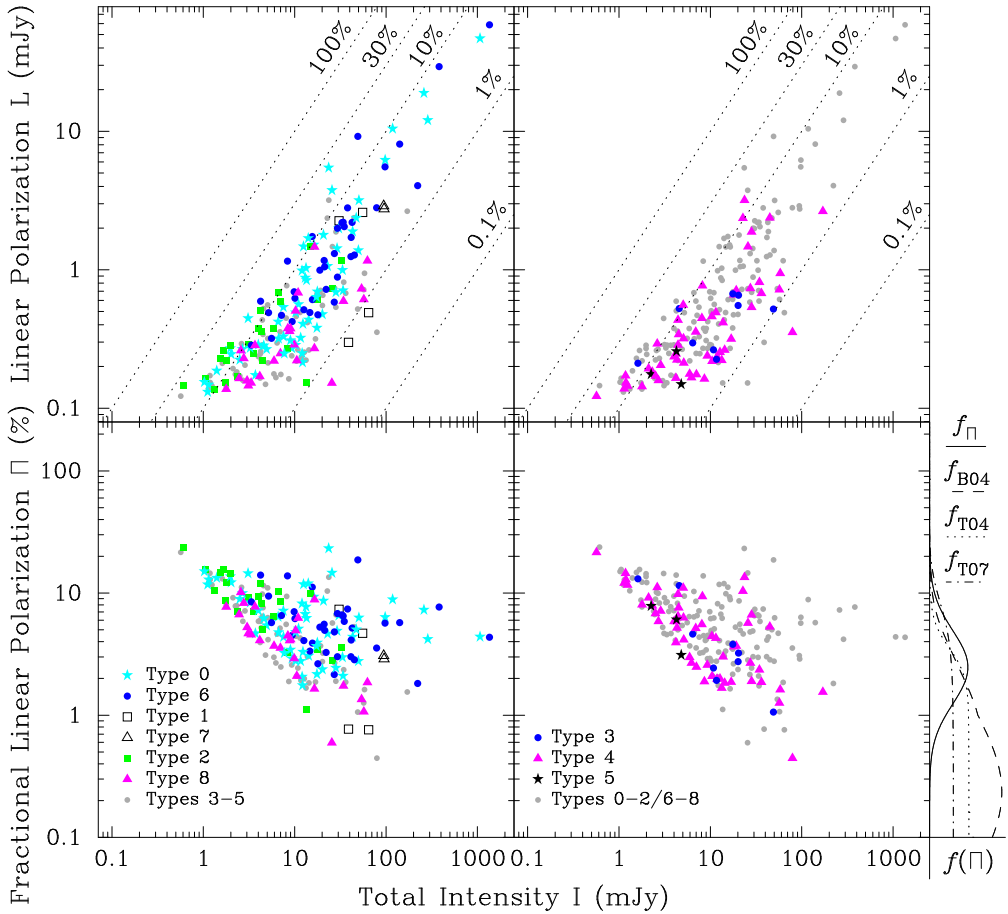}
 \caption{	
 		Polarized ATLAS DR2 components, groups and sources from the left
		column of Fig.~\ref{ch5:fig:fracpolRaw}, highlighted here according to
		the classification scheme presented in \S~6.2 of Paper I (summarised
		in \S~\ref{sec:1} of this work).
		The legends are independent for each column. Note that only Types
		3, 4, 6, 7, or 8 may represent sources.
}
 \label{ch5:fig:fracpolTypes}
\end{figure*}
In the lower-left panel we plot the levels of fractional
polarization exhibited by classical double or triple radio sources (Types 6--8) and their
individual lobes (Types 0--2, respectively). We find that sources classified as Type 6, which
comprise pairs of roughly equally polarized Type 0 lobes, are located throughout most of the
populated parameter space. We find that Type 7 sources, which comprise pairs of Type 1 lobes
where one is clearly less polarized than the other, appear to occupy the same parameter
space populated by Type 6 sources. A selection bias against identifying Type 0/1 lobes,
and thus Type 6/7 sources, is present within the diagonal region of parameter space
populated by polarization upper limits (for visual clarity these limits are not shown
in Fig.~\ref{ch5:fig:fracpolTypes}; see Fig.~\ref{ch5:fig:fracpolRaw}). Type 2 lobes and their
parent Type 8 sources, which represent ambiguous cases in which it is not possible to
differentiate between Types 0/1 or 6/7, are largely confined to this diagonal region.
Given the observed prevalence of Type 6 sources compared with Type 7, it seems likely
that more sensitive observations would result in a majority of Type 8 sources being
reclassified as Type 6. From the lower-right panel of Fig.~\ref{ch5:fig:fracpolTypes} we
find that sources classified as Type 3, which exhibit a single polarized component
situated midway between two total intensity components, appear to populate the same
region of parameter space occupied by Type 8 sources. Similarly, associations
classified as Type 5, which represent cores of triple radio sources, as well as the
remaining unclassified sources denoted by Type 4, also appear to be concentrated
within the diagonal region of parameter space populated by upper limits. We note
that many of the Type 4 associations are likely to represent individual Type 0 or
Type 1 lobes of as-yet unassociated multi-component sources, having been
erroneously assigned to single-component sources in our catalogue (note
\S~6.1 of Paper I; statistics regarding polarized multi-component
sources are presented below). We find that Type 5 associations occupy a parameter
space consistent with Type 6 and Type 7 sources. As the latter represent average
polarization properties for dual-lobed radio sources, it is possible that Type 5
associations also represent dual-lobed structures but with small angular sizes,
such as compact steep-spectrum (CSS) sources \citep{1998PASP..110..493O}. Curiously,
we found that each of the 3 sources with Type 5 cores was found to exhibit unpolarized
outer radio lobes. It is possible that the Type 5 cores represent restarted AGN activity
and that the outer lobes are unpolarized because any large-scale magnetic fields
within them have dissipated over time since their production during an
earlier distinct phase of AGN activity. For example, we may be seeing sources
similar to the double-double radio galaxy J1835$+$620 \citep{1999A&A...348..699L},
though at an earlier stage of evolution where the inner lobes have not yet separated
into two separate lobes (note that fractional polarization levels for the inner lobes
of J1835$+$620 are higher than for the outer lobes).

\subsubsection{Angular Sizes}

\begin{figure*}
 \centering
 \includegraphics[trim = 0mm 0mm 0mm 0mm, clip, angle=-90, width=0.75\textwidth]{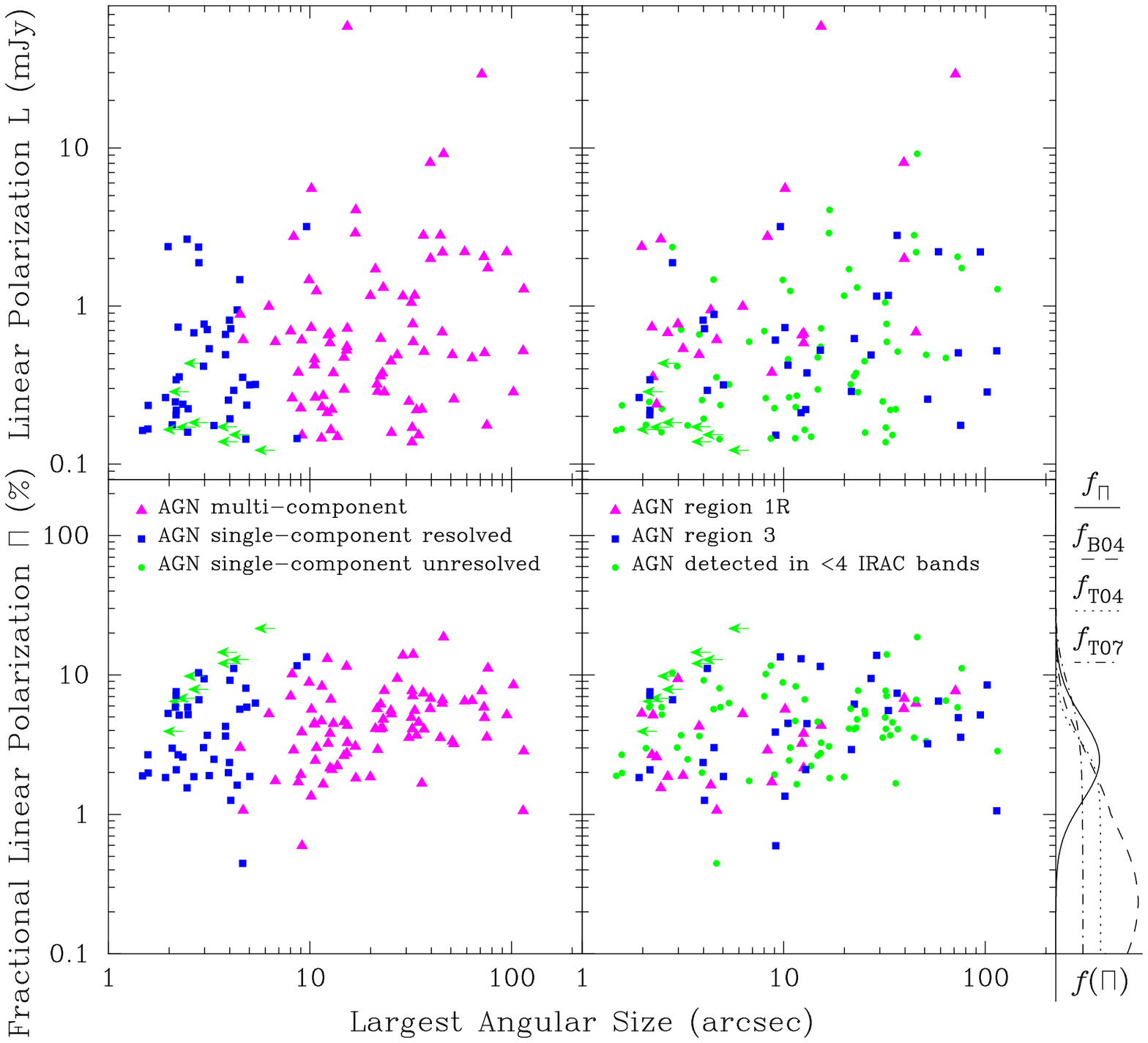}
 \caption{	Linearly polarized flux density (upper panels) and corresponding
		fractional linear polarization (lower panels) versus LAS for all 
		polarized ATLAS DR2 sources. For visual clarity, sources with
		polarization upper limits are not displayed here; they are presented
		separately in Fig.~\ref{ch5:fig:fracpolThetaULs}. Panels in the left column
		highlight sources according to their number of observed constituent
		components. Panels in the right column highlight sources according to the
		properties of their infrared counterparts where available, similar to
		those described for the data presented in the left column of
		Fig.~\ref{ch5:fig:fracpolClass}. Note that all polarized ATLAS DR2 sources
		have been classified as AGNs. The curves
		adjacent to the lower-right panel replicate the fractional polarization
		distributions presented in Fig.~\ref{ch5:fig:PImodels} (see \S~\ref{ch5:SecResSubPI}).
}
 \label{ch5:fig:fracpolTheta}
\end{figure*}

\begin{figure*}
 \centering
 \includegraphics[trim = 0mm 0mm 0mm 0mm, clip, angle=-90, width=0.75\textwidth]{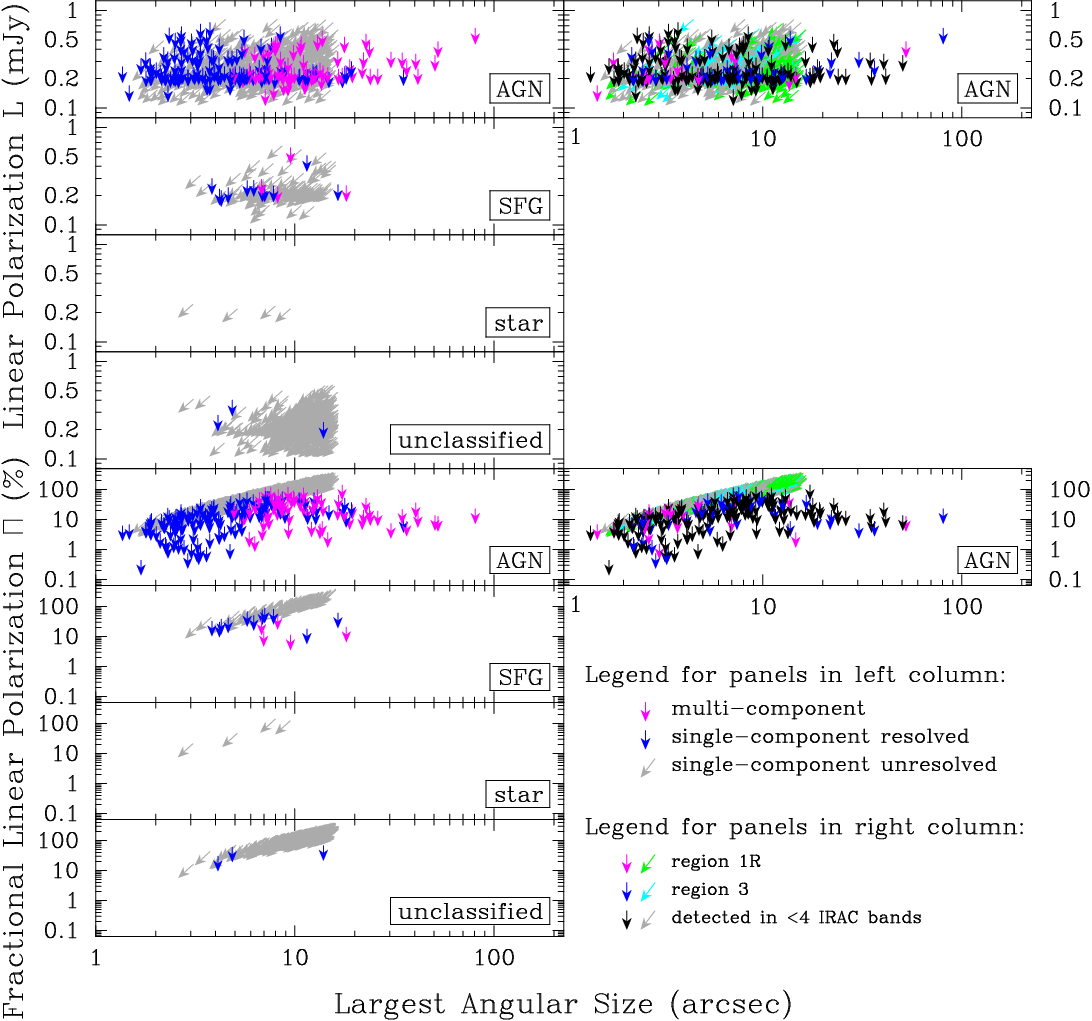}
 \caption{	Polarization upper limits for sources omitted from Fig.~\ref{ch5:fig:fracpolTheta}.
		The 4 upper-left panels each correspond to the upper-left panel from
		Fig.~\ref{ch5:fig:fracpolTheta} (though with different y-axis scaling),
		separated here into 4 source classes for clarity. Similarly, the upper-right
		panel corresponds to the upper-right panel from Fig.~\ref{ch5:fig:fracpolTheta}.
		Panels in the lower-half of the figure are similar to the upper-half panels,
		but for fractional polarization upper limits. Only sources classified as AGNs
		with infrared counterparts in regions 1R and 3 of Fig.~\ref{ch5:fig:NIRCC}, and
		those detected in less than 4 IRAC bands, are displayed in the right-column
		panels. In all panels, vertical arrows represent resolved sources, while
		diagonal arrows represent unresolved sources with deconvolved angular size
		upper limits.
}
 \label{ch5:fig:fracpolThetaULs}
\end{figure*}

\begin{figure}
 \centering
 \includegraphics[trim = 0mm 0mm 0mm 0mm, clip, angle=-90, width=0.45\textwidth]{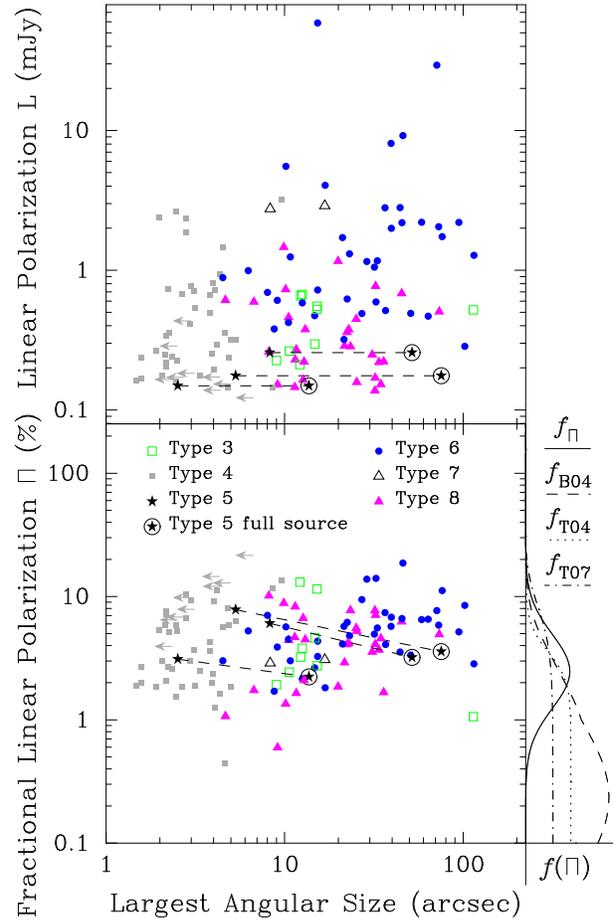}
 \caption{	Reproduction of data from Fig.~\ref{ch5:fig:fracpolTheta},
 		highlighted here according to the classification scheme presented in
		\S~6.2 of Paper~I (summarised in \S~\ref{sec:1} of this work).
		Note that under this scheme, only Types 3,
		4, 6, 7, or 8 may represent sources. Additional data points are
		displayed for Type 5 one-to-one associations (which represent polarized
		cores within classical triple morphologies) with abscissae given by their
		deconvolved angular sizes, and for their parent sources (which include
		outer lobes) with abscissae given by their LASs.
}
 \label{ch5:fig:fracpolTheta2}
\end{figure}

In Fig.~\ref{ch5:fig:fracpolTheta} we plot polarized flux density and fractional
polarization versus largest angular size (LAS) for all polarized ATLAS DR2 sources,
highlighted according to morphology and infrared colour.
The LAS for a single-component source is given by its total intensity deconvolved angular
size or size upper limit, while the LAS for a multi-component source is given by the
maximum angular separation between its constituent total intensity components.
For visual clarity we plot sources with polarization upper limits separately in
Fig.~\ref{ch5:fig:fracpolThetaULs}, also highlighted according to morphology and infrared colour.
Note that the apparent anti-correlations between fractional polarization upper limits
and angular size upper limits for single-component sources throughout
Fig.~\ref{ch5:fig:fracpolThetaULs} are spurious; the restrictiveness of both types of upper
limits are intrinsically anti-correlated with total intensity flux density.
In Fig.~\ref{ch5:fig:fracpolTheta2}
we again plot polarized flux density and fractional
polarization versus LAS for all polarized sources, but now highlighted according to
the polarization$-$total intensity classification scheme from \S~6.2 of Paper I.
For reference, we note that 1\arcsec\ subtends a linear scale of 1.8, 3.3, 6.1, 8.0, and
8.5~kpc at redshifts 0.1, 0.2, 0.5, 1.0, and 2.0, respectively \citep{2006PASP..118.1711W},
assuming a $\Lambda$CDM cosmology with parameters $H_0=71$~km~s$^{-1}$~Mpc$^{-1}$, $\Omega_m=0.27$,
and $\Omega_\Lambda=0.73$. Following an evolutionary relationship for galaxy sizes given by
\begin{equation}
	\tnm{diameter (linear scale)} \propto {H_0}^{-1}
	\left[\Omega_m\left(1+z\right)^3+\Omega_\Lambda\right]^{-\frac{1}{2}} \;,
\end{equation}
and assuming that a typical galaxy has size $\sim20$~kpc at $z=0$
\citep[e.g.][]{2004ApJ...600L.107F}, the corresponding sizes of typical galaxies
at the redshifts above are approximately\footnote{Note that surface brightness dimming, which
is $\propto (1+z)^4$ \citep{1930PNAS...16..511T,2001AJ....121.2271S}, may cause observed angular
sizes of extended sources to be smaller than true sizes, due to faint source edges.} 19, 18, 16, 12, and
7~kpc, respectively, or 10\arcsec, 5\farcsecd6, 2\farcsecd6, 1\farcsecd5, and 0\farcsecd8,
respectively. We summarise our findings from Figs.~\ref{ch5:fig:fracpolTheta}--\ref{ch5:fig:fracpolTheta2}
as follows.

Of the 130 (2091) polarized (unpolarized) sources catalogued in ATLAS DR2 and presented
in Fig.~\ref{ch5:fig:fracpolTheta} (Fig.~\ref{ch5:fig:fracpolThetaULs}), 81 (74) comprise
multiple components in total intensity, 40 (140) comprise a single resolved component
in total intensity, and 9 (1877) comprise a single unresolved component in total intensity.
We note that while components observed in linear polarization in ATLAS DR2 are typically
unresolved (only 29/172 or 17\% of polarized components are resolved; see \S~5 of Paper I),
121/130 or 93\% of sources exhibiting polarized emission are resolved in total intensity.
These statistics support the findings by \citet{2010ApJ...714.1689G} that polarized
1.4~GHz sources tend to have structure at arcsecond scales and that, as a consequence,
their polarized emission is unlikely to be beamed. Combined with our earlier classification
from Fig.~\ref{ch5:fig:rFIR} of all polarized ATLAS sources as AGNs, and our interpretation
from Fig.~\ref{ch5:fig:fracpolTypes} that most or all polarized components are associated
with AGN jets or lobes (rather than cores), the statistics above demonstrate that
(sub-)millijansky polarized sources tend to be extended jet- or lobe-dominated active radio
galaxies. This conclusion is supported by the finding from \citet{2010ApJ...714.1689G}
that polarized sources tend to have steep spectra, which are characteristic of lobes.

In Fig.~\ref{ch5:fig:fracpolThetaULs} we find that ATLAS DR2 sources typically have
LAS~{\footnotesize $\lesssim$}~10\arcsec, suggesting that most sources are located
at $z$~{\footnotesize $\gtrsim$}~0.2. This is consistent with the preliminary redshift
distributions presented by \citet{2006AJ....132.2409N} and \citet{2008AJ....135.1276M}
for ATLAS DR1 sources \citep[see also discussion of radio source redshift distribution
by][]{1988gera.book..641C}.

Focusing on the panels in the lower-left corners of Fig.~\ref{ch5:fig:fracpolTheta}
and Fig.~\ref{ch5:fig:fracpolThetaULs}, we find that single- and multi-component sources
are distributed approximately equally in fractional polarization space;
their fractional polarization upper limits are not restrictive enough to
identify any possible underlying trends. However, having found above that
polarized sources are likely to represent lobed galaxies, it is perhaps surprising
that we do not find a clear correlation between fractional polarization and
LAS due to beam depolarization. Given the $\sim10\arcsec$ resolution of ATLAS, in
general a classical double radio source with dual polarized lobes should exhibit
greater fractional polarization than a similar source with smaller LAS that is
observed as a single-component source. A likely explanation may be that a
significant number of the polarized single-component sources indicated in
Fig.~\ref{ch5:fig:fracpolTheta} are actually individual lobes of as-yet unassociated
multi-component sources (see \S~6.1 of Paper I). Note that all
single-component sources in Fig.~\ref{ch5:fig:fracpolTheta} are classified as
Type 4 in Fig.~\ref{ch5:fig:fracpolTheta2}. Another potential explanation may be
that for dual-lobed sources with small angular size observed as single-component
sources, asymmetric depolarization between the lobes \citep{1988Natur.331..147G,1988Natur.331..149L}
could result in overall source fractional polarization levels similar to those of Type 7
sources (see Fig.~\ref{ch5:fig:fracpolTheta2}), rather than resulting in
significantly beam-depolarized (and thus perhaps unpolarized) sources overall.

The upper limits presented in the left column of Fig.~\ref{ch5:fig:fracpolThetaULs} do
not reveal any clear underlying trends within or between source classes. The
multi-component sources classified as SFGs in Fig.~\ref{ch5:fig:fracpolThetaULs}, which
are also shown located within the AGN parameter space in the lower-right panel of
Fig.~\ref{ch5:fig:rFIR}, require future study. These may
represent composite sources exhibiting both AGN and SFG characteristics, for
example similar to the ultra-luminous infrared galaxy F00183--7111 investigated by
\citet{2012MNRAS.422.1453N} or the more general classes of post-starburst quasars
\citep[e.g.][]{2011ApJ...741..106C}.

Focusing on the right column of Fig.~\ref{ch5:fig:fracpolTheta}, we do not find any
angular size distinctions between polarized sources based on their infrared colours.
Furthermore, we find no underlying trends within the associated upper limit data
from the right column of Fig.~\ref{ch5:fig:fracpolThetaULs}.

Focusing on Fig.~\ref{ch5:fig:fracpolTheta2}, we find that Type 6 sources typically
extend to greater angular sizes than Type 7 sources, though a larger sample size
with proportionally fewer Type 8 classifications is required to confirm this finding.
We also find that each of the 3 polarized cores classified as Type 5 are resolved,
and that they populate the same region of parameter space as Type 4 sources.

\subsection{Differential Component Counts}\label{ch5:SecResSubCnts}

We present Euclidean-normalised differential number-counts derived from the
ATLAS DR2 total intensity and linear polarization component catalogues in
Fig.~\ref{ch5:fig:countsI} and Fig.~\ref{ch5:fig:countsL}, respectively, and in
tabulated form in Appendix~A.
\begin{figure*}
 \centering
 \includegraphics[trim = 0mm 0mm 0mm 0mm, clip, angle=-90, width=0.75\textwidth]{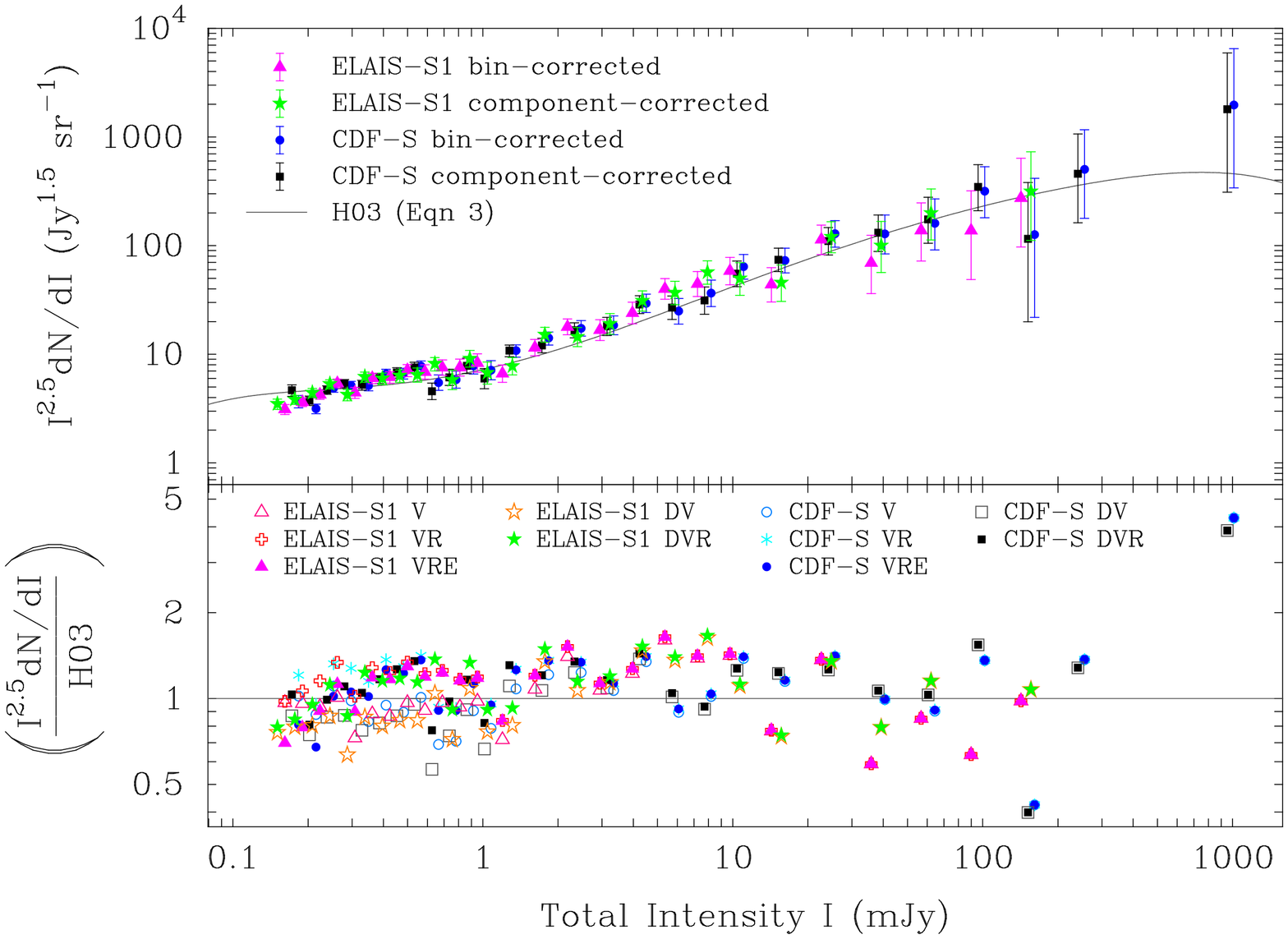}
 \caption{	Euclidean-normalised differential component counts at 1.4~GHz in total
 		intensity for the ATLAS CDF-S and ELAIS-S1 fields. {\it Upper Panel:}
		Fully corrected counts for each field resulting from the two alternative
		Eddington bias correction schemes applied in this work. Errors
		are $1\sigma$ Poissonian following \citet{regener}. The solid curve
		represents the empirical fit from \citet{2003AJ....125..465H}, given
		by Equation~(\ref{ch4:eqn:H03}).
		{\it Lower Panel:} Effects of resolution and Eddington bias corrections
		on differential component counts, relative to the H03 curve. For
		data associated with the bin-correction scheme for Eddington bias, counts
		are indicated following visibility area correction (V), resolution bias
		correction (VR), and finally all three corrections including the
		bin-value Eddington bias correction (VRE). For data associated with the
		component-correction scheme, counts are indicated following individual
		component deboosting and visibility area correction (DV), and finally
		the resolution bias correction (DVR). Note that the VRE and DVR points
		correspond to the points shown in the upper panel.
		}
 \label{ch5:fig:countsI}
\end{figure*}
\begin{figure*}
 \centering
 \includegraphics[trim = 0mm 0mm 0mm 0mm, clip, angle=-90, width=0.75\textwidth]{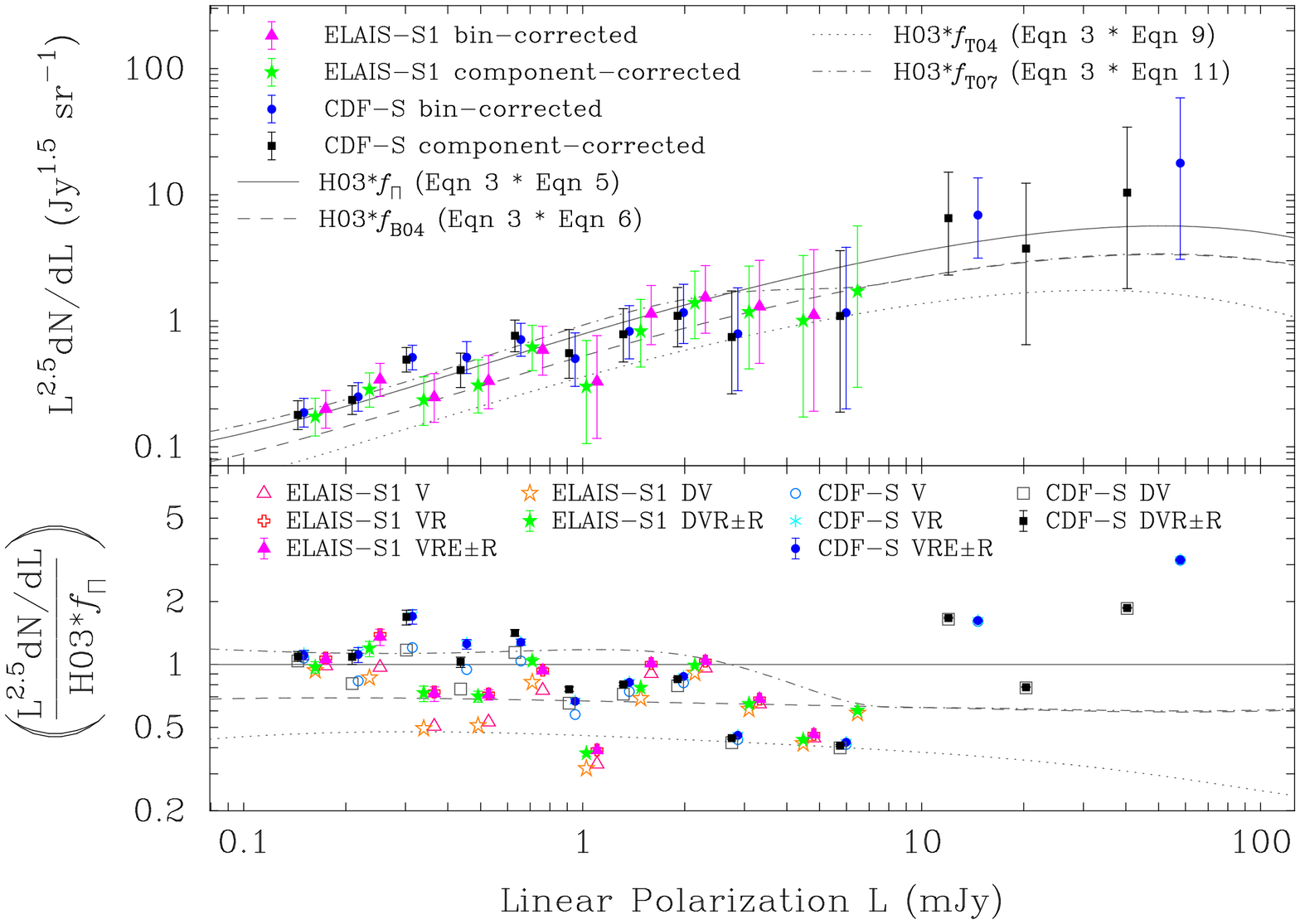}
 \caption{	Euclidean-normalised differential component counts at 1.4~GHz in linear
 		polarization for the ATLAS CDF-S and ELAIS-S1 fields. {\it Upper Panel:}
		Fully corrected counts for each field resulting from the two alternative
		Eddington bias correction schemes applied in this work. Errors
		are $1\sigma$ Poissonian following \citet{regener}. The four curves
		indicate number-count predictions obtained by convolving the total
		intensity H03 curve with the various fractional polarization distributions
		presented in Fig.~\ref{ch5:fig:PImodels} (see \S~\ref{ch5:SecResSubPI}).
		{\it Lower Panel:} Effects of resolution and Eddington bias corrections
		on differential component counts, relative to the $\tnm{H03}*f_{\ms \Pi}$
		curve. The legend is similar to that presented in Fig.~\ref{ch5:fig:countsI}.
		Error bounds on the VRE and DVR points indicate uncertainties regarding
		the resolution bias correction factors (see Fig.~22 in Paper I).
		}
 \label{ch5:fig:countsL}
\end{figure*}
Counts for each bin have been plotted and tabulated at the expected
average flux density, which we denote by $S_\tnm{\tiny AV}$, as given by Equation~(19)
from \citet{1984A&AS...58....1W}. This value takes into account the number-count
slope and becomes important when assigning flux densities for bins containing few
counts or with large widths in flux density space; $S_\tnm{\tiny AV}$ only equals
the bin geometric mean when $\gamma=2$, where $\gamma$ is the slope of the differential
number counts $dN/dS \propto S^{-\gamma}$.

Bin widths for all total intensity counts were selected to be a factor of 0.07~dex for
$I<1$~mJy, 0.13~dex for $1 \le I<10$~mJy, and 0.2~dex otherwise. In linear polarization,
bin widths were selected to be a factor of 0.16~dex for $L<10$~mJy, and 0.3~dex
otherwise. We removed all bins containing components with visibility area corrections
$\ge10$, so as to prevent the number-counts from being dominated by the few components
detected in the most sensitive and potentially least-representative regions of the
ATLAS images. (Note that we did not remove individual offending components
in order to retain the faintest bins, as this would have led to a bias in their resulting
number-counts.) In total intensity this resulted in the removal of the faintest few
bins containing $\sim$30 components from each of the CDF-S component- and
bin-corrected datasets, and $\sim$20 components from each of the ELAIS-S1 component-
and bin-corrected datasets. The maximum visibility area corrections for any components
in the remaining valid CDF-S and ELAIS-S1 bins were $9.9$ and $6.1$, respectively.
In linear polarization, the maximum visibility area corrections for any components
in the CDF-S and ELAIS-S1 datasets were $7.9$ and $3.7$, respectively. As a result,
we did not remove any bins in linear polarization.

Resolution and Eddington bias corrections were calculated in \S~7 of Paper~I. The former
was designed to correct for incompleteness to resolved components with low surface brightness,
and for the redistibution of counts between bins resulting from systematic undervaluation
of flux densities for components classified as unresolved. The latter was designed to correct
for the redistribution of counts between bins due to random measurement errors in the
presence of a non-uniformly distributed component population. These bias corrections were
calculated in Paper~I by assuming that the true underlying differential number counts in
total intensity were given by the sixth-order empirical fit to the Phoenix and FIRST
surveys presented by \citet{2003AJ....125..465H}. This fit, which we denote H03, is given by
\begin{equation}\label{ch4:eqn:H03}
	\log\left[\frac{dN_{\trm{\tiny H03}}/dI}{I^{\ms -2.5}}
	\right] = \sum_{j=0}^{6} a_{j} \left[ \log\left(
	\frac{I}{\trm{mJy}} \right)\right]^{j}\,,
\end{equation}
with $a_{\ms 0}=0.859$, $a_{\ms 1}=0.508$, $a_{\ms 2}=0.376$, $a_{\ms 3}=-0.049$,
$a_{\ms 4}=-0.121$, $a_{\ms 5}=0.057$, and $a_{\ms 6}=-0.008$.
To illustrate the potential boosting effects of an exaggerated population of
faint components, Paper~I also defined a modified H03 distribution, denoted H03M,
in which a Euclidean slope was inserted between 30$-$300$\mu$Jy,
\begin{equation}\label{ch4:eqn:H03M}
	\frac{dN_{\trm{\tiny H03M}}}{dI}\left(I\right)
	= \left\{
	\begin{array}{l l}
		dN_{\trm{\tiny H03}}/dI\left(I\right)
		& \quad \textrm{if $I\ge300~\mu$Jy}\\
		dN_{\trm{\tiny H03}}/dI\left(\textrm{$300~\mu$Jy}\right)
		& \quad \textrm{if $30\le I < 300~\mu$Jy}\\
		dN_{\trm{\tiny H03}}/dI\left(10\times I\right)
		& \quad \textrm{if $I<30~\mu$Jy}\,.\\
	\end{array} \right.
\end{equation}
For bias corrections in linear polarization, we modelled the true underlying
differential number counts $dN/dL$ by convolving the total intensity H03
distribution from Equation~(\ref{ch4:eqn:H03}) with a probability distribution
for fractional linear polarization $f_{\ms \Pi}(\Pi) \equiv f_{\ms \Pi}(L/I)$,
which we denote $\tnm{H03}*f_{\ms \Pi}$. The $f_{\ms \Pi}(\Pi)$ distribution
is presented in Equation~(\ref{ch5:eqn:fracpol}) in \S~\ref{ch5:SecResSubPI}.

The ATLAS DR2 component counts extend down to a flux density of approximately
140~$\mu$Jy in both total intensity and linear polarization. The brightest flux
density bins are sparsely sampled because the ATLAS survey areas are not large
enough to include significant numbers of increasingly rare bright components. In
both Fig.~\ref{ch5:fig:countsI} and Fig.~\ref{ch5:fig:countsL} we find that the number-counts
from the two separate ATLAS fields are consistent within the errors over their
full observed flux density ranges. The impacts of the combined resolution and
Eddington bias corrections on the number-counts appear to be relatively minor. In
total intensity, the two corrections largely cancel each other out, while in linear
polarization the resolution bias corrections dominate. In both total intensity and
linear polarization, the combined corrections affect the underlying visibility
area corrected counts by a factor of {\footnotesize $\lesssim$}~0.5, and do not
affect the counts for $S$~{\footnotesize $\gtrsim$}~3~mJy. We find that differences
between the two independent Eddington bias correction schemes are largely negligible
for both the total intensity and linear polarization number-counts, providing
confidence in these approaches.

In Fig.~\ref{ch5:fig:countsI} we find that the total intensity counts closely
follow the H03 model within a factor of $\sim20\%$, though the ATLAS counts may
begin to systematically drop below the H03 model for $S$~{\footnotesize $\lesssim$}~$0.2$~mJy.
It is likely that the drop is caused by residual
incompleteness in our resolution bias corrections, in turn caused by
uncertainties regarding our assumed true angular size distribution
for $\Theta<3\arcsec$ as discussed in \S~7.1 of Paper I. However,
we note that if we assume that the \citet{2003A&A...403..857B} model
presented in Fig.~19 of Paper~I is the best representation of the
true angular size distribution (without any flux density scaling), then
the faintest bins at $S\approx140$~$\mu$Jy only require an additional
correction factor of at most approximately $+$30\%. The faintest bins
are therefore consistent with the H03 model.

As we do not find any systematic divergence between the ATLAS total intensity
counts and the H03 model at the faintest flux densities (when accounting for the
suspected residual resolution bias described above), we confirm that the H03
model is suitable for predicting 1.4~GHz component counts (and source counts
as described below) down to at least $\sim100$~$\mu$Jy in surveys with
resolution FWHM $\sim10\arcsec$. Should we have found a systematic divergence,
it would have indicated that our predicted Eddington bias corrections were
unrealistic, and that in turn the H03 model underpinning these corrections
formed an increasingly poor representation of the true number-counts for
decreasing flux density. Under this hypothetical situation, an iterative
approach would have been required in order to correctly identify an input
true number-count model so as to bring about convergence with the fully
corrected observed counts. In \S~7.2 of Paper I we predicted the levels
of Eddington bias that would be present within the observed ATLAS counts if
the true counts were given by the H03 or H03M models [the latter model
contains a larger population of components with $S<0.3$~mJy than the former;
see Equation~(\ref{ch4:eqn:H03M})].
We predicted that the H03M model would induce significantly greater Eddington
bias at $S<0.3$~mJy than the H03 model (see Fig.~23 in Paper~I). Therefore, if
the H03 model was used to predict the observed Eddington bias when in fact
the H03M model best represented the true counts, then the observed counts
would exhibit significant positive residual Eddington bias; if vice versa,
the residual bias would be negative. Given that we do not observe a
systematic rise (or fall) at faint flux densities in the fully corrected
ATLAS counts (again accounting for the suspected residual resolution bias
described above), we conclude that the H03M model
is not supported by the ATLAS data. We note that the resolution bias
corrections applied in this work are practically insensitive to changes
between the H03 and H03M models. This is because for any given flux
density bin, the resolution bias corrections are unaffected by the assumed
form of the number-count distribution at fainter flux densities. Therefore,
assuming that our resolution bias corrections are appropriate to begin with,
we can focus on Eddington bias alone in order to draw the conclusions
described above.

Below a flux density of $\sim1$~mJy, we expect the ATLAS total intensity
component counts to be dominated by single-component sources, with negligible
contributions from components within multi-component sources. While we are unable
to explicitly quantify this expectation given present data, we note that
conservatively $<30\%$ of all 2416 ATLAS components reside within
multi-component sources (this fraction takes into account the number of
components estimated to reside within as-yet unassociated multi-component
sources; see \S~6.1 of Paper~I).
We expect that most of these multi-component sources represent FRII sources,
which are known to dominate the source counts at flux densities {\footnotesize
$\gtrsim$}~10~mJy and which diminish significantly below $\sim 1$~mJy
\citep[e.g.][]{2008MNRAS.388.1335W}. At sub-mJy levels, radio sources in general
are expected to have angular sizes $\sim 1\arcsec$; these are likely to be
observed as single-component sources in ATLAS. Therefore, we conclude that
the ATLAS component counts may act as a suitable proxy for source counts at
sub-mJy levels. We note that our characterisation of the faint component/source
population using the H03 model in this work has relied on the similar resolutions
of the ATLAS and Phoenix surveys. Should these resolutions have differed
significantly, so too would have the properties of their observed components.
\citet{2003AJ....125..465H} obtained their model by using a sixth-order fit to
the observed component counts from the Phoenix survey, supplemented at
$S>2.5$~mJy by source counts from the FIRST survey \citep{1997ApJ...475..479W}.
The H03 model was thus intended to characterise source counts at all flux
densities, despite being derived from a component catalogue at faint flux
densities.

\begin{figure}
 \centering
 \includegraphics[trim = 0mm 0mm 0mm 0mm, clip, angle=-90, width=0.48\textwidth]{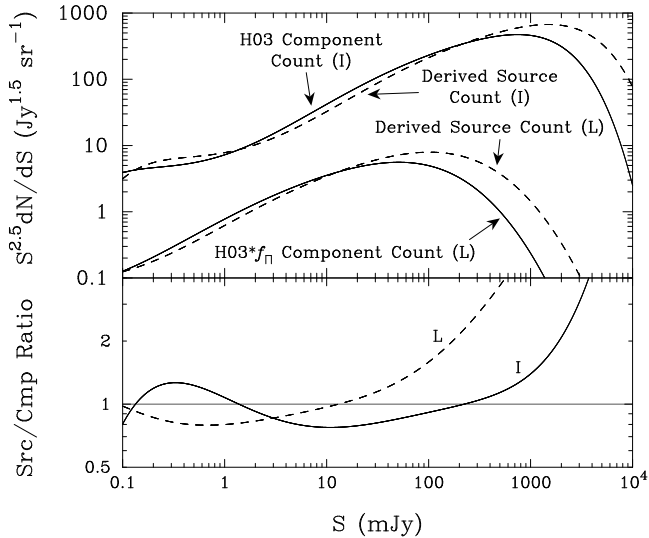}
 \caption{	Idealised relationship between source counts and component counts
 		in total intensity (I) and linear polarization (L). Dashed curves
		in the upper panel are derived from their respective solid curves
		by assuming that all I or L sources comprise two identical I or L
		components with half the flux density of their parents. Curves in
		the lower panel indicate the ratio between source and component
		counts for I and L. See \S~\ref{ch5:SecResSubCnts} for details.
		}
 \label{ch5:fig:ratioSC}
\end{figure}

For $S>2.5$~mJy, the ATLAS total intensity component counts follow the H03 model
and thus the FIRST source counts. We explain this correspondence as follows by
first presenting results that examine how source and component counts are expected
to differ. Given that FRII sources dominate the source counts above $\sim10$~mJy
and that these sources are likely to comprise multiple components within a survey
such as ATLAS, we expect the differential counts for sources to rise and extend
to brighter flux densities than those for components. To roughly illustrate this
behaviour and examine the difference between source and component counts in
general, we considered an idealised scenario in which all sources were assumed
to comprise two identical components, each with half the flux density of their
parent. For illustrative purposes we assumed that the component count distribution
was given by the H03 model. To derive the idealised differential source counts,
we integrated the differential component counts to obtain integral component
counts, divided these integral counts by two, doubled the flux density scale,
and differentiated. For completeness, we also derived differential
source counts in linear polarization by following a similar procedure,
where the relevant differential component counts were assumed to follow the
$\tnm{H03}*f_{\ms \Pi}$ model. We present the resulting total intensity and
linear polarization source counts in Fig.~\ref{ch5:fig:ratioSC}.
We find that the predicted source counts remain within $\sim 30\%$ of the
component counts across the flux density ranges probed by the ATLAS data
in total intensity ($I$~{\footnotesize $\lesssim$}~1~Jy) and linear polarization
($L$~{\footnotesize $\lesssim$}~100~mJy). (Separately, while not shown, we
note that the integral counts for both components and sources within our
rudimentary model are very similar, for both total intensity and linear
polarization.) As expected, at bright flux densities the component counts
drop below the source counts, though these drops occur at brighter flux
densities than relevant to the ATLAS data. Note that in reality, the
differential source and component counts are likely to overlap more closely
than presented in Fig.~\ref{ch5:fig:ratioSC} because of the presence of
single-component sources. Thus we conclude that for surveys with resolution
FWHM $\sim10\arcsec$ similar to Phoenix and ATLAS, the H03 model may be
used to characterize both component and source counts in total intensity
for $S$~{\footnotesize $\lesssim$}~1~Jy.

We conjecture that, as modelled above, the H03 model characterises
component rather than source counts at all flux densities, including at $S>1$~Jy.
To justify this claim, we note that components in the FIRST survey were only
grouped into multi-component sources if they were located within 50\arcsec\
\citep{1997ApJ...475..479W}. From Fig.~\ref{ch5:fig:fracpolTheta} of this work
we can see that a cutoff of 50\arcsec\ is likely to be too small to capture
sources with the most widely-separated components, which are also likely to
be the brightest sources. In addition, flux densities for extended
FIRST components are likely to be underestimated due to insensitivity to
extended emission. Therefore, the FIRST source counts are likely to
be deficient at the brightest flux densities. Incidentally, the FIRST
source counts and thus the H03 model appear to form a suitable hybrid
distribution for describing component counts at all flux densities in surveys
with resolution FWHM $\sim10\arcsec$ such as ATLAS. We may therefore conclude
that the $\tnm{H03}*f_{\ms \Pi}$ model is suitable for characterising
component counts in linear polarization at all flux densities, not just
at $L$~{\footnotesize $\lesssim$}~100~mJy where differences between
polarized component and source counts are likely to diminish as shown in
Fig.~\ref{ch5:fig:ratioSC}. If the H03 model were to better represent source
counts rather than component counts at $I>1$~Jy, then the polarized counts
resulting from convolution with $f_{\ms \Pi}$ would reside ambiguously
between a component and source count distribution for
$L$~{\footnotesize $\gtrsim$}~5~mJy. Thus it would be inappropriate to
estimate integral component or source counts from the $\tnm{H03}*f_{\ms \Pi}$
(or indeed H03) model; this point is relevant to results presented shortly.

In Fig.~\ref{ch5:fig:countsL} we find that the ATLAS linear polarization component
counts steadily decline with decreasing flux density, as generally predicted by all
four models displayed in the background. The solid curve displays our assumed
true component count model, namely $\tnm{H03}*f_{\ms \Pi}$, which we used to
derive the corrections for resolution and Eddington bias. The fully corrected ATLAS
counts closely follow this model within statistical error, indicating consistency
between the model, the corrections, and the observational data. Each of the four
background models in Fig.~\ref{ch5:fig:countsL} were calculated by convolving the
H03 model with a fractional polarization distribution. We note that these
convolutions are only appropriate because, as described above, the H03 model
appears to appropriately characterise the total intensity component counts at
all flux densities relevant to ATLAS. In \S~\ref{ch5:SecResSubPI} we describe each
of the fractional polarization distributions underlying the four background
models, and compare their abilities to predict the ATLAS polarized counts
and polarization data in general.

\begin{figure}
 \centering
 \includegraphics[trim = 0mm 0mm 0mm 0mm, clip, angle=-90, width=0.48\textwidth]{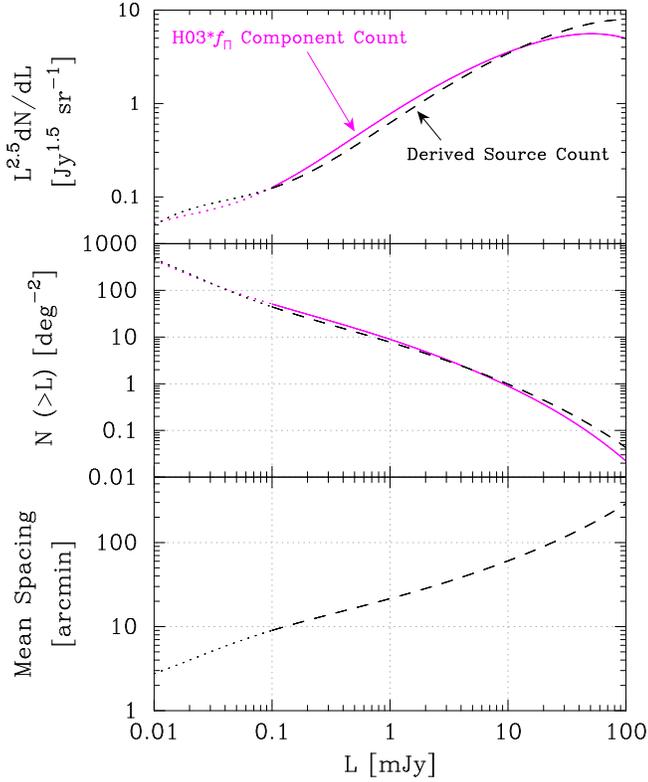}
 \caption{	
 		{\it Upper panel:} Reproduction of linearly polarized component and
		source counts from Fig.~\ref{ch5:fig:ratioSC}. The dotted portions of the
		curves at $L<0.1$~mJy indicate extrapolations beyond the observed ATLAS DR2
		data, assuming an unchanging distribution of fractional polarization
		(though note comments in \S~\ref{ch5:SecResSubPI}). {\it Middle panel:}
		Estimated sky density for components and sources above a given linearly
		polarized flux density. {\it Lower panel:} Mean spacing between linearly
		polarized sources as a function of the faintest detected sources.
		}
 \label{ch5:fig:17}
\end{figure}

The number of polarized components expected per square degree at or brighter than a
given flux density, as constrained by the observed ATLAS component counts, can be
estimated by integrating the $\tnm{H03}*f_{\ms \Pi}$ polarized count distribution
(the solid curve in Fig.~\ref{ch5:fig:countsL}). The resulting integral component
counts are displayed in Fig.~\ref{ch5:fig:17}. We estimate that the sky density
of polarized components for $L\ge200$~$\mu$Jy is 30~deg$^{-2}$, for $L\ge100$~$\mu$Jy
it is 50~deg$^{-2}$, and for $L\ge50$~$\mu$Jy it is 90~deg$^{-2}$. If we make the
rudimentary assumption described earlier regarding Fig.~\ref{ch5:fig:ratioSC} that
every polarized component belongs to a dual-component source with double the flux
density, we can estimate the integral source count distribution; this is displayed
alongside the integral component count distribution in Fig.~\ref{ch5:fig:17}.
We thus estimate that the sky density of polarized sources for $L\ge200$~$\mu$Jy
is $\sim 25$~deg$^{-2}$, and for $L\ge100$~$\mu$Jy it is $\sim 45$~deg$^{-2}$.
We expect that these integral source count estimates are accurate to within
10\%, even if a more suitable model incorporating polarized single-component
sources is utilised.

\subsection{Distribution of Fractional Linear Polarization}\label{ch5:SecResSubPI}

In this section we present a model to describe the distribution of fractional
polarization for AGN sources and their components/groups observed at 1.4~GHz
in surveys with resolution FWHM~{\footnotesize $\gtrsim$}~10\arcsec, as
constrained by the ATLAS DR2 data.

There appears to be a significant overlap between the fractional polarization
properties of all classification Types representing both components/groups
and sources in Fig.~\ref{ch5:fig:fracpolTypes}. Taking into account the presence of
upper limits (see Fig.~\ref{ch5:fig:fracpolRaw}), we find that typical levels of
fractional polarization are concentrated between 0.4\% and 20\%, regardless of
whether the focus is on sources or on their constituent components/groups.
Given this apparent overlap, we assume for simplicity that the
distribution of fractional polarization for both components/groups and
sources can be modelled using the same PDF, which we denote by $f_{\ms \Pi}(\Pi)$.
Before presenting our model for this distribution, we note three caveats.
First, following our conclusions presented in \S~\ref{ch5:SecResIdentM} regarding
potential correlation of the distribution of fractional polarization with total
flux density, we assume
that $f_{\ms \Pi}(\Pi)$ is independent of total intensity flux density. This
assumption may not be suitable for $I$~{\footnotesize $\lesssim$}~$10$~mJy for which
our ATLAS data become sparse. Second, our model for $f_{\ms \Pi}(\Pi)$ may only
be relevant for surveys with resolution FWHM~{\footnotesize $\gtrsim$}~10\arcsec.
Surveys with finer resolution may encounter less beam depolarization across
components, and thus recover higher average levels of fractional polarization
(in \S~5 of Paper~I we found that $<17\%$ of polarized ATLAS components were
resolved). We note that surveys with coarser resolution will incur increased
blending between components within multi-component sources, resulting in a
greater number of low-$\Pi$ sources than observed for ATLAS due to enhanced
beam depolarization. And third, given that all polarized components in ATLAS
DR2 are associated with AGNs, we restrict our model for $f_{\ms \Pi}(\Pi)$ to
the characterisation of AGNs, rather than the characterisation of all radio
sources including SFGs and individual stars. We do not attempt to differentiate
between different types of AGNs or their components within our model,
i.e. FRI/FRII/radio quiet/core/lobe. We discuss fractional polarization levels
for SFGs in \S~\ref{ch5:SecDiscSFG}.

We modelled $f_{\ms \Pi}(\Pi)$ by qualitatively fitting two independent
sets of ATLAS data: (i) the fractional polarizations
of components, groups, and sources displayed in Fig.~\ref{ch5:fig:fracpolRaw},
importantly taking into account upper limits,
and (ii) the differential number-counts for polarized components displayed in
Fig.~\ref{ch5:fig:countsL}. We obtained a concordance fit to these data by
modelling $f_{\ms \Pi}(\Pi)$ using a log-normal distribution,
\begin{equation}\label{ch5:eqn:fracpol}
	f_{\ms \Pi}\left(\Pi\right) =
	\frac{1}{\Pi \sigma_{\ms 10}\ln\!\left(10\right)\sqrt{2\pi}}
	\exp\Bigg\{\frac{-\left[\log_{\ms 10}\!\left(\Pi/\Pi_{\ms 0}\right)\right]^{\ms 2}}
	{2 \sigma_{\ms 10}^{\ms 2}}\Bigg\} \,,
\end{equation}
where the parameters $\Pi_{\ms 0}$ and $\sigma_{\ms 10}$ are the median fractional
polarization and scale parameter, respectively, given by best-fit values
$\Pi_{\ms 0}=4.0\%$ and $\sigma_{\ms 10}=0.3$.
The fit given by Equation~(\ref{ch5:eqn:fracpol}) is consistent with the result
obtained by analysing the fractional polarization data alone, using the
product-limit estimator \citep{kapmei} as implemented within the
{\tt survival}\footnote{http://cran.r-project.org/web/packages/survival/index.html}
package in the {\tt R}\footnote{http://www.r-project.org} environment. The mean level of fractional
polarization for the distribution in Equation~(\ref{ch5:eqn:fracpol}) is given by
$\log_{\ms 10}\mu_{\ms \Pi}=\log_{\ms 10}\left(\Pi_{\ms 0}\right)+0.5\ln\!\left(10\right)\sigma_{\ms 10}^{\ms 2}$,
which equates to $\mu_{\ms \Pi}=5.1\%$. For values of $\Pi_{\ms 0}$ or
$\sigma_{\ms 10}$ larger than the best-fit values above, we found that the
$\tnm{H03}*f_{\ms \Pi}$ model predicted differential counts in excess of
the observed ATLAS counts. For smaller values, the predicted counts were deficient.

\begin{figure}
 \centering
 \includegraphics[trim = 0mm 0mm 0mm 0mm, clip, angle=-90, width=0.46\textwidth]{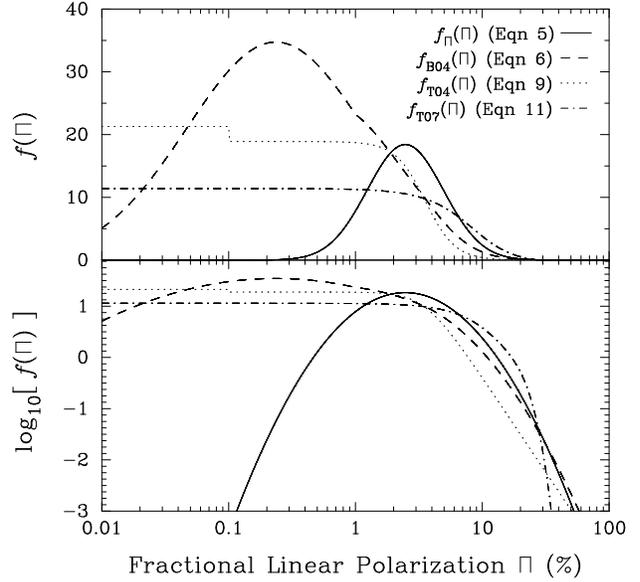}
 \caption{	Proposed models of 1.4~GHz fractional linear polarization; see
 		\S~\ref{ch5:SecResSubPI} for details. The solid curve represents
		a fit to the ATLAS DR2 data, and is the distribution assumed in
		this work. The vertical axis is linear in the upper panel and
		logarithmic in the lower panel. The curves presented here are compared
		to the fractional polarization data in Fig.~\ref{ch5:fig:fracpolRaw}
		(also Fig.'s \ref{ch5:fig:fracpolClass}-\ref{ch5:fig:fracpolTheta}
		and \ref{ch5:fig:fracpolTheta2}) and to the number count data in
		Fig.~\ref{ch5:fig:countsL} (also Fig.~\ref{ch5:fig:countsL2}).
}
 \label{ch5:fig:PImodels}
\end{figure}

We plot Equation~(\ref{ch5:eqn:fracpol}) in Fig.~\ref{ch5:fig:PImodels}.
For comparison we also plot the 1.4~GHz fractional polarization distributions
proposed by \citet{2004NewAR..48.1289B}, \citet{2004MNRAS.349.1267T},
and \citet{2007ApJ...666..201T}. For clarity we explicitly document each
of these distributions, as follows. \citet{2004NewAR..48.1289B} investigated
the distribution of fractional polarization for NVSS sources with $I>80$~mJy,
which they fit using the following quasi log-normal form,
\begin{equation}\label{ch5:eqn:fracpolB04}
	f_{\trm{\tiny B04}}\left(\Pi\right) =
	\frac{a_{\trm{\tiny B04}}}{\Pi \ln\!\left(10\right)}
	\exp\Bigg\{\frac{-\left[\log_{\ms 10}\!\left(\Pi/\Pi_{\trm{\tiny B04}}
	\right)\right]^{\ms 2}}
	{2 \sigma_{\trm{\tiny B04}}^{\ms 2}}\Bigg\} \,,
\end{equation}
where
\begin{equation}\label{ch5:eqn:fracpolB04amp}
	a_{\trm{\tiny B04}} = \left\{
	\begin{array}{l l}
		0.690 & \quad \textrm{if $\log_{\ms 10}\Pi\le-2$}\\
		0.808 & \quad \textrm{if $\log_{\ms 10}\Pi>-2$}\,,\\
	\end{array} \right.
\end{equation}
\begin{equation}\label{ch5:eqn:fracpolB04width}
	\sigma_{\trm{\tiny B04}} = \left\{
	\begin{array}{l l}
		0.700 & \quad \textrm{if $\log_{\ms 10}\Pi\le-2$}\\
		0.550 & \quad \textrm{if $-2<\log_{\ms 10}\Pi\le-1.5$}\\
		0.353 & \quad \textrm{if $\log_{\ms 10}\Pi>-1.5$}\,,\\
	\end{array} \right.
\end{equation}
and where $\log_{\ms 10}\Pi_{\trm{\tiny B04}}=-1.5$. The median and mean
fractional polarization levels of the $f_{\trm{\tiny B04}}$ distribution
are 2.1\% and 3.3\%, respectively. Similarly, \citet{2004MNRAS.349.1267T}
investigated the distribution of fractional polarization for NVSS sources
with $I>100$~mJy, which they fit using the following monotonic form,
\begin{equation}\label{ch5:eqn:fracpolT04}
	f_{\trm{\tiny T04}}\left(\Pi\right) = 1.32 \left\{ a_{\trm{\tiny T04}}
	\left[2.7+0.025\left(100\,\Pi\right)^{\ms 3.7}\right]^{\ms -1}
	+ b_{\trm{\tiny T04}}\right\}
\end{equation}
where $a_{\trm{\tiny T04}}=51$,
\begin{equation}\label{ch5:eqn:fracpolT04bval}
	b_{\trm{\tiny T04}} = \left\{
	\begin{array}{l l}
		2.4 & \quad \textrm{if $\Pi\le0.1$}\\
		0 & \quad \textrm{if $\Pi>0.1$}\,,\\
	\end{array} \right.
\end{equation}
and where we have included a correction factor of 1.32 to ensure that the
distribution is normalised. The median and mean fractional polarization levels
of the $f_{\trm{\tiny T04}}$ distribution are 2.1\% and 2.7\%, respectively.
\citet{2007ApJ...666..201T} fit the distribution of fractional polarization
for sources with $I<30$~mJy in the ELAIS-N1 field by modifying a Gram-Charlier
series of type A \citep[e.g.][]{1993ApJ...407..525V}, resulting in the following
monotonic form,
\begin{equation}\label{ch5:eqn:fracpolT07}
	f_{\trm{\tiny T07}}\left(\Pi\right) = \left\{
	\begin{array}{l l}
		11.06\, \exp\!\left(\frac{-\Pi^{2}}{2\sigma_{\trm{\tiny T07}}^{\ms 2}}\right)
		\Bigg\{1 + \frac{h_{\ms 4}}{\sqrt{24}}\Bigg[
		4\!\left(\frac{\Pi}{\sigma_{\trm{\tiny T07}}^{\ms }}\right)^{\!4} - \\ \hspace{2.7cm}
		12\!\left(\frac{\Pi}{\sigma_{\trm{\tiny T07}}^{\ms }}\right)^{\!2} + 3 \Bigg]\Bigg\}
		& \\ \hspace{4cm} \textrm{if $I<30$~mJy}\\
		f_{\trm{\tiny B04}}\left(\Pi\right) & \\ \hspace{4cm} \textrm{if $I\ge30$~mJy}\,,\\
	\end{array} \right.
\end{equation}
where $\sigma_{\trm{\tiny T07}}=0.07$, $h_{\ms 4}=0.05$, and where we have
included a correction factor of 11.06 to ensure that the distribution is
normalised. For $I\ge30$~mJy, \citet{2007ApJ...666..201T} found that the
ELAIS-N1 data were consistent with the $f_{\trm{\tiny B04}}$ distribution
from Equation~(\ref{ch5:eqn:fracpolB04}). The median and mean fractional
polarization levels of the $f_{\trm{\tiny T07}}$ distribution for
$I<30$~mJy are 4.8\% and 6.0\%, respectively.

The four curves presented in Fig.~\ref{ch5:fig:PImodels} are replicated
in Figs.~\ref{ch5:fig:fracpolRaw}--\ref{ch5:fig:fracpolTheta} and
Fig.~\ref{ch5:fig:fracpolTheta2}. The four curves are also presented in
Fig.'s~\ref{ch5:fig:countsL} and \ref{ch5:fig:countsL2} following convolution
with the H03 differential count model. In Fig.~\ref{ch5:fig:countsL2} we find that the
fractional polarization distributions proposed by \citet{2004NewAR..48.1289B},
\citet{2004MNRAS.349.1267T}, and \citet{2007ApJ...666..201T} are in general agreement
with the observed ATLAS polarized number counts. The models predict polarized counts
that are within a factor of 5 of each other, and they all pass within a few standard
errors of the ATLAS data points. However, we find that these three
distributions are incompatible with the observed distribution of fractional
polarization for ATLAS components, groups, sources, and in particular upper
limits as presented in Fig.~\ref{ch5:fig:fracpolRaw}. The extended tails below
$\Pi<1\%$ for the distributions proposed by \citet{2004NewAR..48.1289B}
and \citet{2004MNRAS.349.1267T} are likely to reflect the various systematic
biases we described earlier in \S~\ref{ch5:SecResIdentM} regarding the NVSS data.
Polarized flux densities for NVSS sources were recorded regardless of whether
or not the measurements met statistical criteria for formal detection. If
upper limits were calculated for the NVSS data following a similar procedure
to that described for the ATLAS data in \S~6.2 of Paper~I, then we
suspect that far fewer detections strictly implying $\Pi<1\%$ would have
been made. We note that the $f_{\ms \Pi}$ model proposed in this work peaks
at $\Pi\approx2.5\%$, which is consistent with the NVSS data for $\Pi>1\%$
from \citet{2002A&A...396..463M} and \citet{2004MNRAS.349.1267T}. The extended
tail below $\Pi<1\%$ in the \citet{2007ApJ...666..201T} model reflects their
assumption that the distribution peaks at $\Pi=0$ and declines monotonically
with increasing $\Pi$. The ATLAS DR2 data do not support this
assumption.

As noted earlier, a caveat of the $f_{\ms \Pi}$ model is that it may not
be suitable for $I$~{\footnotesize $\lesssim$}~$10$~mJy, because the upper
limits presented in Fig.~\ref{ch5:fig:fracpolRaw} do not constrain the behaviour
of the true fractional polarization distribution for low values of $\Pi$.
However, given that the maximum level of fractional polarization exhibited
by ATLAS components and sources appears to be limited to
$\Pi$~{\footnotesize $\lesssim$}~$20\%$, and given that this limit appears
to be uncorrelated with flux density down to at least $\sim1$~mJy (see
comments regarding Fig.~\ref{ch5:fig:fracpolClass} in \S~\ref{ch5:SecResIdentM}), 
we may draw tentative conclusions regarding the true distribution
of fractional polarization for
$1$~{\footnotesize $\lesssim$}~$I$~{\footnotesize $\lesssim$}~$10$~mJy.
The ATLAS DR2 data are consistent with 3 general alternatives. First, the
$f_{\ms \Pi}$ distribution may remain unchanged for $I<10$~mJy. Second,
for decreasing $I$, the mean of $f_{\ms \Pi}$ may decrease while its
dispersion increases so as to maintain an approximately constant level
of fractional polarization for outliers with large $\Pi$. And third, for
decreasing $I$, the mean of $f_{\ms \Pi}$ may increase while its
dispersion decreases. More sensitive observations are required to
distinguish between these alternatives.

\section{Discussion}\label{ch5:SecDisc}

\subsection{Comparison of Component Counts to Other Deep Surveys}\label{ch5:SecDiscSubCounts}

In Fig.~\ref{ch5:fig:ratioSC} of \S~\ref{ch5:SecResSubCnts} we demonstrated that
differences between differential number-counts of components and sources within a
survey such as ATLAS are likely to be negligible below $\sim1$~Jy in total intensity,
and below $\sim100$~mJy in linear polarization. We may therefore directly compare the
ATLAS DR2 component counts with source counts from the literature in both total
intensity and linear polarization. We present these comparisons in the following
two sections.

\subsubsection{Total Intensity}\label{ch5:SecDiscSubCountsI}

\begin{figure*}
 \centering
 \includegraphics[trim = 0mm 0mm 0mm 0mm, clip, angle=-90, width=0.75\textwidth]{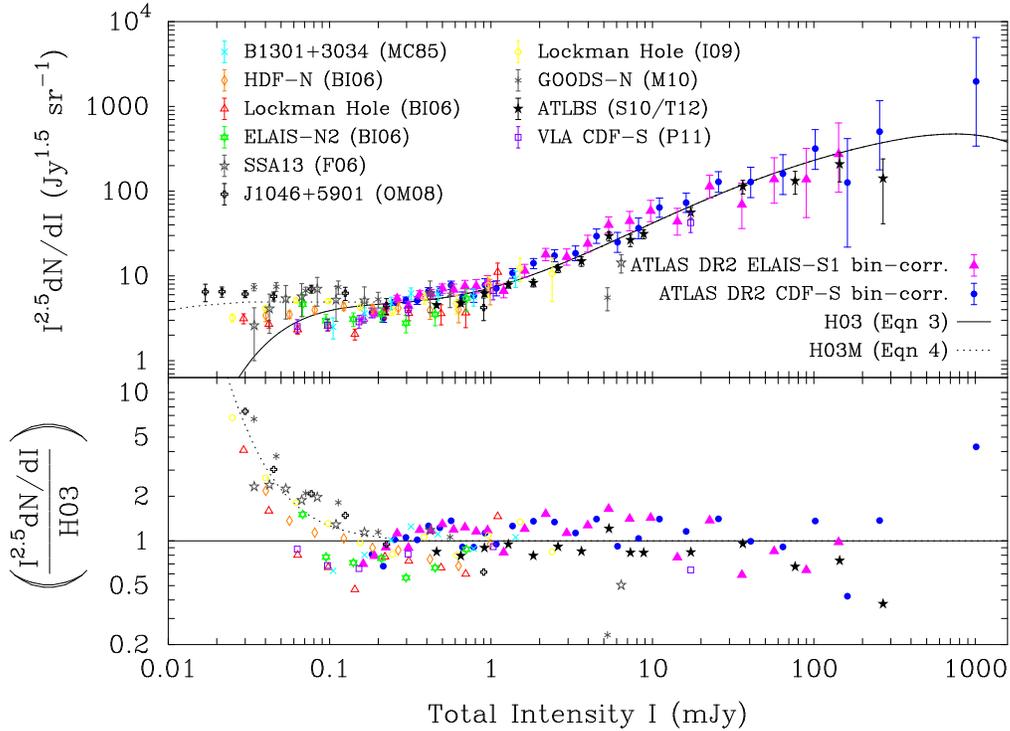}
 \caption{	Comparison of ATLAS DR2 bin-corrected total intensity component counts
 		with 1.4~GHz source counts from other surveys; see \S~\ref{ch5:SecDiscSubCountsI}
		for reference details. The panel layout follows Fig.~\ref{ch5:fig:countsI}.
		The solid and dotted curves represent the H03 and H03M models given by
		Equations~(\ref{ch4:eqn:H03}) and (\ref{ch4:eqn:H03M}), respectively.
		Data from the upper panel are reproduced in the lower panel relative
		to the H03 curve.
		}
 \label{ch5:fig:countsI2}
\end{figure*}

In Fig.~\ref{ch5:fig:countsI2} we compare the ATLAS DR2 bin-corrected total intensity
component counts (from Fig.~\ref{ch5:fig:countsI} or tabulated data from Appendix~A)
with source counts from other 1.4~GHz surveys of comparable sensitivity.
These include the B1301$+$3034 field \citep{1985AJ.....90.1957M}, the
HDF-N, Lockman Hole, and ELAIS-N2 fields \citep{2006MNRAS.371..963B},
the SSA13 field \citep{2006ApJS..167..103F}, the J1046$+$5901 field
\citep[][hereafter OM08]{2008AJ....136.1889O}, a revised survey of the Lockman Hole
\citep{2009MNRAS.397..281I}, the GOODS-N field \citep{2010ApJS..188..178M},
the CDF-S field observed with the VLA \citep{2011ApJ...740...20P}, and
the ATLBS fields with counts at $S>8$~mJy \citep{2010MNRAS.402.2792S}
and deeper counts \citep{2013ApJ...762...16T}.

At high flux densities the source counts are dominated by luminous radio galaxies
and quasars. The flattening of the source counts below 1~mJy is produced by the
emerging dominance of a population of sources comprised of radio-quiet
AGNs (AGNs lacking significant jets and dominated in the radio band by non-thermal
emission), low-power radio-loud AGNs, and star forming galaxies
\citep{2004NewAR..48.1173J,2006MNRAS.372..741S,2008ApJS..177...14S,
2008MNRAS.390..819G,2011ApJ...740...20P}. The extent to which the source counts
flatten is somewhat controversial because counts from deep surveys appear to exhibit
a large degree of scatter, for example as seen in Fig.~\ref{ch5:fig:countsI2} where
there is a factor of 2 variation in the counts below 1~mJy. Measurements at 3~GHz
from the Absolute Radiometer for Cosmology, Astrophysics, and Diffuse Emission (ARCADE)
2 balloon-borne experiment have indicated a temperature for the radio background about
five times that previously expected from known populations of radio sources
\citep{2011ApJ...734....5F,2011ApJ...734....6S}, which if not due to a residual
calibration error \citep{2013ApJ...776...42S} suggest the presence of a new population
of faint ($<10$~$\mu$Jy at 1.4~GHz) or diffuse (few Mpc in extent) extragalactic sources
\citep{2010MNRAS.409.1172S,2011MNRAS.415.3641V,2012ApJ...758...23C,2014ApJ...780..112H}.

Some studies have attributed the large scatter in the faint counts to cosmic
variance, namely to intrinsic differences between survey fields caused
by large scale structure \citep[e.g.][]{2004MNRAS.352..131S}. However, significant
differences in the counts for fields observed in separate studies, such as the
GOODS-N field (located within the HDF-N field) or the Lockman Hole (see
Fig.~\ref{ch5:fig:countsI2}), indicate that data processing and calibration errors may
be entirely responsible for the scatter \citep[e.g.][]{2009MNRAS.397..281I}.
By considering the consistent power-law form of the angular correlation function for
both NVSS and FIRST sources obtained by \citet{2002MNRAS.337..993B} and
\citet{2003A&A...405...53O}, \citet{2010MNRAS.404..532M} estimated the cosmic
variance for millijansky radio sources to be
$\sigma_v^2=2.36\times10^{-3}\;\Omega^{-0.4}$ where $\Omega$ is the survey area in
square degrees. The total variance for each source count bin containing $N$ sources
is then given by $N\left(1+N\sigma_v^2\right)$, which includes Poisson chance. For
a survey with $\Omega>0.5$~deg$^2$ and $N<100$, cosmic variance
contributes $<15\%$ to the total rms uncertainty for each bin (this is
consistent with a similar estimate presented by \citealt{2006MNRAS.372..741S}
and a more detailed analysis by \citealt{2013MNRAS.432.2625H}). The
clustering behaviour of sub-millijansky sources is likely to be similar to that
of millijansky sources \citep{2003A&A...405...53O,2011MNRAS.418.2251F}, or perhaps
even less clustered \citep{2006MNRAS.368..935N}, in which case the cosmic variance
contribution estimated above represents a conservative upper limit. The error bars
for many of the faintest counts in Fig.~\ref{ch5:fig:countsI2} require enlargement
by factors much larger than $\sim15\%$ to become consistent with each other within
a few standard errors. Our experience in constructing source counts for ATLAS suggests
to us that there are a large number of data processing procedures that, if not carefully
implemented, could easily give rise to significant systematic biases of order the observed
scatter in the faint counts. We therefore agree with previous conclusions in the literature
\citep[e.g.][]{2006MNRAS.371..963B,2006MNRAS.372..741S,2007ASPC..380..189C,2009MNRAS.397..281I,
2010A&ARv..18....1D,2012ApJ...758...23C,2013MNRAS.432.2625H} that the observed scatter in
the sub-millijansky counts is likely to be significantly affected by data processing differences
between surveys.

The ATLAS data support the H03 model down to $\sim100$~$\mu$Jy and rule out any
flattening above this level; flattening similar to the H03M model is ruled out
by a lack of residual Eddington bias. However, the DR2 data are not sensitive
enough to support or refute the general trend of flattening reported by deeper
surveys. Recently, \citet{2012ApJ...758...23C} used the probability of deflection
technique [$p(D)$; \citealt{1957PCPS...53..764S}] and a spectral index conversion
to investigate the behaviour of the 1.4~GHz source counts at $2-20$~$\mu$Jy
within a confusion-limited observation of the OM08 J1046$+$5901 field at 3~GHz.
By combining the results from a similar $p(D)$ analysis performed by MC85,
\citet{2012ApJ...758...23C} ruled out any flattening or an upturn in the
1.4~GHz Euclidean counts between 2~$\mu$Jy and 100~$\mu$Jy, such as that reported by OM08
or proposed to account for the ARCADE 2 results \citep{2011ApJ...734....6S,
2010MNRAS.409.1172S,2011MNRAS.415.3641V}.

\subsubsection{Linear Polarization}\label{ch5:SecDiscSubCountsL}

\begin{figure*}
 \centering
 \includegraphics[trim = 0mm 0mm 0mm 0mm, clip, angle=-90, width=0.75\textwidth]{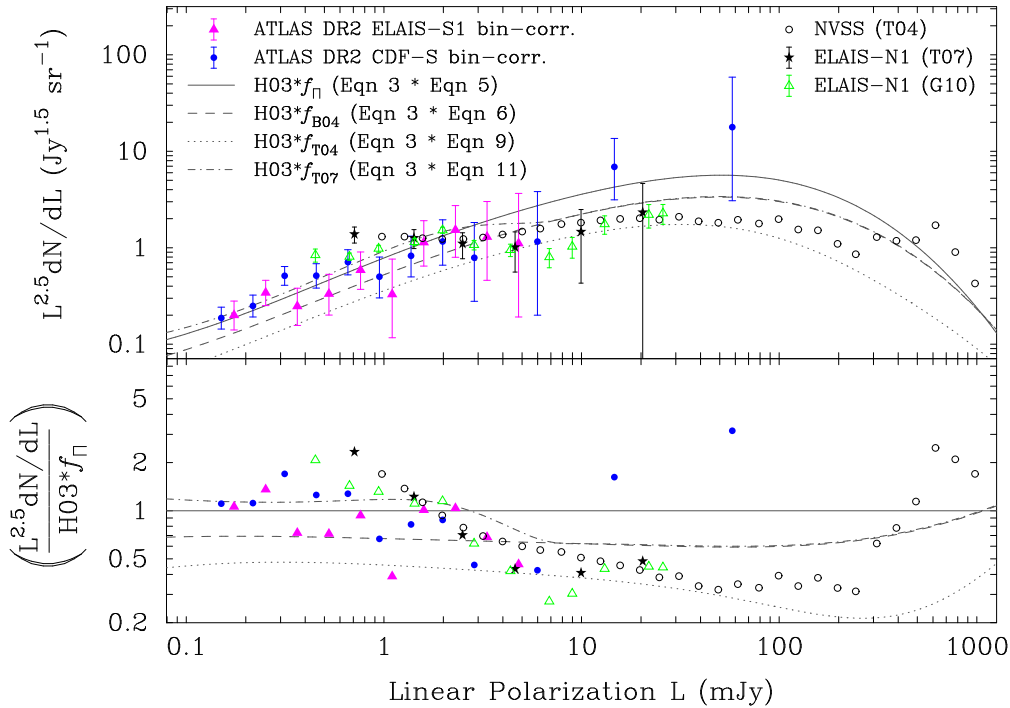}
 \caption{	Comparison of ATLAS DR2 bin-corrected linear polarization component
 		counts with 1.4~GHz source counts from other surveys; see \S~\ref{ch5:SecDiscSubCountsL}
		for details. The four number-count curves and panel layout follow
		Fig.~\ref{ch5:fig:countsL}. Data from the upper panel are reproduced in the
		lower panel relative to the $\tnm{H03}*f_{\ms \Pi}$ curve.
		}
 \label{ch5:fig:countsL2}
\end{figure*}

In Fig.~\ref{ch5:fig:countsL2} we compare the ATLAS DR2 bin-corrected linear
polarization component counts (from Fig.~\ref{ch5:fig:countsL} or tabulated data from
Appendix~A) with the 1.4~GHz polarized source counts from the NVSS \citep{2004MNRAS.349.1267T} and
the ELAIS-N1 field (\citealt{2007ApJ...666..201T}; deeper counts from \citealt{2010ApJ...714.1689G}).
The ATLAS DR2 counts improve upon the \citet{2010ApJ...714.1689G} study by a factor of
$\sim2$ in sensitivity. The observed number-counts from the \citet{2004MNRAS.349.1267T},
\citet{2007ApJ...666..201T}, and \citet{2010ApJ...714.1689G} studies are in general
agreement with the ATLAS counts, though the ATLAS data do not exhibit flattening
at faint levels that might otherwise lead to suggestions of increasing levels
of fractional polarization with decreasing flux density or perhaps the emergence
of a new source population. The flattening of the data from these studies are unlikely
to be real, but rather probably reflective of spurious populations of sources with abnormally
high levels of fractional polarization as described earlier in \S~\ref{ch5:SecResIdentM}.
This explains the difficulty encountered by \citet{2008evn..confE.107O} in attempting
to model the flattening. Similarly, despite the apparent consistency between the
observed counts and the various predicted curves displayed in Fig.~\ref{ch5:fig:countsL2},
in \S~\ref{ch5:SecResSubPI} we found that the fractional polarization distributions
presented by \citet{2004MNRAS.349.1267T}, \citet{2007ApJ...666..201T}, and \citet{2010ApJ...714.1689G}
were inconsistent with the ATLAS data and therefore unlikely to be suitable for population modelling.

The flux density range over which the $\tnm{H03}*f_{\ms \Pi}$ model extends in
Fig.~\ref{ch5:fig:countsL2} corresponds to the brightest regions of the total
intensity counts, in which luminous radio galaxies and quasars dominate. This is
consistent with our independent conclusion from \S~\ref{ch5:SecResIdentM} that the
polarized sources contributing to the ATLAS counts tend to be FRI/II radio galaxies,
and with our earlier findings regarding the infrared colours of polarized sources
from \S~\ref{ch5:SecRes:2}. To fully confirm this picture, luminosity functions
for polarized sources of different classifications need to be constructed
(e.g. FRI/FRII/radio-quiet), which can then be compared with theory
\citep[e.g.][]{2008evn..confE.107O}.

Very recently, while we were finalising this manuscript for resubmission, \citet{2014arXiv1402.3637R}
published a similar study of faint polarized sources. \citet{2014arXiv1402.3637R} presented
1.4~GHz linearly polarized integral (not differential) source counts from the GOODS-N field, observed with the
VLA at $1.6\arcsec$ resolution. Their polarized counts extend to 20~$\mu$Jy, an order of magnitude
deeper than our ATLAS DR2 results. Qualitatively, their results are consistent with our main
finding that the fractional polarization levels of faint sources are not
anti-correlated with total flux density. Quantitatively, however, their results are discrepant
with ours. \citet{2014arXiv1402.3637R} predict that surveys with $10\arcsec$ resolution
will observe a polarized source density of 22~deg$^{-2}$ for $L\ge100$~$\mu$Jy; this is a
factor of 2 lower than the directly observed counts presented in this work. \citet{2014arXiv1402.3637R}
explain that it is difficult to directly compare polarized source counts from surveys with
$1.6\arcsec$ and $10\arcsec$ resolutions, but they consider a factor of 2 difference to be
optimistically large (this factor was used to form their $10\arcsec$ resolution
prediction). Given that \citet{2014arXiv1402.3637R} use peak surface brightness
measurements as a proxy for flux density irrespective of source angular size,
that they do not attempt to correct their data for effects such as resolution bias,
and that they do not present total intensity counts with similar processing as for their
linear polarization counts, it is difficult to assess the robustness of their results here.
Detailed assessment is beyond the scope of this work, and is better suited to future studies
when results from other deep polarization surveys or modelling efforts become available.

\subsection{Star-Forming Galaxies}\label{ch5:SecDiscSFG}

We did not detect any polarized SFGs in this work. The fractional polarization
upper limits for individual SFGs presented in the lower-right panel of
Fig.~\ref{ch5:fig:fracpolClass} indicate that characteristic $\Pi$ levels for the
sub-millijansky ($I$~{\footnotesize $\gtrsim$}~100~$\mu$Jy) SFG population are
likely to be typically less than $\sim60\%$.

Given that SFGs are only expected to begin contributing significantly to the
total intensity source counts at $I$~{\footnotesize $\lesssim$}~100~$\mu$Jy \citep[e.g.
see Fig.~\ref{ch5:fig:fracpolClass}; see also][]{2008MNRAS.388.1335W,2011ApJ...740...20P},
the limit above indicates that the $\tnm{H03}*f_{\ms \Pi}$ model is
unlikely to be affected by the presence of SFGs unless $L$~{\footnotesize $\lesssim$}~60~$\mu$Jy.
Our limit of $\Pi_\tnm{\tiny SFG}<60\%$ is consistent with the fractional polarization
distribution for 1.4~GHz SFGs predicted by \citet{2009ApJ...693.1392S}; see panel~(b)
of Fig.~6 from their work. \citet{2009ApJ...693.1392S} modelled the integrated polarized
emission of spiral galaxies, finding typical fractional polarization levels of
$<5\%$ with overall mean level $<1\%$. The $\tnm{H03}*f_{\ms \Pi}$ number-counts
predicted in this work are therefore likely to represent an upper limit to the true
polarized number-counts at $\mu$Jy levels, due to the diminished mean level of
fractional polarization for faint radio sources with respect to the $f_{\ms \Pi}$
model. More sensitive observations are required to detect polarized emission from
faint SFGs and to quantify their polarization properties.

\subsection{Asymmetric Depolarization in Double Radio Sources}\label{ch5:SecDiscSubDA}

\citet{1988Natur.331..147G} and \citet{1988Natur.331..149L} discovered that
double radio sources depolarize less rapidly with increasing wavelength
on the side with the brighter (or only) radio jet than on the opposite side, providing
strong evidence that the apparent one-sided nature of jets in otherwise symmetric
radio galaxies and quasars is caused by relativistic beaming. This `Laing-Garrington'
effect is typically interpreted as being caused by orientation-induced path-length
differences through a foreground, turbulent, magnetised intragroup or intracluster
medium which surrounds the entire radio source, where the approaching side is seen through
less of this medium \citep{1991MNRAS.250..198G,1991MNRAS.250..726T,1992MNRAS.256..281T,
2003ApJ...597..870E,2008MNRAS.391..521L,2011A&A...529A..13K}. However, this interpretation
is not unambiguous. The asymmetric depolarization effect may be contaminated or even
dominated by depolarization internal to the lobes \citep{2013ApJ...764..162O,2013ApJ...766...48S},
a sheath mixing layer at the interface where relativistic and thermal plasmas meet
\citep{1990ApJ...357..373B,1995MNRAS.273..877J,2004rcfg.proc..153R,2009ApJ...707..114F},
draping of undisturbed intracluster magnetic fields over the surface of a lobe expanding
subsonically \citep{2008ApJ...677..993D} or supersonically \citep{2011MNRAS.413.2525G,
2011MNRAS.418.1621H,2012MNRAS.423.1335G}, or by intrinsic asymmetries in local environment
which act separately or in addition to orientation-induced depolarization
\citep{1991MNRAS.249..343L,1991MNRAS.253..669L,1991ApJ...371..478M,1998MNRAS.300..269I,
2004astro.ph..9761G,2008ApJS..174...74K}.

In reality, it is likely that each of the mechanisms above may contribute, requiring
a `unification scheme' to predict which will dominate for any given source. For
example, \citet{2012MNRAS.423.1335G} describes an emerging picture that differentiates
between properties expected for FRI and FRII sources, and that includes an inner
depolarization region associated with shells of dense thermal plasma around the radio
jets in addition to the undisturbed intergalactic medium surrounding the source. However,
this picture does not yet include variations on the general orientation-induced
depolarization effect due to source environment asymmetries, such as the correlation
between lobe properties and optical line emission described by \citet{1991ApJ...371..478M}.

If the Laing-Garrington effect is caused predominantly by source orientation, rather
than asymmetries in source environment, then we expect the fractional surface density
of sources exhibiting asymmetric depolarization in a volume-limited sample to approximately relate
to the fraction of randomly-oriented sources with $S$~{\footnotesize $\gtrsim$}~10~mJy that are
pointed towards Earth. We justify this expected relationship by noting that FRII sources
are dominated by un-beamed lobe emission rather than jet emission which may be beamed, FRII sources
dominate FRI sources in flux-limited samples for $S$~{\footnotesize $\gtrsim$}~10~mJy
\citep{2008MNRAS.388.1335W,2011ApJ...740...20P}, and the median redshift for FRII
sources is $z\approx0.8$ with relatively small scatter \citep{1988gera.book..641C} such
that a flux-limited sample may crudely approximate a volume-limited sample.

For a jet lying within an angle $\vartheta$ to the line of sight, the fraction of randomly-oriented
sources pointed towards Earth is $1-\cos(\vartheta)$.
Orientation schemes predict that the transition from quasars (typically beamed) to
radio galaxies (typically not beamed) is expected to occur at $\vartheta \sim 45\degree$
\citep{1989ApJ...336..606B,1995PASP..107..803U}. The critical angle to induce
asymmetric depolarization in a double radio source (quasar or radio galaxy) is likely
to be similar \citep[e.g.][]{2001AJ....122...83F}; here we estimate this angle as ranging
between $30\degree$ and $70\degree$, implying fractional surface densities of $15-65\%$ amongst
the general double radio source population. This is, of course, a crude model, not least
because intracluster magnetic field strengths at $z\sim0.8$ \citep[i.e. the expected median
redshift for double radio sources;][]{1988gera.book..641C} are only expected to be a
few percent of their $z=0$ values \citep{2002A&A...387..383D,2009ApJ...698L..14X}.
A magnetised cluster atmosphere is clearly a prerequisite for depolarization, though
a separate depolarizing medium close to the radio jets as described by
\citet{2012MNRAS.423.1335G} may null this point.

As far as we are aware, no census of asymmetric depolarization has been performed
for radio sources in a blind survey; studies to date have typically compiled samples
of targeted observations \citep[e.g.][]{2001AJ....122...83F, 2008ApJS..174...74K}.
The ATLAS data are suitable for this purpose. We note that it is difficult to
estimate what the fractional surface density of asymmetrically depolarized sources
might be if environmental asymmetries were to dominate, rather than
orientation \citep[significant merger activity is certainly expected in clusters;
e.g.][]{2004ApJ...607..125T}. Therefore, we simply focus on whether the ATLAS
data are consistent with an orientation scheme or not.

To avoid selection effects relating to visibility area and the detectability of
sources with low fractional polarization, we selected only the 40 ATLAS sources with
total intensity $S>30$~mJy. We expect that each of these sources is a radio-loud
AGN with dual-lobe structure dominated by un-beamed lobe emission (though not all
need exhibit multiple components due to viewing angle and source size). The breakdown
of these 40 sources according to the polarization$-$total intensity classification scheme
(see \S~\ref{sec:1}) is as follows. There was 1 polarized source classified as Type 3 (midway polarized),
though it is unclear if this demonstrates a depolarization asymmetry or not. There were
7 polarized single-component sources classified as Type 4. In an attempt to account for the
possibility that many of these `unclassified' sources represent as-yet unassociated
lobes of multi-component sources (see comments in \S~\ref{ch5:SecResIdentM}), we assumed
that perhaps only 1 of the detected Type 4 sources was likely to truly represent
a polarized single-component source. We have interpreted this as a
dual-lobed asymmetrically-depolarized source with jet axis close to the line
of sight, such that only a single component is effectively seen. There were 15 polarized dual-lobed
sources classified as Type 6 (no asymmetric depolarization). There were 2 polarized
dual-lobed sources classified as Type 7 (clear indication of asymmetric depolarization);
for reference, these sources were displayed earlier in Fig.~17 of Paper~I.
There were 4 polarized dual-lobed sources classified as Type 8 (unclear whether
asymmetrically depolarized or not). We assumed that approximately one quarter
of these sources (i.e. 1 source) would likely demonstrate asymmetric
depolarization if more sensitive observations were obtained. This assumption is
consistent with the finding from Fig.~\ref{ch5:fig:fracpolTheta2} that most
Type 8 sources have LASs reflective of Type 6 sources, rather than the
smaller LASs observed for the Type 7 sources. Finally, our sample included
11 unpolarized sources (Type 9), each with fractional polarization upper
limits below 1\% (see Fig.~\ref{ch5:fig:fracpolClass}). We do not interpret
these sources as being asymmetrically depolarized.

We note that while it is possible that some of the 11 unpolarized sources
represent unassociated lobes of multi-component sources, at least
some of them must be truly isolated, single-component sources with $\Pi<1\%$.
For example, the brightest unpolarized source displayed in
Fig.~\ref{ch5:fig:fracpolClass} is the source C3, which is barely-resolved in
ATLAS DR2 (deconvolved angular size $1.69\arcsec\pm1.65\arcsec$) and has
a $1.4-2.3$~GHz spectral index of $\alpha=-0.4$ \citep{2012A&A...544A..38Z}.
This source is therefore consistent with identification as a CSO \citep{1996ApJ...460..634R};
CSOs are known to exhibit flat radio spectra \citep{2006MNRAS.368.1411A}
and low fractional polarization due to strong depolarization
\citep{2003PASA...20...12C}.

The statistics above suggest that between $1+2+1=4$ and $1+7+2+4=14$ of the
40 ATLAS DR2 sources in our flux density limited sample exhibit depolarization
asymmetry, i.e., $10-35\%$. This fraction falls within the theoretical range
estimated above, demonstrating that the Laing-Garrington effect appears consistent
with orientation dependence, at least within the rudimentary confines of our
analysis. Future high-resolution polarization studies are clearly required
to form more robust conclusions.

\section{Conclusions}\label{ch5:SecConc}

In this work we have presented results and discussion for ATLAS DR2.
Our key results are summarised as follows. For convenience
we use the term `millijansky' loosely below to indicate flux densities in
the range $0.1-1000$~mJy.
\begin{enumerate}[(i)]

\item Radio emission from polarized millijansky sources is most likely powered
by AGNs, where the active nuclei are embedded within host galaxies with mid-infrared
spectra dominated by old-population (10~Gyr) starlight or continuum
produced by dusty tori. We find no evidence for polarized SFGs or individual
stars to the sensitivity limits of our data - all polarized ATLAS sources are
classified as AGNs.

\item The ATLAS data indicate that fractional polarization levels for sources with
starlight-dominated mid-infrared hosts and those with continuum-dominated mid-infrared
hosts are similar.

\item The morphologies and angular sizes of polarized ATLAS components and sources
are consistent with the interpretation that polarized emission in millijansky sources
originates from the jets or lobes of extended AGNs, where coherent large-scale magnetic
fields are likely to be present. We find that the majority of polarized ATLAS sources
are resolved in total intensity, even though the majority of components in linear polarization
are unresolved. This is consistent with the interpretation that large-scale magnetic fields that do
not completely beam depolarize are present in these sources, despite the relatively
poor resolutions of the ATLAS data.

\item We do not find any components or sources with fractional polarization levels
greater than 24\%, in contrast with previous studies of faint polarized sources.
We attribute this finding to our improved data analysis procedures.

\item The ATLAS data are consistent with a distribution of fractional
polarization at 1.4~GHz that is independent of flux density down to $I\sim10$~mJy,
and perhaps even down to 1~mJy when considering the upper envelope of the distribution.
This result is in contrast to the findings from previous deep 1.4~GHz
polarization surveys \citep[with the very recent exception of][]{2014arXiv1402.3637R},
and is consistent with results at higher frequencies ($\ge4.8$~GHz). The anti-correlation
observed in previous 1.4~GHz studies is due to two effects: a selection bias, and
spurious high fractional polarization detections. Both of these effects can become
more prevalent at faint total flux densities. We find that components and sources
can be characterised using the same distribution of fractional linear polarization,
with a median level of 4\%. We have presented a new lognormal model to describe
the distribution of fractional polarization for 1.4~GHz components and sources,
specific to AGNs, in surveys with resolution FWHMs $\sim10\arcsec$.

\item No polarized SFGs were detected in ATLAS DR2 down to the linear polarization
detection threshold of $\sim200$~$\mu$Jy. The ATLAS data constrain typical fractional
polarization levels for the $I$~{\footnotesize $\gtrsim$}~100~$\mu$Jy SFG population
to be $\Pi_\tnm{\tiny SFG}<60\%$.

\item Differences between differential number-counts of components and of
sources in 1.4~GHz surveys with resolution FWHM $\sim10\arcsec$ are not likely
to be significant ({\footnotesize $\lesssim$}~20\%) at millijansky levels.

\item The ATLAS total intensity differential source counts do not exhibit
any unexpected flattening down to the survey limit $\sim100\mu$Jy.

\item The ATLAS linearly polarized differential component counts do not exhibit
any flattening below $\sim1$~mJy, unlike previous findings which have led to
suggestions of increasing levels of fractional polarization with decreasing
flux density or the emergence of a new source population. The polarized counts
down to $\sim100$~$\mu$Jy are consistent with being drawn from the total
intensity counts at flux densities where luminous FR-type radio galaxies
and quasars dominate.

\item Constrained by the ATLAS data, we estimate that the surface density of linearly
polarized components in a 1.4~GHz survey with resolution FWHM $\sim10\arcsec$ is
50~deg$^{-2}$ for $L_{\tnm{\tiny cmp}}\ge100$~$\mu$Jy, and 90~deg$^{-2}$ for
$L_{\tnm{\tiny cmp}}\ge50$~$\mu$Jy. We estimate that the surface density for
polarized sources is $\sim45$~deg$^{-2}$ for $L_{\tnm{\tiny src}}\ge100$~$\mu$Jy,
assuming that most polarized components belong to dual-component sources
(e.g. FR-type) at these flux densities.

\item We find that the statistics of ATLAS sources exhibiting asymmetric
depolarization are consistent with the interpretation that the
Laing-Garrington effect is due predominantly to source orientation
within a surrounding magnetoionic medium. To our knowledge, this work
represents the first attempt to investigate asymmetric depolarization
in a blind survey.

\end{enumerate}

\section*{Acknowledgments}

We thank Walter Max-Moerbeck for insightful discussions. We thank the anonymous
referee for helpful comments that led to the improvement of this manuscript.
C.~A.~H. acknowledges the support of an Australian Postgraduate Award,
a CSIRO OCE Scholarship, and a Jansky Fellowship from the National Radio
Astronomy Observatory. B.~M.~G. and R.~P.~N. acknowledge the support of the Australian
Research Council Centre of Excellence for All-sky Astrophysics (CAASTRO), through
project number CE110001020. The Australia Telescope Compact Array is part of the Australia
Telescope National Facility which is funded by the Commonwealth of Australia for
operation as a National Facility managed by CSIRO. This paper includes
archived data obtained through the Australia Telescope Online Archive
({\tt http://atoa.atnf.csiro.au}).
% http://www.atnf.csiro.au/research/publications/policy.html
% http://www.atnf.csiro.au/research/publications/Acknowledgements.html

\appendix

\section{Differential Component Counts}

This appendix presents 1.4~GHz Euclidean-normalised differential number-counts,
in tabulated form, derived from the ATLAS DR2 total intensity and linear polarization
component catalogues from Appendix~A of Paper~I.

The tabulated results have been organised as follows according
to emission type, Eddington bias correction scheme, and ATLAS field.
Tables~\ref{tbl:countsICDFSie} and \ref{tbl:countsIELAISie} (each with 7 columns)
present total intensity number-counts for the CDF-S and ELAIS-S1 fields, respectively, using
`component-corrected' data whereby individual components were deboosted prior
to the application of visibility area and resolution bias corrections. Similarly,
Tables~\ref{tbl:countsICDFS} and \ref{tbl:countsIELAIS} (each with 9 columns) present total intensity
number-counts for the two ATLAS fields, but now using `bin-corrected' data
whereby non-deboosted components were corrected for visibility
area, resolution bias, and Eddington bias. Tables~\ref{tbl:countsLCDFSie} and
\ref{tbl:countsLELAISie} (each with 11 columns) present linear polarization number-counts for the CDF-S
and ELAIS-S1 fields, respectively, using component-corrected data.
Tables~\ref{tbl:countsLCDFS} and \ref{tbl:countsLELAIS} (each with 17 columns)
present bin-corrected linear polarization number-counts for the CDF-S and ELAIS-S1
fields,respectively.

Columns for the tables above are organised as follows. Tables describing component-corrected
total intensity data give for each bin the flux density range ($\Delta S$),
expected average flux density ($S_\tnm{\tiny AV}$), raw number of deboosted
components ($N_\tnm{\tiny raw}^\tnm{\tiny D}$), effective number of deboosted
components following visibility area correction only ($N_\tnm{\tiny eff}^\tnm{\tiny DV}$),
effective number of deboosted components following both visibility area and resolution
bias corrections ($N_\tnm{\tiny eff}^\tnm{\tiny DVR}$), Euclidean-normalised counts
following visibility area correction only ($S^{2.5}dN_\tnm{\tiny eff}^\tnm{\tiny DV}/dS$),
and Euclidean-normalised counts following both visibility area and resolution bias
corrections ($S^{2.5}dN_\tnm{\tiny eff}^\tnm{\tiny DVR}/dS$). Columns for tables
describing bin-corrected total intensity data are similar, but without the superscript D
which indicates use of deboosted component data. The bin-corrected tables contain two
additional columns: effective number of components following combined visibility area,
resolution bias, and Eddington bias corrections ($N_\tnm{\tiny eff}^\tnm{\tiny VRE}$),
and an associated column for their Euclidean-normalised counts
($S^{2.5}dN_\tnm{\tiny eff}^\tnm{\tiny VRE}/dS$). The tables describing linear
polarization data are similar to those for total intensity data, but with additional
columns cataloguing the effective number of components or Euclidean-normalised
counts resulting from the resolution bias corrections associated with the lower (-R) or
upper (+R) bounds described in \S~7.1 of Paper~I and displayed in Fig.~22 of Paper~I.
Thus the additional columns have been assigned descriptors with superscripts
V-R, V+R, V-RE, and V+RE. Errors associated with the fully corrected Euclidean-normalised
counts in the figures and tables above are $1\sigma$ Poissonian and were calculated
following \citet{regener}.

\begin{table*}
 \centering
 \caption{ATLAS 1.4~GHz DR2 total intensity component-corrected counts for CDF-S field.
 	  See Appendix~A for column details.}
 \label{tbl:countsICDFSie}
 \begin{tabular}{@{}ccccccccc@{}}
 \hline
	\multicolumn{1}{c}{$\Delta I$} &
	\multicolumn{1}{c}{$I_\tnm{\tiny AV}$} &
	\multicolumn{1}{c}{$N_\tnm{\tiny raw}^\tnm{\tiny D}$} &
	\multicolumn{1}{c}{$N_\tnm{\tiny eff}^\tnm{\tiny DV}$} &
	\multicolumn{1}{c}{$N_\tnm{\tiny eff}^\tnm{\tiny DVR}$} &
	\multicolumn{1}{c}{$I^{2.5}dN_\tnm{\tiny eff}^\tnm{\tiny DV}/dI$} &
	\multicolumn{1}{c}{$I^{2.5}dN_\tnm{\tiny eff}^\tnm{\tiny DVR}/dI$} \\
	\multicolumn{1}{c}{(mJy)} &
	\multicolumn{1}{c}{(mJy)} & & & &
	\multicolumn{1}{c}{(Jy$^{1.5}\,$sr$^{-1}$)} &
	\multicolumn{1}{c}{(Jy$^{1.5}\,$sr$^{-1}$)} \\
	\multicolumn{1}{c}{(1)} &
	\multicolumn{1}{c}{(2)} &
	\multicolumn{1}{c}{(3)} &
	\multicolumn{1}{c}{(4)} &
	\multicolumn{1}{c}{(5)} &
	\multicolumn{1}{c}{(6)} &
	\multicolumn{1}{c}{(7)} \\
 \hline
$0.159-0.187$ & 0.172 & 66 & 309.5 & 367.9 & 3.93 & $4.67^{\ms +0.57}_{\ms -0.57}$ \\
$0.187-0.219$ & 0.202 & 98 & 214.0 & 232.2 & 3.46 & $3.75^{\ms +0.38}_{\ms -0.38}$ \\
$0.219-0.258$ & 0.238 & 125 & 198.6 & 228.3 & 4.08 & $4.70^{\ms +0.42}_{\ms -0.42}$ \\
$0.258-0.303$ & 0.279 & 118 & 162.0 & 204.7 & 4.24 & $5.36^{\ms +0.49}_{\ms -0.49}$ \\
$0.303-0.356$ & 0.328 & 94 & 115.9 & 156.7 & 3.86 & $5.23^{\ms +0.54}_{\ms -0.54}$ \\
$0.356-0.418$ & 0.386 & 89 & 99.6 & 141.9 & 4.23 & $6.03^{\ms +0.64}_{\ms -0.64}$ \\
$0.418-0.491$ & 0.453 & 82 & 86.1 & 125.4 & 4.66 & $6.78^{\ms +0.75}_{\ms -0.75}$ \\
$0.491-0.577$ & 0.532 & 77 & 77.5 & 110.0 & 5.34 & $7.58^{\ms +0.86}_{\ms -0.86}$ \\
$0.577-0.678$ & 0.626 & 38 & 38.0 & 52.0 & 3.34 & $4.57^{\ms +0.87}_{\ms -0.74}$ \\
$0.678-0.797$ & 0.735 & 40 & 41.5 & 54.8 & 4.65 & $6.13^{\ms +1.1}_{\ms -0.97}$ \\
$0.797-0.936$ & 0.864 & 43 & 43.2 & 55.0 & 6.15 & $7.84^{\ms +1.4}_{\ms -1.2}$ \\
$0.936-1.10$ & 1.01 & 26 & 26.7 & 32.9 & 4.85 & $5.98^{\ms +1.4}_{\ms -1.2}$ \\
$1.10-1.48$ & 1.28 & 66 & 66.4 & 78.6 & 9.15 & $10.8^{\ms +1.3}_{\ms -1.3}$ \\
$1.48-2.00$ & 1.72 & 49 & 49.3 & 55.8 & 10.6 & $12.0^{\ms +2.0}_{\ms -1.7}$ \\
$2.00-2.70$ & 2.33 & 45 & 45.0 & 49.2 & 15.2 & $16.6^{\ms +2.9}_{\ms -2.5}$ \\
$2.70-3.64$ & 3.14 & 32 & 32.0 & 34.1 & 17.0 & $18.1^{\ms +3.8}_{\ms -3.2}$ \\
$3.64-4.91$ & 4.24 & 33 & 33.0 & 34.5 & 27.4 & $28.7^{\ms +5.9}_{\ms -5.0}$ \\
$4.91-6.63$ & 5.71 & 20 & 20.0 & 20.6 & 26.1 & $26.9^{\ms +7.5}_{\ms -6.0}$ \\
$6.63-8.94$ & 7.71 & 15 & 15.0 & 15.3 & 30.6 & $31.3^{\ms +10}_{\ms -8.0}$ \\
$8.94-12.1$ & 10.4 & 17 & 17.0 & 17.2 & 54.4 & $55.2^{\ms +17}_{\ms -13}$ \\
$12.1-19.1$ & 15.2 & 20 & 20.0 & 20.2 & 73.5 & $74.2^{\ms +21}_{\ms -16}$ \\
$19.1-30.3$ & 24.1 & 15 & 15.0 & 15.1 & 110 & $110^{\ms +36}_{\ms -28}$ \\
$30.3-48.0$ & 38.2 & 9 & 9.0 & 9.0 & 131 & $132^{\ms +60}_{\ms -43}$ \\
$48.0-76.1$ & 60.5 & 6 & 6.0 & 6.0 & 174 & $175^{\ms +100}_{\ms -69}$ \\
$76.1-121$ & 95.9 & 6 & 6.0 & 6.0 & 347 & $348^{\ms +210}_{\ms -140}$ \\
$121-191$ & 152 & 1 & 1.0 & 1.0 & 115 & $115^{\ms +270}_{\ms -95}$ \\
$191-303$ & 240 & 2 & 2.0 & 2.0 & 459 & $459^{\ms +610}_{\ms -300}$ \\
$761-1206$ & 952 & 1 & 1.0 & 1.0 & 1800 & $1800^{\ms +4100}_{\ms -1500}$ \\
 \hline
 \end{tabular}
\end{table*}

\begin{table*}
 \centering
 \caption{ATLAS 1.4~GHz DR2 total intensity component-corrected counts for ELAIS-S1 field.
 	  See Appendix~A for column details.}
 \label{tbl:countsIELAISie}
 \begin{tabular}{@{}ccccccccc@{}}
 \hline
	\multicolumn{1}{c}{$\Delta I$} &
	\multicolumn{1}{c}{$I_\tnm{\tiny AV}$} &
	\multicolumn{1}{c}{$N_\tnm{\tiny raw}^\tnm{\tiny D}$} &
	\multicolumn{1}{c}{$N_\tnm{\tiny eff}^\tnm{\tiny DV}$} &
	\multicolumn{1}{c}{$N_\tnm{\tiny eff}^\tnm{\tiny DVR}$} &
	\multicolumn{1}{c}{$I^{2.5}dN_\tnm{\tiny eff}^\tnm{\tiny DV}/dI$} &
	\multicolumn{1}{c}{$I^{2.5}dN_\tnm{\tiny eff}^\tnm{\tiny DVR}/dI$} \\
	\multicolumn{1}{c}{(mJy)} &
	\multicolumn{1}{c}{(mJy)} & & & &
	\multicolumn{1}{c}{(Jy$^{1.5}\,$sr$^{-1}$)} &
	\multicolumn{1}{c}{(Jy$^{1.5}\,$sr$^{-1}$)} \\
	\multicolumn{1}{c}{(1)} &
	\multicolumn{1}{c}{(2)} &
	\multicolumn{1}{c}{(3)} &
	\multicolumn{1}{c}{(4)} &
	\multicolumn{1}{c}{(5)} &
	\multicolumn{1}{c}{(6)} &
	\multicolumn{1}{c}{(7)} \\
 \hline
$0.139-0.163$ & 0.151 & 88 & 245.9 & 256.4 & 3.34 & $3.49^{\ms +0.37}_{\ms -0.37}$ \\
$0.163-0.192$ & 0.177 & 108 & 208.2 & 221.0 & 3.60 & $3.82^{\ms +0.37}_{\ms -0.37}$ \\
$0.192-0.225$ & 0.208 & 109 & 169.6 & 200.5 & 3.74 & $4.42^{\ms +0.42}_{\ms -0.42}$ \\
$0.225-0.265$ & 0.244 & 108 & 147.0 & 189.3 & 4.13 & $5.31^{\ms +0.51}_{\ms -0.51}$ \\
$0.265-0.311$ & 0.287 & 71 & 86.5 & 118.5 & 3.09 & $4.24^{\ms +0.50}_{\ms -0.50}$ \\
$0.311-0.366$ & 0.337 & 85 & 94.7 & 135.9 & 4.31 & $6.18^{\ms +0.67}_{\ms -0.67}$ \\
$0.366-0.430$ & 0.396 & 68 & 71.6 & 103.1 & 4.15 & $5.98^{\ms +0.72}_{\ms -0.72}$ \\
$0.430-0.505$ & 0.465 & 60 & 61.1 & 85.9 & 4.51 & $6.34^{\ms +0.82}_{\ms -0.82}$ \\
$0.505-0.593$ & 0.547 & 50 & 50.2 & 68.4 & 4.73 & $6.44^{\ms +1.0}_{\ms -0.91}$ \\
$0.593-0.697$ & 0.643 & 51 & 51.8 & 68.2 & 6.21 & $8.17^{\ms +1.1}_{\ms -1.1}$ \\
$0.697-0.818$ & 0.755 & 29 & 29.9 & 38.0 & 4.56 & $5.80^{\ms +1.3}_{\ms -1.1}$ \\
$0.818-0.962$ & 0.887 & 37 & 38.0 & 46.8 & 7.39 & $9.10^{\ms +1.8}_{\ms -1.5}$ \\
$0.962-1.13$ & 1.04 & 22 & 22.8 & 27.2 & 5.64 & $6.74^{\ms +1.8}_{\ms -1.4}$ \\
$1.13-1.52$ & 1.31 & 35 & 35.9 & 41.3 & 6.75 & $7.77^{\ms +1.5}_{\ms -1.3}$ \\
$1.52-2.06$ & 1.77 & 46 & 46.4 & 51.3 & 13.7 & $15.1^{\ms +2.6}_{\ms -2.2}$ \\
$2.06-2.77$ & 2.39 & 29 & 29.1 & 31.2 & 13.4 & $14.4^{\ms +3.2}_{\ms -2.7}$ \\
$2.77-3.74$ & 3.22 & 25 & 25.0 & 26.3 & 18.1 & $19.0^{\ms +4.6}_{\ms -3.8}$ \\
$3.74-5.05$ & 4.35 & 26 & 26.4 & 27.3 & 29.9 & $31.0^{\ms +7.4}_{\ms -6.0}$ \\
$5.05-6.81$ & 5.87 & 20 & 20.2 & 20.7 & 35.9 & $36.8^{\ms +10}_{\ms -8.1}$ \\
$6.81-9.18$ & 7.92 & 20 & 20.0 & 20.4 & 55.7 & $56.7^{\ms +16}_{\ms -13}$ \\
$9.18-12.4$ & 10.7 & 11 & 11.2 & 11.4 & 49.1 & $49.6^{\ms +20}_{\ms -15}$ \\
$12.4-19.6$ & 15.6 & 9 & 9.0 & 9.1 & 45.1 & $45.5^{\ms +21}_{\ms -15}$ \\
$19.6-31.1$ & 24.8 & 12 & 12.0 & 12.0 & 120 & $120^{\ms +46}_{\ms -34}$ \\
$31.1-49.3$ & 39.2 & 5 & 5.0 & 5.0 & 99.5 & $99.7^{\ms +67}_{\ms -43}$ \\
$49.3-78.2$ & 62.2 & 5 & 5.0 & 5.0 & 198 & $198^{\ms +130}_{\ms -86}$ \\
$124-196$ & 156 & 2 & 2.0 & 2.0 & 315 & $315^{\ms +420}_{\ms -200}$ \\
 \hline
 \end{tabular}
\end{table*}

\begin{table*}
 \centering
 \caption{ATLAS 1.4~GHz DR2 total intensity bin-corrected counts for CDF-S field.
  	  See Appendix~A for column details.}
 \label{tbl:countsICDFS}
 \begin{tabular}{@{}ccccccccc@{}}
 \hline
	\multicolumn{1}{c}{$\Delta I$} &
	\multicolumn{1}{c}{$I_\tnm{\tiny AV}$} &
	\multicolumn{1}{c}{$N_\tnm{\tiny raw}$} &
	\multicolumn{1}{c}{$N_\tnm{\tiny eff}^\tnm{\tiny V}$} &
	\multicolumn{1}{c}{$N_\tnm{\tiny eff}^\tnm{\tiny VR}$} &
	\multicolumn{1}{c}{$N_\tnm{\tiny eff}^\tnm{\tiny VRE}$} &
	\multicolumn{1}{c}{$I^{2.5}dN_\tnm{\tiny eff}^\tnm{\tiny V}/dI$} &
	\multicolumn{1}{c}{$I^{2.5}dN_\tnm{\tiny eff}^\tnm{\tiny VR}/dI$} &
	\multicolumn{1}{c}{$I^{2.5}dN_\tnm{\tiny eff}^\tnm{\tiny VRE}/dI$} \\
	\multicolumn{1}{c}{(mJy)} &
	\multicolumn{1}{c}{(mJy)} & & & & &
	\multicolumn{1}{c}{(Jy$^{1.5}\,$sr$^{-1}$)} &
	\multicolumn{1}{c}{(Jy$^{1.5}\,$sr$^{-1}$)} &
	\multicolumn{1}{c}{(Jy$^{1.5}\,$sr$^{-1}$)} \\
	\multicolumn{1}{c}{(1)} &
	\multicolumn{1}{c}{(2)} &
	\multicolumn{1}{c}{(3)} &
	\multicolumn{1}{c}{(4)} &
	\multicolumn{1}{c}{(5)} &
	\multicolumn{1}{c}{(6)} &
	\multicolumn{1}{c}{(7)} &
	\multicolumn{1}{c}{(8)} &
	\multicolumn{1}{c}{(9)} \\
 \hline
$0.169-0.199$ & 0.183 & 59 & 334.9 & 396.2 & 265.7 & 4.65 & 5.51 & $3.69^{\ms +0.48}_{\ms -0.48}$ \\
$0.199-0.233$ & 0.215 & 96 & 232.1 & 251.8 & 178.4 & 4.11 & 4.46 & $3.16^{\ms +0.32}_{\ms -0.32}$ \\
$0.233-0.274$ & 0.253 & 146 & 234.6 & 280.8 & 216.5 & 5.29 & 6.33 & $4.88^{\ms +0.40}_{\ms -0.40}$ \\
$0.274-0.322$ & 0.297 & 123 & 167.9 & 218.3 & 180.3 & 4.82 & 6.27 & $5.17^{\ms +0.47}_{\ms -0.47}$ \\
$0.322-0.378$ & 0.349 & 94 & 114.8 & 158.6 & 140.5 & 4.20 & 5.80 & $5.14^{\ms +0.53}_{\ms -0.53}$ \\
$0.378-0.445$ & 0.410 & 96 & 106.4 & 153.7 & 141.6 & 4.95 & 7.16 & $6.59^{\ms +0.67}_{\ms -0.67}$ \\
$0.445-0.522$ & 0.482 & 79 & 82.1 & 118.5 & 112.6 & 4.87 & 7.03 & $6.68^{\ms +0.75}_{\ms -0.75}$ \\
$0.522-0.614$ & 0.566 & 76 & 76.3 & 106.9 & 103.2 & 5.76 & 8.08 & $7.79^{\ms +0.89}_{\ms -0.89}$ \\
$0.614-0.721$ & 0.665 & 43 & 43.3 & 58.5 & 57.1 & 4.17 & 5.63 & $5.49^{\ms +0.97}_{\ms -0.83}$ \\
$0.721-0.847$ & 0.781 & 36 & 37.2 & 48.5 & 47.7 & 4.56 & 5.94 & $5.84^{\ms +1.1}_{\ms -0.97}$ \\
$0.847-0.995$ & 0.918 & 40 & 40.2 & 50.6 & 50.0 & 6.28 & 7.90 & $7.81^{\ms +1.4}_{\ms -1.2}$ \\
$0.995-1.17$ & 1.08 & 29 & 29.7 & 36.2 & 35.9 & 5.90 & 7.20 & $7.14^{\ms +1.6}_{\ms -1.3}$ \\
$1.17-1.58$ & 1.36 & 61 & 61.4 & 71.9 & 71.6 & 9.27 & 10.9 & $10.8^{\ms +1.4}_{\ms -1.4}$ \\
$1.58-2.13$ & 1.83 & 53 & 53.3 & 59.8 & 59.6 & 12.6 & 14.2 & $14.1^{\ms +1.9}_{\ms -1.9}$ \\
$2.13-2.87$ & 2.47 & 43 & 43.0 & 46.7 & 46.7 & 16.0 & 17.3 & $17.3^{\ms +3.1}_{\ms -2.6}$ \\
$2.87-3.87$ & 3.34 & 30 & 30.0 & 31.8 & 31.8 & 17.4 & 18.5 & $18.5^{\ms +4.0}_{\ms -3.4}$ \\
$3.87-5.22$ & 4.50 & 31 & 31.0 & 32.3 & 32.3 & 28.3 & 29.4 & $29.5^{\ms +6.3}_{\ms -5.3}$ \\
$5.22-7.05$ & 6.07 & 17 & 17.0 & 17.5 & 17.5 & 24.3 & 25.0 & $25.0^{\ms +7.6}_{\ms -6.0}$ \\
$7.05-9.50$ & 8.19 & 16 & 16.0 & 16.3 & 16.3 & 35.8 & 36.6 & $36.6^{\ms +12}_{\ms -9.0}$ \\
$9.50-12.8$ & 11.1 & 18 & 18.0 & 18.3 & 18.3 & 63.1 & 64.0 & $64.0^{\ms +19}_{\ms -15}$ \\
$12.8-20.3$ & 16.2 & 18 & 18.0 & 18.2 & 18.2 & 72.4 & 73.2 & $73.1^{\ms +22}_{\ms -17}$ \\
$20.3-32.2$ & 25.6 & 16 & 16.0 & 16.1 & 16.1 & 128 & 129 & $129^{\ms +41}_{\ms -32}$ \\
$32.2-51.0$ & 40.6 & 8 & 8.0 & 8.0 & 8.0 & 128 & 128 & $128^{\ms +63}_{\ms -44}$ \\
$51.0-80.9$ & 64.3 & 5 & 5.0 & 5.0 & 5.0 & 159 & 161 & $161^{\ms +110}_{\ms -69}$ \\
$80.9-128$ & 102 & 5 & 5.0 & 5.0 & 5.0 & 317 & 317 & $317^{\ms +210}_{\ms -140}$ \\
$128-203$ & 161 & 1 & 1.0 & 1.0 & 1.0 & 126 & 126 & $126^{\ms +290}_{\ms -100}$ \\
$203-322$ & 256 & 2 & 2.0 & 2.0 & 2.0 & 503 & 503 & $503^{\ms +660}_{\ms -330}$ \\
$809-1282$ & 1012 & 1 & 1.0 & 1.0 & 1.0 & 1970 & 1970 & $1970^{\ms +4500}_{\ms -1600}$ \\
 \hline
 \end{tabular}
\end{table*}

\begin{table*}
 \centering
 \caption{ATLAS 1.4~GHz DR2 total intensity bin-corrected counts for ELAIS-S1 field.
  	  See Appendix~A for column details.}
 \label{tbl:countsIELAIS}
 \begin{tabular}{@{}ccccccccc@{}}
 \hline
	\multicolumn{1}{c}{$\Delta I$} &
	\multicolumn{1}{c}{$I_\tnm{\tiny AV}$} &
	\multicolumn{1}{c}{$N_\tnm{\tiny raw}$} &
	\multicolumn{1}{c}{$N_\tnm{\tiny eff}^\tnm{\tiny V}$} &
	\multicolumn{1}{c}{$N_\tnm{\tiny eff}^\tnm{\tiny VR}$} &
	\multicolumn{1}{c}{$N_\tnm{\tiny eff}^\tnm{\tiny VRE}$} &
	\multicolumn{1}{c}{$I^{2.5}dN_\tnm{\tiny eff}^\tnm{\tiny V}/dI$} &
	\multicolumn{1}{c}{$I^{2.5}dN_\tnm{\tiny eff}^\tnm{\tiny VR}/dI$} &
	\multicolumn{1}{c}{$I^{2.5}dN_\tnm{\tiny eff}^\tnm{\tiny VRE}/dI$} \\
	\multicolumn{1}{c}{(mJy)} &
	\multicolumn{1}{c}{(mJy)} & & & & &
	\multicolumn{1}{c}{(Jy$^{1.5}\,$sr$^{-1}$)} &
	\multicolumn{1}{c}{(Jy$^{1.5}\,$sr$^{-1}$)} &
	\multicolumn{1}{c}{(Jy$^{1.5}\,$sr$^{-1}$)} \\
	\multicolumn{1}{c}{(1)} &
	\multicolumn{1}{c}{(2)} &
	\multicolumn{1}{c}{(3)} &
	\multicolumn{1}{c}{(4)} &
	\multicolumn{1}{c}{(5)} &
	\multicolumn{1}{c}{(6)} &
	\multicolumn{1}{c}{(7)} &
	\multicolumn{1}{c}{(8)} &
	\multicolumn{1}{c}{(9)} \\
 \hline
$0.149-0.175$ & 0.161 & 90 & 286.2 & 289.1 & 206.9 & 4.32 & 4.36 & $3.12^{\ms +0.33}_{\ms -0.33}$ \\
$0.175-0.206$ & 0.190 & 117 & 229.5 & 254.0 & 189.2 & 4.41 & 4.88 & $3.63^{\ms +0.34}_{\ms -0.34}$ \\
$0.206-0.242$ & 0.223 & 116 & 179.9 & 221.6 & 173.8 & 4.40 & 5.42 & $4.25^{\ms +0.39}_{\ms -0.39}$ \\
$0.242-0.284$ & 0.262 & 116 & 156.0 & 206.7 & 173.6 & 4.86 & 6.44 & $5.41^{\ms +0.50}_{\ms -0.50}$ \\
$0.284-0.334$ & 0.308 & 76 & 90.6 & 126.8 & 111.9 & 3.59 & 5.03 & $4.44^{\ms +0.51}_{\ms -0.51}$ \\
$0.334-0.392$ & 0.361 & 81 & 88.9 & 129.1 & 119.3 & 4.49 & 6.53 & $6.03^{\ms +0.67}_{\ms -0.67}$ \\
$0.392-0.460$ & 0.425 & 68 & 70.9 & 101.0 & 95.7 & 4.56 & 6.50 & $6.16^{\ms +0.75}_{\ms -0.75}$ \\
$0.460-0.541$ & 0.499 & 64 & 64.9 & 90.0 & 86.9 & 5.32 & 7.37 & $7.13^{\ms +0.89}_{\ms -0.89}$ \\
$0.541-0.636$ & 0.586 & 50 & 50.3 & 67.6 & 66.0 & 5.25 & 7.06 & $6.89^{\ms +1.1}_{\ms -0.97}$ \\
$0.636-0.747$ & 0.689 & 43 & 44.5 & 57.7 & 56.7 & 5.92 & 7.67 & $7.54^{\ms +1.3}_{\ms -1.1}$ \\
$0.747-0.877$ & 0.809 & 36 & 36.0 & 45.2 & 44.6 & 6.10 & 7.65 & $7.56^{\ms +1.5}_{\ms -1.3}$ \\
$0.877-1.03$ & 0.951 & 31 & 32.0 & 38.9 & 38.6 & 6.90 & 8.38 & $8.32^{\ms +1.8}_{\ms -1.5}$ \\
$1.03-1.39$ & 1.20 & 34 & 34.9 & 40.7 & 40.6 & 5.71 & 6.67 & $6.64^{\ms +1.3}_{\ms -1.1}$ \\
$1.39-1.88$ & 1.62 & 39 & 40.1 & 44.9 & 44.8 & 10.3 & 11.5 & $11.5^{\ms +2.2}_{\ms -1.8}$ \\
$1.88-2.53$ & 2.18 & 41 & 41.0 & 44.4 & 44.4 & 16.5 & 17.9 & $17.9^{\ms +3.2}_{\ms -2.8}$ \\
$2.53-3.41$ & 2.94 & 25 & 25.1 & 26.5 & 26.5 & 15.8 & 16.7 & $16.7^{\ms +4.1}_{\ms -3.3}$ \\
$3.41-4.60$ & 3.97 & 23 & 23.4 & 24.3 & 24.3 & 23.1 & 24.0 & $24.0^{\ms +6.1}_{\ms -5.0}$ \\
$4.60-6.21$ & 5.36 & 25 & 25.2 & 25.8 & 25.8 & 39.0 & 40.1 & $40.0^{\ms +9.7}_{\ms -7.9}$ \\
$6.21-8.38$ & 7.22 & 18 & 18.0 & 18.3 & 18.3 & 43.7 & 44.5 & $44.5^{\ms +13}_{\ms -10}$ \\
$8.38-11.3$ & 9.74 & 15 & 15.2 & 15.4 & 15.4 & 58.0 & 58.7 & $58.6^{\ms +19}_{\ms -15}$ \\
$11.3-17.9$ & 14.3 & 10 & 10.0 & 10.1 & 10.1 & 43.7 & 44.0 & $44.0^{\ms +19}_{\ms -14}$ \\
$17.9-28.4$ & 22.6 & 13 & 13.0 & 13.0 & 13.1 & 113 & 114 & $114^{\ms +41}_{\ms -31}$ \\
$28.4-45.0$ & 35.8 & 4 & 4.0 & 4.0 & 4.0 & 69.4 & 69.5 & $69.5^{\ms +55}_{\ms -33}$ \\
$45.0-71.3$ & 56.7 & 4 & 4.0 & 4.0 & 4.0 & 138 & 138 & $138^{\ms +110}_{\ms -66}$ \\
$71.3-113$ & 89.9 & 2 & 2.0 & 2.0 & 2.0 & 138 & 138 & $138^{\ms +180}_{\ms -89}$ \\
$113-179$ & 142 & 2 & 2.0 & 2.0 & 2.0 & 274 & 274 & $274^{\ms +360}_{\ms -180}$ \\
 \hline
 \end{tabular}
\end{table*}

\begin{table*}
 \centering
 \caption{ATLAS 1.4~GHz DR2 linear polarization component-corrected counts for CDF-S field.
  	  See Appendix~A for column details.}
 \label{tbl:countsLCDFSie}
 \begin{tabular}{@{}ccccccccccc@{}}
 \hline
	\multicolumn{1}{c}{$\Delta L$} &
	\multicolumn{1}{c}{$L_\tnm{\tiny AV}$} &
	\multicolumn{1}{c}{$N_\tnm{\tiny raw}^\tnm{\tiny D}$} &
	\multicolumn{1}{c}{$N_\tnm{\tiny eff}^\tnm{\tiny DV}$} &
	\multicolumn{1}{c}{$N_\tnm{\tiny eff}^\tnm{\tiny DV-R}$} &
	\multicolumn{1}{c}{$N_\tnm{\tiny eff}^\tnm{\tiny DVR}$} &
	\multicolumn{1}{c}{$N_\tnm{\tiny eff}^\tnm{\tiny DV+R}$} &
	\multicolumn{1}{c}{$L^{2.5}dN_\tnm{\tiny eff}^\tnm{\tiny DV}/dL$} &
	\multicolumn{1}{c}{$L^{2.5}dN_\tnm{\tiny eff}^\tnm{\tiny DV-R}/dL$} &
	\multicolumn{1}{c}{$L^{2.5}dN_\tnm{\tiny eff}^\tnm{\tiny DVR}/dL$} &
	\multicolumn{1}{c}{$L^{2.5}dN_\tnm{\tiny eff}^\tnm{\tiny DV+R}/dL$} \\
	\multicolumn{1}{c}{(mJy)} &
	\multicolumn{1}{c}{(mJy)} & & & & & &
	\multicolumn{1}{c}{(Jy$^{1.5}\,$sr$^{-1}$)} &
	\multicolumn{1}{c}{(Jy$^{1.5}\,$sr$^{-1}$)} &
	\multicolumn{1}{c}{(Jy$^{1.5}\,$sr$^{-1}$)} &
	\multicolumn{1}{c}{(Jy$^{1.5}\,$sr$^{-1}$)} \\
	\multicolumn{1}{c}{(1)} &
	\multicolumn{1}{c}{(2)} &
	\multicolumn{1}{c}{(3)} &
	\multicolumn{1}{c}{(4)} &
	\multicolumn{1}{c}{(5)} &
	\multicolumn{1}{c}{(6)} &
	\multicolumn{1}{c}{(7)} &
	\multicolumn{1}{c}{(8)} &
	\multicolumn{1}{c}{(9)} &
	\multicolumn{1}{c}{(10)} &
	\multicolumn{1}{c}{(11)} \\
 \hline
$0.120-0.173$ & 0.144 & 18 & 40.1 & 39.9 & 42.1 & 44.6 & 0.170 & $0.170^{\ms +0.050}_{\ms -0.040}$ & $0.179^{\ms +0.053}_{\ms -0.042}$ & $0.189^{\ms +0.056}_{\ms -0.044}$ \\
$0.173-0.251$ & 0.209 & 18 & 23.8 & 29.3 & 31.9 & 34.3 & 0.175 & $0.216^{\ms +0.064}_{\ms -0.050}$ & $0.236^{\ms +0.070}_{\ms -0.055}$ & $0.253^{\ms +0.075}_{\ms -0.059}$ \\
$0.251-0.362$ & 0.302 & 24 & 26.6 & 35.0 & 38.3 & 41.2 & 0.341 & $0.449^{\ms +0.11}_{\ms -0.091}$ & $0.491^{\ms +0.12}_{\ms -0.100}$ & $0.529^{\ms +0.13}_{\ms -0.11}$ \\
$0.362-0.524$ & 0.436 & 13 & 13.5 & 17.1 & 18.2 & 19.2 & 0.302 & $0.383^{\ms +0.14}_{\ms -0.10}$ & $0.407^{\ms +0.15}_{\ms -0.11}$ & $0.428^{\ms +0.15}_{\ms -0.12}$ \\
$0.524-0.757$ & 0.631 & 15 & 15.8 & 18.9 & 19.6 & 20.3 & 0.613 & $0.734^{\ms +0.24}_{\ms -0.19}$ & $0.761^{\ms +0.25}_{\ms -0.19}$ & $0.788^{\ms +0.26}_{\ms -0.20}$ \\
$0.757-1.09$ & 0.912 & 7 & 7.0 & 8.0 & 8.2 & 8.4 & 0.475 & $0.541^{\ms +0.29}_{\ms -0.20}$ & $0.553^{\ms +0.30}_{\ms -0.20}$ & $0.568^{\ms +0.31}_{\ms -0.21}$ \\
$1.09-1.58$ & 1.32 & 6 & 6.0 & 6.6 & 6.7 & 6.8 & 0.702 & $0.768^{\ms +0.46}_{\ms -0.30}$ & $0.781^{\ms +0.47}_{\ms -0.31}$ & $0.798^{\ms +0.48}_{\ms -0.32}$ \\
$1.58-2.29$ & 1.90 & 5 & 5.0 & 5.3 & 5.4 & 5.5 & 1.02 & $1.08^{\ms +0.73}_{\ms -0.47}$ & $1.09^{\ms +0.74}_{\ms -0.47}$ & $1.11^{\ms +0.75}_{\ms -0.48}$ \\
$2.29-3.31$ & 2.75 & 2 & 2.0 & 2.1 & 2.1 & 2.1 & 0.706 & $0.735^{\ms +0.97}_{\ms -0.47}$ & $0.743^{\ms +0.98}_{\ms -0.48}$ & $0.755^{\ms +1.00}_{\ms -0.49}$ \\
$4.78-6.91$ & 5.75 & 1 & 1.0 & 1.0 & 1.0 & 1.0 & 1.06 & $1.08^{\ms +2.5}_{\ms -0.90}$ & $1.09^{\ms +2.5}_{\ms -0.90}$ & $1.10^{\ms +2.5}_{\ms -0.91}$ \\
$9.98-14.4$ & 12.0 & 2 & 2.0 & 2.0 & 2.0 & 2.0 & 6.42 & $6.48^{\ms +8.5}_{\ms -4.2}$ & $6.51^{\ms +8.6}_{\ms -4.2}$ & $6.57^{\ms +8.7}_{\ms -4.2}$ \\
$14.4-28.8$ & 20.3 & 1 & 1.0 & 1.0 & 1.0 & 1.0 & 3.71 & $3.72^{\ms +8.6}_{\ms -3.1}$ & $3.74^{\ms +8.6}_{\ms -3.1}$ & $3.77^{\ms +8.7}_{\ms -3.1}$ \\
$28.8-57.4$ & 40.4 & 1 & 1.0 & 1.0 & 1.0 & 1.0 & 10.4 & $10.4^{\ms +24}_{\ms -8.6}$ & $10.4^{\ms +24}_{\ms -8.6}$ & $10.5^{\ms +24}_{\ms -8.7}$ \\
 \hline
 \end{tabular}
\end{table*}

\begin{table*}
 \centering
 \caption{ATLAS 1.4~GHz DR2 linear polarization component-corrected counts for ELAIS-S1 field.
  	  See Appendix~A for column details.}
 \label{tbl:countsLELAISie}
 \begin{tabular}{@{}ccccccccccc@{}}
 \hline
	\multicolumn{1}{c}{$\Delta L$} &
	\multicolumn{1}{c}{$L_\tnm{\tiny AV}$} &
	\multicolumn{1}{c}{$N_\tnm{\tiny raw}^\tnm{\tiny D}$} &
	\multicolumn{1}{c}{$N_\tnm{\tiny eff}^\tnm{\tiny DV}$} &
	\multicolumn{1}{c}{$N_\tnm{\tiny eff}^\tnm{\tiny DV-R}$} &
	\multicolumn{1}{c}{$N_\tnm{\tiny eff}^\tnm{\tiny DVR}$} &
	\multicolumn{1}{c}{$N_\tnm{\tiny eff}^\tnm{\tiny DV+R}$} &
	\multicolumn{1}{c}{$L^{2.5}dN_\tnm{\tiny eff}^\tnm{\tiny DV}/dL$} &
	\multicolumn{1}{c}{$L^{2.5}dN_\tnm{\tiny eff}^\tnm{\tiny DV-R}/dL$} &
	\multicolumn{1}{c}{$L^{2.5}dN_\tnm{\tiny eff}^\tnm{\tiny DVR}/dL$} &
	\multicolumn{1}{c}{$L^{2.5}dN_\tnm{\tiny eff}^\tnm{\tiny DV+R}/dL$} \\
	\multicolumn{1}{c}{(mJy)} &
	\multicolumn{1}{c}{(mJy)} & & & & & &
	\multicolumn{1}{c}{(Jy$^{1.5}\,$sr$^{-1}$)} &
	\multicolumn{1}{c}{(Jy$^{1.5}\,$sr$^{-1}$)} &
	\multicolumn{1}{c}{(Jy$^{1.5}\,$sr$^{-1}$)} &
	\multicolumn{1}{c}{(Jy$^{1.5}\,$sr$^{-1}$)} \\
	\multicolumn{1}{c}{(1)} &
	\multicolumn{1}{c}{(2)} &
	\multicolumn{1}{c}{(3)} &
	\multicolumn{1}{c}{(4)} &
	\multicolumn{1}{c}{(5)} &
	\multicolumn{1}{c}{(6)} &
	\multicolumn{1}{c}{(7)} &
	\multicolumn{1}{c}{(8)} &
	\multicolumn{1}{c}{(9)} &
	\multicolumn{1}{c}{(10)} &
	\multicolumn{1}{c}{(11)} \\
 \hline
$0.135-0.195$ & 0.162 & 11 & 25.1 & 24.5 & 26.1 & 27.7 & 0.167 & $0.163^{\ms +0.065}_{\ms -0.048}$ & $0.173^{\ms +0.070}_{\ms -0.051}$ & $0.184^{\ms +0.074}_{\ms -0.055}$ \\
$0.195-0.282$ & 0.235 & 13 & 17.7 & 22.4 & 24.5 & 26.5 & 0.205 & $0.259^{\ms +0.093}_{\ms -0.071}$ & $0.284^{\ms +0.10}_{\ms -0.078}$ & $0.307^{\ms +0.11}_{\ms -0.084}$ \\
$0.282-0.408$ & 0.340 & 7 & 7.9 & 10.6 & 11.7 & 12.6 & 0.158 & $0.213^{\ms +0.11}_{\ms -0.079}$ & $0.234^{\ms +0.13}_{\ms -0.086}$ & $0.253^{\ms +0.14}_{\ms -0.093}$ \\
$0.408-0.589$ & 0.491 & 6 & 6.4 & 8.2 & 8.8 & 9.3 & 0.222 & $0.288^{\ms +0.17}_{\ms -0.11}$ & $0.307^{\ms +0.18}_{\ms -0.12}$ & $0.325^{\ms +0.19}_{\ms -0.13}$ \\
$0.589-0.852$ & 0.710 & 8 & 8.0 & 9.7 & 10.1 & 10.6 & 0.485 & $0.590^{\ms +0.29}_{\ms -0.20}$ & $0.615^{\ms +0.30}_{\ms -0.21}$ & $0.640^{\ms +0.32}_{\ms -0.22}$ \\
$0.852-1.23$ & 1.03 & 2 & 2.4 & 2.8 & 2.8 & 2.9 & 0.254 & $0.292^{\ms +0.38}_{\ms -0.19}$ & $0.300^{\ms +0.40}_{\ms -0.19}$ & $0.310^{\ms +0.41}_{\ms -0.20}$ \\
$1.23-1.78$ & 1.48 & 4 & 4.0 & 4.4 & 4.5 & 4.6 & 0.732 & $0.807^{\ms +0.64}_{\ms -0.39}$ & $0.824^{\ms +0.65}_{\ms -0.39}$ & $0.846^{\ms +0.67}_{\ms -0.40}$ \\
$1.78-2.57$ & 2.14 & 4 & 4.0 & 4.3 & 4.3 & 4.4 & 1.27 & $1.36^{\ms +1.1}_{\ms -0.65}$ & $1.38^{\ms +1.1}_{\ms -0.66}$ & $1.41^{\ms +1.1}_{\ms -0.68}$ \\
$2.57-3.72$ & 3.10 & 2 & 2.0 & 2.1 & 2.1 & 2.2 & 1.10 & $1.15^{\ms +1.5}_{\ms -0.75}$ & $1.17^{\ms +1.5}_{\ms -0.76}$ & $1.20^{\ms +1.6}_{\ms -0.77}$ \\
$3.72-5.37$ & 4.47 & 1 & 1.0 & 1.0 & 1.0 & 1.1 & 0.959 & $0.987^{\ms +2.3}_{\ms -0.82}$ & $1.000^{\ms +2.3}_{\ms -0.83}$ & $1.02^{\ms +2.3}_{\ms -0.84}$ \\
$5.37-7.77$ & 6.46 & 1 & 1.0 & 1.0 & 1.0 & 1.0 & 1.67 & $1.70^{\ms +3.9}_{\ms -1.4}$ & $1.71^{\ms +3.9}_{\ms -1.4}$ & $1.74^{\ms +4.0}_{\ms -1.4}$ \\
 \hline
 \end{tabular}
\end{table*}

\begin{table*}
 \centering
 \caption{ATLAS 1.4~GHz DR2 linear polarization bin-corrected counts for CDF-S field -- Part I of II.
  	  See Appendix~A for column details.}
 \label{tbl:countsLCDFS}
 \begin{tabular}{@{}ccccccccccc@{}}
 \hline
	\multicolumn{1}{c}{$\Delta L$} &
	\multicolumn{1}{c}{$L_\tnm{\tiny AV}$} &
	\multicolumn{1}{c}{$N_\tnm{\tiny raw}$} &
	\multicolumn{1}{c}{$N_\tnm{\tiny eff}^\tnm{\tiny V}$} &
	\multicolumn{1}{c}{$N_\tnm{\tiny eff}^\tnm{\tiny V-R}$} &
	\multicolumn{1}{c}{$N_\tnm{\tiny eff}^\tnm{\tiny VR}$} &
	\multicolumn{1}{c}{$N_\tnm{\tiny eff}^\tnm{\tiny V+R}$} &
	\multicolumn{1}{c}{$N_\tnm{\tiny eff}^\tnm{\tiny V-RE}$} &
	\multicolumn{1}{c}{$N_\tnm{\tiny eff}^\tnm{\tiny VRE}$} &
	\multicolumn{1}{c}{$N_\tnm{\tiny eff}^\tnm{\tiny V+RE}$} &
	\multicolumn{1}{c}{$L^{2.5}dN_\tnm{\tiny eff}^\tnm{\tiny V}/dL$} \\
	\multicolumn{1}{c}{(mJy)} &
	\multicolumn{1}{c}{(mJy)} & & & & & & & & &
	\multicolumn{1}{c}{(Jy$^{1.5}\,$sr$^{-1}$)} \\
	\multicolumn{1}{c}{(1)} &
	\multicolumn{1}{c}{(2)} &
	\multicolumn{1}{c}{(3)} &
	\multicolumn{1}{c}{(4)} &
	\multicolumn{1}{c}{(5)} &
	\multicolumn{1}{c}{(6)} &
	\multicolumn{1}{c}{(7)} &
	\multicolumn{1}{c}{(8)} &
	\multicolumn{1}{c}{(9)} &
	\multicolumn{1}{c}{(10)} &
	\multicolumn{1}{c}{(11)} \\
 \hline
$0.125-0.181$ & 0.150 & 18 & 40.1 & 40.2 & 42.3 & 44.6 & 39.4 & 41.5 & 43.7 & 0.181 \\
$0.181-0.261$ & 0.218 & 18 & 23.8 & 29.8 & 32.6 & 35.1 & 29.1 & 31.8 & 34.3 & 0.187 \\
$0.261-0.377$ & 0.315 & 24 & 26.6 & 34.9 & 38.0 & 40.8 & 34.4 & 37.5 & 40.3 & 0.363 \\
$0.377-0.546$ & 0.455 & 15 & 16.3 & 20.5 & 21.8 & 22.9 & 20.4 & 21.6 & 22.7 & 0.387 \\
$0.546-0.789$ & 0.657 & 14 & 14.0 & 16.7 & 17.3 & 17.9 & 16.6 & 17.2 & 17.8 & 0.578 \\
$0.789-1.14$ & 0.950 & 6 & 6.0 & 6.9 & 7.0 & 7.2 & 6.8 & 7.0 & 7.2 & 0.433 \\
$1.14-1.65$ & 1.37 & 6 & 6.0 & 6.5 & 6.6 & 6.8 & 6.5 & 6.6 & 6.8 & 0.747 \\
$1.65-2.38$ & 1.98 & 5 & 5.0 & 5.3 & 5.4 & 5.5 & 5.3 & 5.4 & 5.5 & 1.08 \\
$2.38-3.44$ & 2.87 & 2 & 2.0 & 2.1 & 2.1 & 2.1 & 2.1 & 2.1 & 2.1 & 0.751 \\
$4.98-7.19$ & 5.99 & 1 & 1.0 & 1.0 & 1.0 & 1.0 & 1.0 & 1.0 & 1.0 & 1.13 \\
$10.4-20.7$ & 14.7 & 3 & 3.0 & 3.0 & 3.0 & 3.1 & 3.0 & 3.0 & 3.1 & 6.83 \\
$41.4-82.6$ & 57.9 & 1 & 1.0 & 1.0 & 1.0 & 1.0 & 1.0 & 1.0 & 1.0 & 17.7 \\
 \hline
 \end{tabular}
\end{table*}

\begin{table*}
 \centering
 \contcaption{ATLAS 1.4~GHz DR2 linear polarization bin-corrected counts for CDF-S field -- Part II of II.
  	  See Appendix~A for column details.}
 \begin{tabular}{@{}cccccc@{}}
 \hline
	\multicolumn{1}{c}{$L^{2.5}dN_\tnm{\tiny eff}^\tnm{\tiny V-R}/dL$} &
	\multicolumn{1}{c}{$L^{2.5}dN_\tnm{\tiny eff}^\tnm{\tiny VR}/dL$} &
	\multicolumn{1}{c}{$L^{2.5}dN_\tnm{\tiny eff}^\tnm{\tiny V+R}/dL$} &
	\multicolumn{1}{c}{$L^{2.5}dN_\tnm{\tiny eff}^\tnm{\tiny V-RE}/dL$} &
	\multicolumn{1}{c}{$L^{2.5}dN_\tnm{\tiny eff}^\tnm{\tiny VRE}/dL$} &
	\multicolumn{1}{c}{$L^{2.5}dN_\tnm{\tiny eff}^\tnm{\tiny V+RE}/dL$} \\
	\multicolumn{1}{c}{(Jy$^{1.5}\,$sr$^{-1}$)} &
	\multicolumn{1}{c}{(Jy$^{1.5}\,$sr$^{-1}$)} &
	\multicolumn{1}{c}{(Jy$^{1.5}\,$sr$^{-1}$)} &
	\multicolumn{1}{c}{(Jy$^{1.5}\,$sr$^{-1}$)} &
	\multicolumn{1}{c}{(Jy$^{1.5}\,$sr$^{-1}$)} &
	\multicolumn{1}{c}{(Jy$^{1.5}\,$sr$^{-1}$)} \\
	\multicolumn{1}{c}{(12)} &
	\multicolumn{1}{c}{(13)} &
	\multicolumn{1}{c}{(14)} &
	\multicolumn{1}{c}{(15)} &
	\multicolumn{1}{c}{(16)} &
	\multicolumn{1}{c}{(17)} \\
 \hline
0.182 & 0.191 & 0.201 & $0.178^{\ms +0.053}_{\ms -0.042}$ & $0.187^{\ms +0.055}_{\ms -0.044}$ & $0.197^{\ms +0.058}_{\ms -0.046}$ \\
0.234 & 0.256 & 0.276 & $0.228^{\ms +0.068}_{\ms -0.053}$ & $0.250^{\ms +0.074}_{\ms -0.058}$ & $0.269^{\ms +0.080}_{\ms -0.063}$ \\
0.476 & 0.519 & 0.557 & $0.469^{\ms +0.12}_{\ms -0.095}$ & $0.512^{\ms +0.13}_{\ms -0.10}$ & $0.550^{\ms +0.14}_{\ms -0.11}$ \\
0.488 & 0.517 & 0.543 & $0.484^{\ms +0.16}_{\ms -0.12}$ & $0.513^{\ms +0.17}_{\ms -0.13}$ & $0.539^{\ms +0.18}_{\ms -0.14}$ \\
0.687 & 0.712 & 0.736 & $0.686^{\ms +0.24}_{\ms -0.18}$ & $0.710^{\ms +0.25}_{\ms -0.19}$ & $0.735^{\ms +0.25}_{\ms -0.19}$ \\
0.491 & 0.502 & 0.515 & $0.490^{\ms +0.29}_{\ms -0.19}$ & $0.501^{\ms +0.30}_{\ms -0.20}$ & $0.514^{\ms +0.31}_{\ms -0.20}$ \\
0.814 & 0.827 & 0.844 & $0.812^{\ms +0.49}_{\ms -0.32}$ & $0.825^{\ms +0.49}_{\ms -0.33}$ & $0.843^{\ms +0.50}_{\ms -0.33}$ \\
1.15 & 1.16 & 1.18 & $1.15^{\ms +0.78}_{\ms -0.50}$ & $1.16^{\ms +0.79}_{\ms -0.50}$ & $1.18^{\ms +0.80}_{\ms -0.51}$ \\
0.779 & 0.788 & 0.800 & $0.779^{\ms +1.0}_{\ms -0.50}$ & $0.787^{\ms +1.0}_{\ms -0.51}$ & $0.800^{\ms +1.1}_{\ms -0.52}$ \\
1.15 & 1.16 & 1.17 & $1.15^{\ms +2.6}_{\ms -0.95}$ & $1.16^{\ms +2.7}_{\ms -0.96}$ & $1.17^{\ms +2.7}_{\ms -0.97}$ \\
6.86 & 6.90 & 6.96 & $6.86^{\ms +6.7}_{\ms -3.7}$ & $6.89^{\ms +6.7}_{\ms -3.8}$ & $6.96^{\ms +6.8}_{\ms -3.8}$ \\
17.7 & 17.8 & 17.9 & $17.7^{\ms +41}_{\ms -15}$ & $17.8^{\ms +41}_{\ms -15}$ & $17.9^{\ms +41}_{\ms -15}$ \\
 \hline
 \end{tabular}
\end{table*}

\begin{table*}
 \centering
 \caption{ATLAS 1.4~GHz DR2 linear polarization bin-corrected counts for ELAIS-S1 field -- Part I of II.
  	  See Appendix~A for column details.}
 \label{tbl:countsLELAIS}
 \begin{tabular}{@{}ccccccccccc@{}}
 \hline
	\multicolumn{1}{c}{$\Delta L$} &
	\multicolumn{1}{c}{$L_\tnm{\tiny AV}$} &
	\multicolumn{1}{c}{$N_\tnm{\tiny raw}$} &
	\multicolumn{1}{c}{$N_\tnm{\tiny eff}^\tnm{\tiny V}$} &
	\multicolumn{1}{c}{$N_\tnm{\tiny eff}^\tnm{\tiny V-R}$} &
	\multicolumn{1}{c}{$N_\tnm{\tiny eff}^\tnm{\tiny VR}$} &
	\multicolumn{1}{c}{$N_\tnm{\tiny eff}^\tnm{\tiny V+R}$} &
	\multicolumn{1}{c}{$N_\tnm{\tiny eff}^\tnm{\tiny V-RE}$} &
	\multicolumn{1}{c}{$N_\tnm{\tiny eff}^\tnm{\tiny VRE}$} &
	\multicolumn{1}{c}{$N_\tnm{\tiny eff}^\tnm{\tiny V+RE}$} &
	\multicolumn{1}{c}{$L^{2.5}dN_\tnm{\tiny eff}^\tnm{\tiny V}/dL$} \\
	\multicolumn{1}{c}{(mJy)} &
	\multicolumn{1}{c}{(mJy)} & & & & & & & & &
	\multicolumn{1}{c}{(Jy$^{1.5}\,$sr$^{-1}$)} \\
	\multicolumn{1}{c}{(1)} &
	\multicolumn{1}{c}{(2)} &
	\multicolumn{1}{c}{(3)} &
	\multicolumn{1}{c}{(4)} &
	\multicolumn{1}{c}{(5)} &
	\multicolumn{1}{c}{(6)} &
	\multicolumn{1}{c}{(7)} &
	\multicolumn{1}{c}{(8)} &
	\multicolumn{1}{c}{(9)} &
	\multicolumn{1}{c}{(10)} &
	\multicolumn{1}{c}{(11)} \\
 \hline
$0.145-0.210$ & 0.175 & 11 & 25.1 & 25.8 & 27.4 & 29.0 & 25.5 & 27.0 & 28.7 & 0.186 \\
$0.210-0.303$ & 0.252 & 14 & 19.0 & 24.7 & 27.2 & 29.5 & 24.1 & 26.6 & 28.8 & 0.244 \\
$0.303-0.438$ & 0.365 & 7 & 7.7 & 10.3 & 11.2 & 12.1 & 10.1 & 11.1 & 11.9 & 0.172 \\
$0.438-0.633$ & 0.527 & 6 & 6.3 & 8.1 & 8.6 & 9.1 & 8.0 & 8.6 & 9.0 & 0.246 \\
$0.633-0.915$ & 0.762 & 7 & 7.0 & 8.4 & 8.7 & 9.1 & 8.4 & 8.7 & 9.0 & 0.473 \\
$0.915-1.32$ & 1.10 & 2 & 2.4 & 2.7 & 2.8 & 2.9 & 2.7 & 2.8 & 2.9 & 0.282 \\
$1.32-1.91$ & 1.59 & 5 & 5.0 & 5.5 & 5.6 & 5.7 & 5.5 & 5.6 & 5.7 & 1.02 \\
$1.91-2.76$ & 2.30 & 4 & 4.0 & 4.3 & 4.3 & 4.4 & 4.2 & 4.3 & 4.4 & 1.42 \\
$2.76-3.99$ & 3.32 & 2 & 2.0 & 2.1 & 2.1 & 2.2 & 2.1 & 2.1 & 2.2 & 1.23 \\
$3.99-5.77$ & 4.80 & 1 & 1.0 & 1.0 & 1.0 & 1.1 & 1.0 & 1.0 & 1.1 & 1.07 \\
 \hline
 \end{tabular}
\end{table*}

\begin{table*}
 \centering
 \contcaption{ATLAS 1.4~GHz DR2 linear polarization bin-corrected counts for ELAIS-S1 field -- Part II of II.
  	  See Appendix~A for column details.}
 \begin{tabular}{@{}ccccccc@{}}
 \hline
	\multicolumn{1}{c}{$L^{2.5}dN_\tnm{\tiny eff}^\tnm{\tiny V-R}/dL$} &
	\multicolumn{1}{c}{$L^{2.5}dN_\tnm{\tiny eff}^\tnm{\tiny VR}/dL$} &
	\multicolumn{1}{c}{$L^{2.5}dN_\tnm{\tiny eff}^\tnm{\tiny V+R}/dL$} &
	\multicolumn{1}{c}{$L^{2.5}dN_\tnm{\tiny eff}^\tnm{\tiny V-RE}/dL$} &
	\multicolumn{1}{c}{$L^{2.5}dN_\tnm{\tiny eff}^\tnm{\tiny VRE}/dL$} &
	\multicolumn{1}{c}{$L^{2.5}dN_\tnm{\tiny eff}^\tnm{\tiny V+RE}/dL$} \\
	\multicolumn{1}{c}{(Jy$^{1.5}\,$sr$^{-1}$)} &
	\multicolumn{1}{c}{(Jy$^{1.5}\,$sr$^{-1}$)} &
	\multicolumn{1}{c}{(Jy$^{1.5}\,$sr$^{-1}$)} &
	\multicolumn{1}{c}{(Jy$^{1.5}\,$sr$^{-1}$)} &
	\multicolumn{1}{c}{(Jy$^{1.5}\,$sr$^{-1}$)} &
	\multicolumn{1}{c}{(Jy$^{1.5}\,$sr$^{-1}$)} \\
	\multicolumn{1}{c}{(12)} &
	\multicolumn{1}{c}{(13)} &
	\multicolumn{1}{c}{(14)} &
	\multicolumn{1}{c}{(15)} &
	\multicolumn{1}{c}{(16)} &
	\multicolumn{1}{c}{(17)} \\
 \hline
0.191 & 0.202 & 0.215 & $0.189^{\ms +0.076}_{\ms -0.056}$ & $0.200^{\ms +0.080}_{\ms -0.059}$ & $0.212^{\ms +0.085}_{\ms -0.063}$ \\
0.317 & 0.350 & 0.379 & $0.310^{\ms +0.11}_{\ms -0.082}$ & $0.342^{\ms +0.12}_{\ms -0.090}$ & $0.371^{\ms +0.13}_{\ms -0.098}$ \\
0.230 & 0.251 & 0.270 & $0.227^{\ms +0.12}_{\ms -0.084}$ & $0.248^{\ms +0.13}_{\ms -0.091}$ & $0.266^{\ms +0.14}_{\ms -0.098}$ \\
0.315 & 0.334 & 0.353 & $0.313^{\ms +0.19}_{\ms -0.12}$ & $0.332^{\ms +0.20}_{\ms -0.13}$ & $0.350^{\ms +0.21}_{\ms -0.14}$ \\
0.568 & 0.590 & 0.613 & $0.566^{\ms +0.30}_{\ms -0.21}$ & $0.588^{\ms +0.32}_{\ms -0.22}$ & $0.611^{\ms +0.33}_{\ms -0.23}$ \\
0.322 & 0.330 & 0.340 & $0.321^{\ms +0.42}_{\ms -0.21}$ & $0.329^{\ms +0.43}_{\ms -0.21}$ & $0.339^{\ms +0.45}_{\ms -0.22}$ \\
1.12 & 1.14 & 1.17 & $1.11^{\ms +0.75}_{\ms -0.48}$ & $1.14^{\ms +0.77}_{\ms -0.49}$ & $1.17^{\ms +0.79}_{\ms -0.50}$ \\
1.50 & 1.53 & 1.56 & $1.50^{\ms +1.2}_{\ms -0.72}$ & $1.53^{\ms +1.2}_{\ms -0.73}$ & $1.56^{\ms +1.2}_{\ms -0.75}$ \\
1.28 & 1.30 & 1.32 & $1.28^{\ms +1.7}_{\ms -0.83}$ & $1.30^{\ms +1.7}_{\ms -0.84}$ & $1.33^{\ms +1.7}_{\ms -0.86}$ \\
1.10 & 1.11 & 1.13 & $1.10^{\ms +2.5}_{\ms -0.91}$ & $1.11^{\ms +2.6}_{\ms -0.92}$ & $1.13^{\ms +2.6}_{\ms -0.93}$ \\
 \hline
 \end{tabular}
\end{table*}

\bsp

\label{lastpage}

\end{document}